\documentclass[12pt,english]{article}
\pdfoutput=1

\usepackage[T1]{fontenc}
\usepackage[utf8]{inputenc}
\usepackage{lmodern, amsmath,amssymb,extarrows, graphicx, pifont, adjustbox, bm, xcolor}
\usepackage{amsfonts}
\usepackage{geometry}
\usepackage{comment,mdframed,ulem}
\usepackage{mathtools}
\usepackage{float}
\usepackage{slashed}
\usepackage{cite}
\usepackage{ragged2e}
\usepackage{etoolbox}
\usepackage{array}
\usepackage{enumitem}
\usepackage{cancel}
\usepackage{braket}

\usepackage{tikz-feynman} 
\tikzfeynmanset{
every edge/.style={thick},
graviton/.style={decorate, decoration={snake, amplitude=.4mm, segment length=1.5mm, pre length=.5mm, post length=.5mm}, double}
} 
\usepackage{simplewick}   
\usepackage{subcaption} 

\apptocmd{\thebibliography}{\justifying\setlength{\leftskip}{7.4mm}}{}{}
\makeatletter\g@addto@macro\bfseries{\boldmath}\makeatother

\usepackage{stackengine}
\usepackage{esint}
\usepackage[unicode=true,pdfusetitle,
 bookmarks=true,bookmarksnumbered=false,bookmarksopen=false,
 breaklinks=false,pdfborder={0 0 1},backref=false,colorlinks=true]
 {hyperref}
\hypersetup{pdfauthor={Clifford Cheung},
 citecolor=black,linkcolor=black,urlcolor=black}

\newcommand{\appendixref}[1]{\hyperref[#1]{appendix~\ref{#1}}}
\def\equationautorefname~#1\null{eq.\,(#1)\null}
\usepackage{breakurl}
\usepackage[hang,flushmargin]{footmisc} 
\allowdisplaybreaks

\usepackage{changepage}

\newcommand{\be}{\begin{equation}}
\newcommand{\ee}{\end{equation}}
\newcommand{\bea}{\begin{eqnarray}}
\newcommand{\eea}{\end{eqnarray}}

\newcommand{\eq}[2]{\be\begin{aligned}#1 \label{#2}\end{aligned}\ee}

\newcommand{\Eq}[1]{Eq.~(\ref{#1})}
\newcommand{\Eqs}[2]{Eqs.~(\ref{#1}) and (\ref{#2})}
\newcommand{\Sec}[1]{Sec.~\ref{#1}}

\newcommand{\App}[1]{App.~\ref{#1}}

\newcommand{\OO}{\mathcal{O}}

\newcommand{\UU}{U}    
\newcommand{\z}{z}     
\newcommand{\D}{{\cal D} }
\newcommand{\A}{{\cal A} }
\usepackage[scr=boondoxo,scrscaled=1.05]{mathalfa}
\newcommand{\hh}{{\mathscr h} }
\newcommand{\GG}{{\mathscr C} }
\newcommand{\dressedmetric}{{\mathcal G} }

\newcolumntype{P}[1]{>{\centering\arraybackslash}p{#1}}

\begin{document}

\interfootnotelinepenalty=10000
\baselineskip=18pt
\hfill CALT-TH 2026-021

\vspace{2cm}
\thispagestyle{empty}
\begin{center}
{\LARGE \bf
On Perturbatively Dressed Observables}\\
\bigskip\vspace{1cm}
\begin{center}{\large Clifford Cheung${}^{a}$, Allic Sivaramakrishnan${}^{a,b}$, \smallskip \\ Jordan Wilson-Gerow${}^{c}$,  Lihang Zhou${}^{a}$}\end{center}
{
\it ${}^a$Walter Burke Institute for Theoretical Physics and \\[-1mm]
Leinweber Forum for Theoretical Physics, \\[-1mm]
California Institute of Technology, Pasadena, CA 91125 \\[1.5mm]
${}^b$Department of Physics, Yale University, New Haven, CT 06511 \\
${}^c$Department of Physics, Carnegie Mellon University, Pittsburgh, PA 15213
}

 \end{center}

\bigskip
\centerline{\large\bf Abstract}
\begin{quote} \small

A central lesson of gravity is that local observables are ill-defined.  Coordinates themselves are a redundancy of description, so any particular point in spacetime is only meaningful once defined relationally by clocks, rulers, or asymptotic data. Despite extensive formal work on this subject, explicit calculations of the resulting gravitationally-dressed observables are more scarce. In this paper we perturbatively compute dressed matrix elements of local operators in electrodynamics and general relativity, including both potential and radiative photons and gravitons. Our expressions indicate that dressing is not ornamental: it universally induces kinematic singularities that can substantively reshape observables. We further show how dressing is mathematically equivalent to gauge fixing, as demonstrated by a dynamical temporal gauge in which the gauge-fixing vector is itself a geodesic fluid.

\end{quote}
	
\setcounter{footnote}{0}

\setcounter{tocdepth}{3}
\newpage
\tableofcontents

\newpage

\section{Introduction}

It is often said that gravity forbids observations that are arbitrarily localized in space and time.  This slogan is often paired with an evocative thought experiment: a hapless lab technician exhausts tremendous resources pinpointing the position of an object to sub-Planckian accuracy.  Inevitably, the  uncertainty principle intervenes, boomeranging  their hard-earned precision into a fireball of energy.  The technician and their entire apparatus collapse into a black hole, eternally hindering all present and future plans.  

This story may be vivid but it is also misleading.  In particular it suggests a problem with local operators that is buried deep within the nonperturbative regime of gravity.  This is not actually the case.  In fact, any operator located at a place or collection of places will suffer a far less dramatic complication.  Indeed, any spacetime point ``$x$'' is meaningless except in relation to an established set of coordinates.  Hence, all local operators transform actively under changes of coordinates, or diffeomorphisms.  For example a scalar field transforms as
\eq{
\phi(x) \rightarrow \phi(f(x)).
}{}
So, a local operator in gravity is no different in spirit than a nonsinglet operator in gauge theory. The label ``$x$'' is a free, uncontracted index.  As a corollary, it is broadly accepted that the only physical observables in gravity correspond to asymptotic quantities, which naturally reside at infinity where coordinate transformations do not act.

While this appeal to coordinate dependence is technically correct, it is also suspiciously overreaching.  We are told to discard an infinite class of naively sensible observations simply because they require coordinates to be defined.
So the fact that the  lab technician  of our parable must choose between Cartesian or polar coordinates now serves as a justification for forbidding local measurements altogether.  Is coordinate dependence a mere technical nuisance, or a true obstruction to what can be experimentally observed in principle?

The answer is the latter.  In fact, the problem of local observables is starkly unavoidable once we include the effects of quantum mechanics and gravity together.   The rules of quantum theory stipulate that physical fields are described by operators.  Furthermore, any operator is defined by its matrix elements on sets of physical states---including gravitons.   It is easy to show the matrix elements of local operators are nonsensical even in perturbation theory.

Case in point, consider the matrix elements of a scalar field operator $\phi(x)$.  While $\phi(x)$ is not diffeomorphism invariant, its vacuum expectation value, $\langle 0 | \phi(x) |0\rangle$, is safely constant on account of translation invariance.  To diagnose the problem we must go a bit deeper by computing the leading matrix element that is actually sensitive to dynamical gravity.  For simplicity, we work in an effective field theory of gravity describing gravitons propagating and interacting nonlinearly about a flat space background.  Transforming from position space to momentum space, $\phi(p) = \int d^D x \, e^{ipx} \phi(x)$, we obtain
\eq{
\langle 0 | \phi(p_1) |\phi(p_2) h(p_3)\rangle = - \frac{ \epsilon_{3\mu\nu}  p_1^\mu  p_2^\nu}{p_1^2}  \delta^{(D)}(p_1+p_2+p_3), 
}{single_field_insertion}
which is the tree-level matrix element  between the vacuum and a physical state comprised of an external on-shell scalar particle and graviton. 

The expression in \Eq{single_field_insertion} is inconsistent.  Lorentz invariance of the asymptotic states requires that all physical observables are invariant under diffeomorphisms of the graviton polarization,
$\epsilon_{3\mu\nu} \rightarrow \epsilon_{3\mu\nu} + p_{3\mu} q _\nu +  p_{3\nu}q _\mu$.
However, it is straightforward to see that  $\langle 0 | \phi(p_1) |\phi(p_2) h(p_3)\rangle \rightarrow \langle 0 | \phi(p_1) |\phi(p_2) h(p_3)\rangle- p_1^{\mu}q_{\mu}$ is {\it not invariant},
thus indicating the commingling of unphysical gravitational modes.  Consequently, the matrix elements of the scalar field operator, and thus the scalar field itself, are ill-defined even in perturbation theory.  In fact, the matrix element in \Eq{single_field_insertion} is only consistent in the $p_1=0$ soft limit, corresponding to the highly nonlocal zero mode $\int d^Dx  \, \phi(x)$.

Of course, these formal manipulations have no bearing whatsoever on the fact that actual experiments like LIGO \cite{advancedligo}
 and precision tabletop tests all perform measurements that have directly probed gravity, all the while being located somewhere other than asymptotic infinity.  At a conceptual level, the resolution to this  puzzle has been known for decades \cite{Witten1963GravitationAI}: seemingly local measurements are in truth nonlocal because they are defined {\it relationally} with respect to some other landmark that is itself tethered to asymptotic infinity. Intuitively, relational observables are augmented versions of naive local observables that properly encode the fact that measurements are synchronized to clocks and rulers which themselves interact dynamically with gravity. This solution is broadly known as ``gravitational dressing'', and is a generalization of the Wilson line dressing familiar from gauge theory.  In fact, many modern topics in quantum gravity, including holographic bulk reconstruction  \cite{Lewkowycz:2016ukf, Giddings:2018umg}, black hole information \cite{Chandrasekaran:2022eqq, DeVuyst:2024fxc, Herderschee:2022ntr}, and cosmological observables \cite{Ankur:2026ylr, Chandrasekaran:2022cip, Chen:2024rpx, Maldacena:2024spf, Abdalla:2025gzn, Harlow:2025pvj}, intersect with the study of gravitationally-dressed observables. But defining dressed observables, let alone computing them, remains challenging even in linearized quantum gravity \cite{Giddings:2025xym}.

The very simplest way to dress a local observable is to define a dynamical worldline degree of freedom whose internal ruler and clock offer an invariant notion of where and when.  Concretely, consider a worldline parameterized by $z^\mu(\tau)$ where $\tau$ is the proper time.  Under diffeomorphisms, this worldline coordinate transforms as
\eq{
z^\mu(\tau) \rightarrow f^{-1\mu}(z(\tau)),
}{}
which in turn implies that the dressed scalar operator,
\eq{
\phi(z(\tau)) \rightarrow  \phi(z(\tau)),
}{}
is completely diffeomorphism invariant.  So too are the $n$-point correlators of dressed scalars,
\eq{
\langle 0| \phi(z_1(\tau_1)) \phi(z_2(\tau_2))  \cdots \phi(z_n(\tau_n))  |0\rangle.
}{dressed_correlator}
Instead of bookkeeping a discrete inventory of dynamical worldlines, we will find it far simpler to consider a continuous family of worldlines parameterized by a fluid velocity field $u^\mu(x)$ \cite{Brown:1994py}. Physically, this fluid defines the local velocity of a worldline at a point, 
 \eq{
 \frac{dz^\mu (\tau) }{d\tau} = u^\mu(z(\tau)),
 }{}
and for geodesic worldlines it satisfies the fluid equation of motion, 
\eq{
u^\nu(x) \nabla_\nu u^\mu(x) =0.
}{}
We will dress our local observables with integral curves defined by this fluid.   Note that more general dressings can be considered if we impart the fluid with richer dynamics. While the present work is restricted to a pressureless dust, one could, alternatively, consider a fluid satisfying the Navier-Stokes equations, or realized by a collection of relativistic scalar fields.

The formal properties of \Eq{dressed_correlator} have been studied extensively in numerous works.  However, explicit mathematical expressions for  dressed observables are, for whatever reason, much more rarely calculated. Here notable exceptions are the seminal works of ~\cite{Giddings:2005id, Donnelly:2015hta, Donnelly:2016rvo} and ~\cite{Frob:2017apy, Frob:2021mpb, Frob:2017lnt, Frob:2021ore, Frob:2023awn, Frob:2023gng, Frob:2023vay, Frob:2022ciq, Dittrich:2006ee, Dittrich:2007jx}, which incorporated the effects of clock fields, gravitational dressing, and even graviton loops.  In this paper, we contribute to this program by calculating and studying the dressed matrix elements of local operators in flat space. For ease of nomenclature, we will use the term ``dressed correlators'' as a shorthand for these quantities, including both vacuum expectation values and matrix elements. By analyzing our resulting expressions, we confirm several broad takeaways:

\begin{itemize}[leftmargin=1cm, rightmargin=1cm]

\item {\it ``Dressing is gauge fixing.''}   In a theory of dynamical gravity, one can always choose a coordinate frame in which the fluid velocity is static and the corresponding worldlines are static.  In this approach, the entire effect of dressing is to gauge fix the graviton.  This elegant correspondence is explicitly manifested in our perturbative calculations.  In particular, our dressed correlators are exactly equal to bare correlators computed in a {\it dynamical temporal gauge} in which the temporal gauge-fixing parameter evolves geodesically in spacetime. 

\item {\it ``Dressing is not on the side.''} The very nomenclature of gravitational dressing suggests that it is nothing more than ornament on naive local observables.  This is false because dressing worldlines have to be dynamical in order to properly compensate diffeomorphisms.  As active degrees of freedom, these worldlines necessarily couple to gravitons.  Consequently, there exist dressed correlators for which an appropriately chosen external graviton pushes one of those worldlines on resonance, generating a physical kinematic singularity.  At a technical level, this is manifested by nonanalyticities at the locations of the worldlines.  We argue that these singularities are unavoidable and generically induce huge corrections to the matrix elements of local operators.

\end{itemize}

\noindent In \Sec{sec:gaugetheory}, we warm up with a discussion of dressing in the context of scalar quantum electrodynamics (QED).  We compute several dressed observables at tree  and loop level, verifying the equivalence between gauge dressing and gauge fixing.  Afterwards, we then move on to the case of gravitationally interacting scalars in \Sec{sec:gravity}.  For the convenience of the reader, we summarize our mathematical conventions in \App{app:conventions}, give all the integral identities we use at tree level in \App{app:tree_fourier} and at one loop in \App{app:one_loop}, and then unpack the dynamical temporal gauge condition in \App{app:Gammauu_vs_uh}.

\section{Gauge Theory}\label{sec:gaugetheory}

As a warm-up, let us consider the case of gauge theory. To be concrete, our analysis will focus on the case of massless scalar QED, whose Lagrangian is
\eq{
\mathcal{L}
&= -\frac{1}{4} F_{\mu\nu}F^{\mu\nu}
  + (D_\mu\phi)^\dagger(D^\mu\phi) ,  
}{L_SQED}
where $D_\mu  = \partial_\mu + i g A_\mu $. From this Lagrangian we derive the interaction vertices,
\begin{equation}
\vcenter{\hbox{
    \begin{tikzpicture}
    \tikzfeynmanset{
    every edge/.style={thick}
    }
    \begin{feynman}
      \vertex[] (x1) {}; 
      \vertex[right=2.6 of x1] (x2) {};
      \vertex (m) at ( $(x1)!0.5!(x2)$ );
      \vertex[below=0.9 of m] (f) {\(\mu\)};

      \diagram*{
        (x1) --[fermion, reversed momentum={[arrow shorten=0.2]\(p_1\)}] (m) --[fermion, momentum={[arrow shorten=0.2]\(p_2\)}](x2),
        (f) --[photon] (m), 
      };
    \end{feynman}
    \end{tikzpicture}}} = -ig(p_2^{\mu}-p_{1}^{\mu})\,, \qquad\qquad\qquad 
    \vcenter{\hbox{
    \begin{tikzpicture}
    \tikzfeynmanset{
    every edge/.style={thick}
    }
    \begin{feynman}
      \vertex (x1) ; 
      \vertex[right=2.6 of x1] (x2) ;
      \vertex (m) at ( $(x1)!0.5!(x2)$ );
      \vertex (f1) at ($(m)+(-1,-1)$) {\(\mu_1\)};
      \vertex (f2) at ($(m)+(1,-1)$) {\(\mu_2\)};

      \diagram*{
        (x1) --[fermion] (m) --[fermion](x2),
        (f1) --[photon] (m), 
        (f2) --[photon] (m) , 
      };
    \end{feynman}
    \end{tikzpicture}}} = 2 i g^2 \eta^{\mu_1\mu_2}\,.
    \label{eq:QED-vertices}
\end{equation}
Here the arrows denote the flow of particle number, which exits $\phi^\dagger$ and enters $\phi$. In what follows we will introduce gauge-invariant ``dressed'' operators in scalar QED, and discuss their relationship to gauge fixing.  We then perform a number of perturbative calculations involving these operators. Our plan of attack will be to first compute bulk correlators of charged operators, which will obviously not be gauge invariant.\footnote{We use the word “bulk” to emphasize that the operators are inserted at generic points in spacetime, rather than on the asymptotic boundary, as they would be in scattering amplitudes or AdS boundary correlators.}  Afterwards, we incorporate gauge dressing in order to see very precisely how the new contributions restore gauge invariance, even in perturbation theory.

\subsection{Dressed Operators}

Our implementation of dressing requires the existence of auxiliary worldlines.  These objects parallel transport the gauge charge of bulk operators to the boundary.  In anticipation of the gravitational case, we can view these as probes that measure the relative locations of fields in the internal manifold spanned by the gauge symmetry, relative to infinity.
Naturally, we must foliate all of spacetime with such worldlines to consider bulk insertions at any point.  To characterize this ensemble of worldlines, define a fluid velocity field $u^\mu(x)$ which satisfies the geodesic equation of motion,
\eq{
  u^\nu(x)\partial_\nu u^\mu(x)=0.
}{fluid_geodesic_eq} 
As noted earlier, this field describes the local velocity of any given worldline trajectory,
\eq{
\frac{d\z^\mu(\tau)}{d\tau} =  u^\mu(\z(\tau)),
}{}
which defines the integral curves $\z^\mu(\tau)$.  Note that \Eq{fluid_geodesic_eq} has solutions that have nonzero velocity.  However, since the acceleration is vanishing,
\eq{
\frac{d^2\z^\mu (\tau)}{d\tau^2}= 0 ,
}{}
all integral curves correspond to straight-line trajectories.

\subsubsection{Dressed Scalars}

Under a gauge transformation, a charged scalar shifts perturbatively by
\eq{
\delta \phi(x) = ig \Lambda(x) \phi(x).
}{}
It is then a textbook exercise to construct a dressed scalar $\Phi(x)$, which is a nonlocal composite of the gauge covariant scalar $\phi(x)$ and the gauge field $A_\mu(x)$.  By design this object will be gauge invariant, so $\delta \Phi(x)=0$.  As usual, this dressed scalar is
\eq{
\Phi(x) &= W(x)\phi(x),
}{gauge_dressed_scalar}
where we have defined the Wilson line,
\eq{
W(x)&=   \exp\left[ig\int_{-\infty}^{0}ds\, u^\mu(x) A_\mu(x+su(x))\right].
}{Wilson_line}
This object originates at infinity and terminates at the point $x$ with constant velocity $u(x)$ along the curve.  A gauge transformation shifts the log of the Wilson line by
\eq{
\delta \log W(x) =  -ig\int_{-\infty}^{0}ds\, u^\mu(x) \partial_\mu\Lambda(x+su(x))
  =-ig\int_{-\infty}^{0}ds\, \frac{d\Lambda(x+su(x))}{ds}=-ig\Lambda(x),
}
{logW_variation}
where we assumed that the $\Lambda(x)$ vanishes at infinity. The gauge variation is $\delta \Phi(x) = \delta W(x) \phi(x) + W(x) \delta \phi(x)=0$, so  the dressed field is gauge invariant.

In QED, the bulk coordinate $x$ is perfectly well defined and gauge invariant. Thus, we are allowed to describe the act of dressing in either direction: by firing a Wilson line inward from infinity to intersect the fixed point $x$, or equivalently by starting at $x$ and extending the Wilson line back out to infinity. Since the choice of curve is arbitrary, we will use straight lines, both for simplicity and to parallel the gravitational construction.

For later use in perturbative calculations, we write down the expression for $\Phi(x)$ with the leading-order dressing correction,
\eq{
\Phi(x)
&= \phi(x) \left( 1 + ig\int_{-\infty}^{0}ds\, u^\mu(x) A_\mu(x+su(x)) + \cdots \right)\\
&= \phi(x) \left(1 + \int_{k} e^{-ikx} A_\mu(k) \frac{-g u^\mu(x)}{k\!\cdot\!u(x)} + \cdots \right).
}{}
The convergence of the integral over $s$ required an $i\varepsilon$ prescription in the Fourier phase. Since the Wilson line originates in the past, the prescription is $e^{-ikus}\to e^{-ikus+\varepsilon s}$.  Therefore the denominator\footnote{All momentum-dependent denominators will be implicitly understood as having a $+i\varepsilon$ added unless otherwise specified. For example, $(p^2)^{-1}\rightarrow (p^2+i\varepsilon)^{-1}$ and $(\pm k\!\cdot\!u)^{-1}\rightarrow (\pm k\!\cdot\! u+i\varepsilon)^{-1}$, as defined above.} should be understood as $(k\!\cdot\!u + i\varepsilon)^{-1}$.  Note that this is precisely an eikonal propagator, which reflects the fact that the worldline describes the propagation of a hard charged particle from past infinity. The $i\varepsilon$ prescription here is part of specifying the desired Wilson line, but in our calculations is distinct from any $i\varepsilon$ prescription associated with $A_\mu$. 

The $\Phi^\dagger$ insertion is the same except with a flip of the charge $g$. These can be represented by the following Feynman vertices for worldline single-photon emission/absorption,
\begin{equation}
\vcenter{\hbox{
    \begin{tikzpicture}
    \tikzfeynmanset{
    every edge/.style={thick}
    }
    \begin{feynman}
      \vertex[dot, label=90:\(x\)] (x1) {}; 
      \vertex[right=1.3 of x1] (x2) {};
      \vertex[below=2.6 of x1] (w1) {\(u\)};
      \vertex[below=2.6 of x2] (w2) {};
      \vertex (g1) at ( $(x1)!0.5!(w1)$ );
      \vertex (g2) at ( $(x2)!0.5!(w2)$ );
    
      \diagram*{
        (x2) --[fermion, momentum={[arrow shorten=0.3] \(p\)}] (x1),
        (x1) --[scalar, blue] (w1),
        (g2)  --[photon, momentum={[arrow shorten=0.3] \(k,\mu\)}] (g1),
      };
    \end{feynman}
    \end{tikzpicture}}} = \int_{p,k}e^{-ipx-ikx}\frac{-g u^{\mu}}{k\!\cdot\!u}\,,
    \qquad\qquad
\vcenter{\hbox{
    \begin{tikzpicture}
    \tikzfeynmanset{
    every edge/.style={thick}
    }
    \begin{feynman}
      \vertex[dot, label=90:\(x\)] (x1) {}; 
      \vertex[right=1.3 of x1] (x2) {};
      \vertex[below=2.6 of x1] (w1) {\(u\)};
      \vertex[below=2.6 of x2] (w2) {};
      \vertex (g1) at ( $(x1)!0.5!(w1)$ );
      \vertex (g2) at ( $(x2)!0.5!(w2)$ );
    
      \diagram*{
        (x1) --[fermion, rmomentum'={[arrow shorten=0.3] \(p\)}] (x2),
        (x1) --[scalar, blue] (w1),
        (g2)  --[photon, momentum={[arrow shorten=0.3] \(k,\mu\)}] (g1),
      };
    \end{feynman}
    \end{tikzpicture}}} = \int_{p,k}e^{-ipx-ikx}\frac{g u^{\mu}}{k\!\cdot\!u}\,.
    \label{eq:WL-photon-feynman}
\end{equation}
Note that reversing any momentum arrow introduces a minus sign in the corresponding momentum, according to the convention in \Eq{eq:momentum-arrow}.

\subsubsection{Dressed Gauge Bosons}

In parallel with the scalar, it will be convenient to define a dressed gauge field,  
\begin{equation}
\A_\mu(x)=A_\mu(x) +\frac{i}{g} \partial_\mu \log W(x).
\label{gauge_dressed_photon}
\end{equation}
Given the gauge variation $\delta A_\mu(x) = -\partial_\mu \Lambda(x)$, together with \Eq{logW_variation}, we see that  $\delta \A_\mu(x)=0$, so the dressed gauge field is invariant.  Contracting the dressed gauge field with the fluid velocity field, we obtain
\eq{
u^\mu(x)\A_\mu(x)=
u^\mu(x) A_\mu(x) -u^\mu(x) \partial_\mu \int_{-\infty}^{0}ds\, u^\nu(x) A_\nu(x+su(x))
=0,
}{}
which precisely vanishes because the integrand is a total derivative and the terms cancel.  Here we have assumed as usual that the gauge field vanishes at infinity.  Remarkably, this implies that $\A_\mu(x)$ satisfies the inhomogeneous temporal gauge condition
\eq{
u^\mu(x)\A_\mu(x)=0,
}{}
which we refer to as $u(x)$ gauge.
This generalizes the usual temporal gauge by allowing the reference vector to be a spacetime-dependent fluid velocity field.   This suggests the relationship
\eq{
\A_\mu(x) = A_\mu(x)\big|_{u(x)\textrm{-gauge}},
}{}
which we will prove shortly.

This very basic observation has an important implication: the correlators of $\A_\mu(x) $ are literally just the correlators of $A_\mu(x) $ in $u(x)$ gauge.  For example, consider the matrix element of the dressed gauge boson,
\begin{equation}
\begin{aligned}
    \big\langle 0 \big| A_\mu(x) \big| A(k) \big\rangle_{u(x)\text{-gauge}} 
    & =\big\langle 0 \big| \A_\mu(x) \big| A(k) \big\rangle \\
    & = \big\langle 0 \,\big|\, A_\mu(x)-\partial_\mu\int_{-\infty}^0 ds\,u(x)^\nu  A_\nu(x+u(x)s) \,\big|\, A(k) \big\rangle \\
    & = \epsilon_\mu e^{-ikx}-i\partial_\mu\left(\frac{\epsilon\!\cdot\! u(x)}{k\!\cdot\! u(x)}e^{-ikx}\right).
    \label{eq:dr ph 1pt}
\end{aligned}
\end{equation}
This expression should be interpreted as the external wavefunction for the gauge boson in $u(x)$ gauge. It will also be useful to compute the vacuum correlator\footnote{Here, and throughout the paper, we understand operators as insertions in an in-out path integral. They are thus implicitly time-ordered, unless otherwise stated.}  of dressed gauge bosons, 
\begin{multline}
    \big\langle 0\big|{A}_{\mu_1}(x_1){A}_{\mu_2}(x_2)\big|0\big\rangle_{u(x)\text{-gauge}} 
     = \big\langle 0\big|\A_{\mu_1}(x_1)\A_{\mu_2}(x_2)\big|0\big\rangle \\
    \shoveleft{ = \Big\langle 0 \,\Big|\, 
    \big[A_{\mu_1}(x_1)-\frac{\partial}{\partial x_1^{\mu_1}}\int_{-\infty}^0 ds_1\,u^{\nu_1}(x_1) \, A_{\nu_1}(x_1+u(x_1) s_1)\big] }\\
    \times
    \big[A_{\mu_2}(x_2)-\frac{\partial}{\partial x_2^{\mu_2}}\int_{-\infty}^0 ds_2\,u^{\nu_2}(x_2) \, A_{\nu_2}(x_2+u(x_2) s_2)\big]
    \, \Big|\, 0\Big\rangle.
\end{multline}

Since the above expression is gauge invariant by definition, it can be computed in any gauge. Using a general $R_\xi$ gauge to evaluate the Wick contractions, we obtain 
\begin{equation}
\begin{aligned}
\big\langle 0\big|{A}_{\mu_1}(x_1){A}_{\mu_2}(x_2)\big|0\big\rangle_{u(x)-\text{gauge}} 
=\int_k\frac{-i}{k^2}
\bigg[&\eta_{\mu_1\mu_2}e^{-ik(x_1-x_2)}
-\frac{\partial}{\partial x_1^{\mu_1}}\Big(\frac{iu_{\mu_2}(x_1)}{k\!\cdot\!u(x_1)}e^{-ik(x_1-x_2)}\Big)\\
&-\frac{\partial}{\partial x_2^{\mu_2}}\Big(\frac{iu_{\mu_1}(x_2)}{-k\!\cdot\!u(x_2)}e^{-ik(x_1-x_2)}\Big)\\
&-\frac{\partial}{\partial x_1^{\mu_1}}\frac{\partial}{\partial x_2^{\mu_2}}\Big(\frac{u(x_1)\!\cdot\! u(x_2)}{[k\!\cdot\!u(x_1)][-k\!\cdot\!u(x_2)]}e^{-ik(x_1-x_2)}\Big)\bigg],
\end{aligned}
\label{eq:dr ph 2pt}
\end{equation}
which is independent of the gauge parameter $\xi$, as expected. This expression should be interpreted as the propagator for the gauge boson in $u(x)$ gauge.

Here we did not gauge fix by inserting an explicit gauge-fixing term to the action and inverting the kinetic term.\footnote{For constant $u^{\mu}$, we have explicitly checked that the two approaches agree precisely, even including the subtle $i\varepsilon$ terms in the $k\!\cdot\!u$ denominators.} With the gauge choice $u^\mu(x)A_\mu(x)=0$, the coordinate dependence in $u^\mu(x)$ makes that approach challenging. Instead, we have directly  constructed the gauge-invariant dressed photon $\A_\mu(x)$ and computed its correlators. Since $\A_\mu(x)$ reduces to $A_\mu(x)$ in that gauge by construction, these correlators automatically give the $u(x)$-gauge-fixed external polarization and propagator.

\subsubsection{Dressing as Gauge Fixing}

The equivalence of dressing and gauge fixing can be derived at the level of the Lagrangian.  The act of dressing is closely related to a gauge transformation that maps the undressed scalar $\phi$ to the dressed scalar $\Phi$.  The covariant derivative of the former is
\eq{
D_\mu \phi = D_\mu (W^{-1}\Phi) = W^{-1} (\partial_\mu + i g A_\mu -W^{-1}\partial_\mu W ) \phi =W^{-1} (\partial_\mu + i g \A_\mu ) \Phi \, .
}{}
Defining the gauge invariant derivative $\D_\mu = \partial_\mu + i g \A_\mu$, the Lagrangian in \Eq{L_SQED} maps to
\eq{
\mathcal{L}
&\rightarrow -\frac{1}{4} F_{\mu\nu}F^{\mu\nu}
  + (\D_\mu\Phi)^\dagger(\D^\mu\Phi) \, .  
}{L_SQED_dressed}
Here we have simply applied a gauge transformation to switch from the variables $\phi$ and $A$ to the variables $\Phi$ and $\A$. As usual, a gauge transformation must leave all physical observables, like on-shell scattering amplitudes or dressed correlators, completely unchanged.  However, this operation does reorganize the calculation.  The dressed insertions of $\phi$ become unornamented insertions of $\Phi$, at the cost of transforming $A$ to $\A$, which is the very definition of $u(x)$ gauge.\footnote{In later sections we always refer to the dynamical fields as $\phi,A$, regardless of gauge. Whether or not the fields are in $u(x)$ gauge should be clear from the context.}  In this gauge, the Wilson line defined in \Eq{Wilson_line} is trivial, so the dressing disappears.
We thus conclude that dressing is gauge fixing.

Last but not least, one might reasonably worry that this change of coordinates could induce Jacobians that would modify the dynamics at loop level.  However, gauge transformations are not anomalous, so the Jacobians are trivial.

\subsection{Tree-Level Correlator}\label{sec:QEDtreecorrelator}

A central goal in this paper is to demonstrate a variety of explicit calculations involving dressed local operators in both QED and gravity.  The analogous calculations with undressed operators are gauge dependent and lack a clear physical interpretation. Our focus will be the  matrix elements of the time-ordered product $\Phi^{\dagger}(x_1)\Phi(x_2)$. The vacuum expectation value of this operator is sensitive to the effects of dressing at $\OO(g^2)$, which corresponds to the one-loop computation in \Sec{sec:QEDloop}. In this section we instead study the tree-level matrix element that is sensitive to the dressing at leading order, $\langle \Omega | \Phi^{\dagger}(x_1) \Phi(x_2) | A(k)\rangle$, which involves an on-shell radiation photon. This matrix element is related to the expectation value of a plane wave photon coherent state, as we will elaborate on in \Sec{sec:causalitygravity}.

\subsubsection{Without Dressing}
First, let us calculate the matrix element $\langle \Omega | \phi^\dagger(x_1)\phi(x_2)| A(k)\rangle$.  At leading order in the interactions, corresponding to ${\cal O}(g)$, this object is just the tree-level correlator extracted from the bulk three-point interaction vertex,  
\eq{
\begin{array}{c}
\begin{tikzpicture}
    \tikzfeynmanset{
    every edge/.style={thick}
    }
    \begin{feynman}
      \vertex[dot, label=90:\(\phi^\dagger(x_1)\)] (x1) {}; 
      \vertex[right=2 of x1, dot, label=90:\(\phi(x_2)\)] (x2) {};
      \vertex (m) at ( $(x1)!0.5!(x2)$ );
      \vertex[below=1.6 of m] (f);

      \diagram*{
        (x1) --[fermion] (m) --[fermion] (x2),
        (f) --[photon, momentum={[arrow shorten=0.2]\(k\)}] (m)
      };
    \end{feynman}
\end{tikzpicture}
\end{array}\,,
}{diag:QEDtreeUndressed}
yielding the matrix element,
\eq{
\big\langle \Omega \big| \phi^\dagger(x_1)\phi(x_2) \big| A(k)\big\rangle
&= ig \int_{p_1,p_2}e^{-ip_1x_1}e^{-ip_2x_2} \,  \frac{ (p_2-p_1)\cdot\epsilon}{p_1^2\,p_2^2}\, \delta(p_1+p_2-k),
}{QED_undressed} 
where we have used that $\langle 0 | A_\mu(x)| A(k)\rangle = \epsilon_\mu(k)e^{-ikx}$.  Computing the integral, we obtain 
\eq{
\big\langle \Omega\big| \phi^\dagger(x_1)\phi(x_2)\big| A(k)\big\rangle
& =  -g \frac{\Gamma\left(\tfrac{D-2}{2}\right)}{4\pi^{D/2}}\frac{(e^{-i k  x_1}-e^{-i k  x_2})}{(-x_{12}^2+i\varepsilon)^{\frac{D-2}{2}}}\frac{\epsilon\!\cdot\!x_{12}}{k\!\cdot\!x_{12}}\,,
}{}
where we have defined the relative coordinate $x_{12}^\mu= x_1^\mu-x_2^\mu$. 

Under the gauge variation $\delta A_\mu =-\partial_\mu\Lambda$, the external polarization varies by
$\delta \epsilon_\mu = i\tilde{\Lambda} k_\mu$.  At the same time, this shifts the momentum space correlator
\eq{
\big\langle \Omega \big| \tilde\phi^\dagger(p_1)\tilde \phi(p_2) \big| A(k)\big\rangle &= \int_{x_1,x_2}\,e^{ip_1x_1}e^{ip_2x_2}  \big\langle \Omega \big| \phi^\dagger(x_1)\phi(x_2) \big| A(k)\big\rangle ,
}{}
by the following gauge variation:
\eq{
\delta_{\Lambda} \big\langle \Omega \big|  \tilde\phi^\dagger(p_1)\tilde \phi(p_2) \big| A(k)\big\rangle = 
- g \tilde{\Lambda}\left(\frac{1}{p_1^2}-\frac{1}{p_2^2}\right)\,
\delta(p_1 + p_2 - k).
}{}
For general $p_1$ and $p_2$, this expression is nonzero, so the correlator is not gauge invariant.    Of course, if we amputate the scalar legs to obtain an on-shell scattering amplitude, then this multiplies the expression by $p_1^2 p_2^2$, which restores gauge invariance in the on-shell limit. 

In position space the gauge variation is
\eq{
\delta_{\Lambda} \big\langle \Omega\big| \phi^\dagger(x_1)\phi(x_2)\big| A(k)\big\rangle = 
   -ig \tilde{\Lambda} \frac{\Gamma\left(\tfrac{D-2}{2}\right)}{4\pi^{D/2}}\frac{(e^{-i k  x_1}-e^{-i k  x_2})}{(-x_{12}^2+i\varepsilon)^{\frac{D-2}{2}}}\,,
}{eq:QED3pt-nonD}
which is generically nonvanishing. We can recover an invariant expression if we send $x_1$ and $x_2$ to null infinity, where they create particle excitations at some point on the celestial sphere.  In this case the stationary phase integral forces these particles' momenta to satisfy $p_{1,2}\propto x_{1,2}$. Conservation of momentum  implies that $k=p_1+p_2$, in which case $k\!\cdot\!x_{12}\propto (x_1^2 - x_2^2) = 0$.  The phases in \Eq{eq:QED3pt-nonD} then cancel and gauge invariance is restored.

\subsubsection{With Dressing}\label{subsec:QED_dressed_tree}
Correlators of the dressed scalars are gauge invariant by construction.  We can see this in the simple perturbative example of the matrix element $\langle \Omega |\Phi^\dagger(x_1) \Phi(x_2)| A(k)\rangle$ at tree level. Expanding the Wilson lines to $\mathcal{O}(g)$, we obtain
\eq{
& \big\langle \Omega \big| \Phi^\dagger(x_1) \Phi(x_2) \big| A(k) \big\rangle
    =\big\langle \Omega \,\big|\, \phi^\dagger(x_1)\,e^{-ig\int_{-\infty}^{0}ds\,u_1^\mu A_\mu(x_1+su_1)}\,\phi(x_2)\,e^{ig\int_{-\infty}^{0}ds\,u_2^\mu A_\mu(x_2+su_2)} \,\big|\, A(k) \big\rangle\\
    &=\underbrace{\big\langle \Omega \big| \phi^\dagger(x_1)\phi(x_2) \big| A(k) \big\rangle}_{\overset{\;}{\mathrm{I}}}
    + \underbrace{(-ig)\, \big\langle \Omega \,\big|\, \phi^\dagger(x_1)\, \phi(x_2)\,\int_{-\infty}^{0} ds\,u_1^\mu A_\mu(x_1+su_1)\,\big|\, A(k) \big\rangle}_{\overset{\;}{\mathrm{II}}} \\
    &\quad + \underbrace{ig\, \big\langle \Omega \,\big|\, \phi^\dagger(x_1)\, \phi(x_2)\,\int_{-\infty}^{0} ds\, u_2^\mu  A_\mu(x_2+su_2)\,\big|\, A(k) \big\rangle}_{\overset{\;}{\mathrm{III}}} ,
}{}
where we have defined the shorthand notation $u_1 = u(x_1)$ and $u_2=u(x_2)$.  Each contribution to the tree-level correlator corresponds to one of the following Feynman diagrams,

\eq{
\tikzfeynmanset{
    every edge/.style={thick}
    }
\begin{array}{ccc}
\vcenter{\hbox{
\begin{tikzpicture}
    \begin{feynman}
      \vertex[dot, label=90:\(\phi^\dagger(x_1)\)] (x1) {}; 
      \vertex[right=2 of x1, dot, label=90:\(\phi(x_2)\)] (x2) {};
      \vertex[below=2.4 of x1] (w1) {\(u_1\)};
      \vertex[below=2.4 of x2] (w2) {\(u_2\)};

      \vertex (m) at ( $(x1)!0.5!(x2)$ );
      \vertex[below=1.6 of m] (f);

      \diagram*{
        (x1) --[fermion] (m) --[fermion] (x2),
        (x1) --[scalar, blue] (w1),
        (x2) --[scalar, blue] (w2),
        (f) --[photon, momentum={[arrow shorten=0.2]\(k\)}] (m)
      };
    \end{feynman}
\end{tikzpicture}
}}
&
\vcenter{\hbox{
\begin{tikzpicture}
  \begin{feynman}
      \vertex[dot, label=90:\(\phi^\dagger(x_1)\)] (x1) {}; 
      \vertex[right=2 of x1, dot, label=90:\(\phi(x_2)\)] (x2) {};
      \vertex[below=2.4 of x1] (w1) {\(u_1\)};
      \vertex[below=2.4 of x2] (w2) {\(u_2\)};

      \vertex (g) at ( $(x1)!0.5!(w1)$ );
      \vertex[left=1 of g] (f);
    
      \diagram*{
        (x1) --[fermion] (x2),
        (x1) --[scalar, blue] (w1),
        (x2) --[scalar, blue] (w2),
        (f)  --[photon, momentum'={[arrow shorten=0.2] \(k\)}] (g),
      };
    \end{feynman}
\end{tikzpicture}
}}
&
\vcenter{\hbox{
\begin{tikzpicture}
  \begin{feynman}
      \vertex[dot, label=90:\(\phi^\dagger(x_1)\)] (x1) {}; 
      \vertex[right=2 of x1, dot, label=90:\(\phi(x_2)\)] (x2) {};
      \vertex[below=2.4 of x1] (w1) {\(u_1\)};
      \vertex[below=2.4 of x2] (w2) {\(u_2\)};
      
      \vertex (g) at ( $(x2)!0.5!(w2)$ );
      \vertex[right=1 of g] (f);
   
      \diagram*{
        (x1) --[fermion]  (x2),
        (x1) --[scalar, blue] (w1),
        (x2) --[scalar, blue] (w2),
        (f)  --[photon, momentum={[arrow shorten=0.2] \(k\)}] (g),
      };
    \end{feynman}
\end{tikzpicture}
}}
\\[2em]
\textbf{(I)} & \textbf{(II)} & \textbf{(III)}
\end{array}\,,
}{diag:QEDtreeDressed}
where the scalar fields $\phi^\dagger(x_1)$ and $\phi(x_2)$ carry momenta $p_1$ and $p_2$ that are implicitly defined to be flowing towards them, according to the convention in Eq.~\eqref{eq:momentum-arrow}. 

Term (I) is just the undressed correlator from \Eq{QED_undressed},
\eq{
(\mathrm{I})
 = ig\int_{p_1,p_2}e^{-ip_1x_1}e^{-ip_2x_2}\, \frac{(p_2-p_1)\!\cdot\!\epsilon}{p_1^2\,p_2^2}\, \delta(p_1+p_2-k),
}{}
while  (II) and (III) are corrections due to dressing. Since this is the first time we calculate a dressing correction, we will show our steps explicitly. To calculate to $\OO(g)$ we replace $\langle \Omega|$ by $\langle 0|$ since no interaction vertex insertion is needed. Using Eq.~\eqref{eq:WL-photon-feynman}, Eq.~\eqref{eq:X-X-contraction} and Eq.~\eqref{eq:photon-plr}, we obtain
\eq{
(\text{II})
&\equiv
\big\langle 0 \,\big|\,
  \phi^\dagger(x_1)\, \phi(x_2)\,
  \int_{-\infty}^{0} - i g A_\mu(x_1 + s u_1)\, u_1^\mu\, ds
  \,\big|\, A(k) \big\rangle\\
&= \int_{p_1,p_2}e^{-ip_1x_1}e^{-ip_2x_2}\underbrace{\left\langle\tilde{\phi}^{\dagger}(-p_1)\tilde{\phi}(p_2)\right\rangle}_{i\delta(p_1+p_2)/p_2^2}  \int_{q} e^{-iqx_1} \frac{g u_1^\mu}{q\!\cdot\!u_1} \underbrace{\left\langle 0\middle| \tilde{A}_\mu(q) \middle| A(k)\right\rangle}_{\epsilon_\mu(k)\delta(q-k)}\\
&= \int_{p_1,p_2}e^{-ip_1x_1}e^{-ip_2x_2}e^{-ikx_1}g\frac{u_1\!\cdot\!\epsilon}{u_1\!\cdot\! k}\times \frac{i}{p_2^2} \delta(p_1+p_2)\\
&= i g \int_{p_1,p_2}\, e^{-i p_1 x_1 - i p_2 x_2}\,
\left[\frac{1}{p_2^2}\,
\frac{u_1\!\cdot\!\epsilon}{u_1\!\cdot\! k}\right]\,
\delta(p_1+p_2 - k),
}{}
where in the last step we shifted the integration variable by $p_1\to p_1-k$. Similarly,
\eq{
(\text{III})
= i g \int_{p_1,p_2}\, e^{-i p_1 x_1 - i p_2 x_2}\,
\left[\frac{-1}{p_1^2}\,
\frac{u_2\!\cdot\!\epsilon}{u_2\!\cdot\! k}\right]
\delta(p_1+p_2 - k).
}{}
Adding up these contributions we obtain the dressed correlator to order $\mathcal{O}(g)$,
\eq{
\big \langle \Omega \big| \Phi^\dagger(x_1) \Phi(x_2) \big| A(k) \big\rangle
&= (\text{I}) + (\text{II}) + (\text{III})\\
&\hspace{-20pt}= \begin{aligned}[t]
  & i g \int_{p_1,p_2}\, e^{-i p_1 x_1 - i p_2 x_2}\,
\frac{\delta(p_1 + p_2 - k)}{p_1^2\, p_2^2}\Bigg[
(p_2- p_1)\!\cdot\!\epsilon
+ p_1^2\,\frac{u_1\!\cdot\!\epsilon}{u_1\!\cdot\!k}
- p_2^2\,\frac{u_2\!\cdot\!\epsilon}{u_2\!\cdot\!k}
\Bigg] .
\end{aligned}
}{QED_dressed_tree}
The gauge variation of the polarization, $\delta \epsilon_\mu = i\tilde{\Lambda} k_\mu$, shifts the term in  brackets by a factor proportional to
$  (p_2 - p_1)\!\cdot\!k + (p_1^2 - p_2^2) = 0$, so the dressed correlator is indeed gauge invariant. Transforming to position space in \App{app:tree_fourier}, we find that
\eq{
  &\big\langle \Omega\big| \Phi^\dagger(x_1) \Phi(x_2) \big| A(k)\big\rangle=\\  
  &= -g\frac{\Gamma\left(\tfrac{D-2}{2}\right)}{4\pi^{D/2}}\frac{1}{(-x_{12}^{2}+i\varepsilon)^{\tfrac{D-2}{2}}}
  \left[
  \left(\frac{\epsilon \!\cdot\!x_{12}}{k \!\cdot\!x_{12}} - \frac{\epsilon \!\cdot\!u_{1}}{k \!\cdot\!u_{1}}\right) e^{-i k x_{1}}
  -
  \left(\frac{\epsilon \!\cdot\!x_{12}}{k \!\cdot\!x_{12}} - \frac{\epsilon \!\cdot\!u_{2}}{k \!\cdot\!u_{2}}\right) e^{-i k x_{2}}
  \right].
}{eq:treeQEDpositionspace}
The terms in each set of parentheses are separately gauge invariant. 

In fact, our result can be written in a manifestly gauge-invariant manner  using the spatial projector
\eq{
    P^{(i)}_{\mu \nu}=\eta_{\mu \nu}-\frac{k_{\mu}u_{i\,\nu}}{k\!\cdot\!u_{i}+i\varepsilon}\,.
}{eq:spatialprojector}
This projector acts from the left to eliminate directions along $u^{\mu}$,
\eq{
u^{\mu}_{i}P_{\mu\nu}^{(i)}=0\,,
}{}
and from the right is transverse to $k$,
\eq{
P^{(i)}_{\mu\nu}k^{\nu}=0\,.
}{}
Hence, the combination $P_{\mu\nu}^{(i)}\epsilon^{\nu}(k)$ is manifestly gauge invariant. In terms of this projector, the  matrix element of the dressed operators becomes
\eq{
  &\big\langle \Omega\big|\Phi^\dagger(x_1) \Phi(x_2) \big| A(k)\big\rangle=\\  
  &= -g\frac{\Gamma\left(\tfrac{D-2}{2}\right)}{4\pi^{D/2}}\frac{1}{(-x_{12}^{2}+i\varepsilon)^{\tfrac{D-2}{2}}}\frac{1}{x_{12}\!\cdot\!k}
  \left[
  \left(x_{12}\!\cdot\!P^{(1)}\!\cdot\!\epsilon\right) e^{-i k x_{1}}
  -\left(x_{12}\!\cdot\! P^{(2)}\!\cdot\! \epsilon\right)e^{-i k x_{2}}
  \right]\,.
}{eq:QEDtreedressedprojectors}
Despite appearances, there is no pole at $x_{12}\!\cdot\!k=0$ because the phases align as $k\!\cdot\!x_{1}\rightarrow k\!\cdot\!x_{2}$. In contrast, the soft poles at $k\!\cdot\!u_{1,2}=0$ are indeed physical. 

\subsubsection{Via Gauge Fixing}\label{eq:treeQEDviagaugefixing}

As shown in \Eq{L_SQED_dressed}, all observables constructed from the dressed scalar $\Phi$ are simply  $\phi$ observables in $u(x)$ gauge.  This offers an extremely simple way to compute dressed observables.  In particular, the tree-level dressed correlator is computed by the single diagram in \Eq{diag:QEDtreeUndressed} evaluated in $u(x)$ gauge,
\eq{
&\big\langle \Omega\big|\Phi^\dagger(x_1) \Phi(x_2)\big|A(k) \big\rangle
=\big\langle \Omega\big|\phi^\dagger(x_1)\phi(x_2)\big|A(k) \big\rangle_{u(x)\textrm{-gauge}}\\
&= ig \int_{p_1,p_2}
\int_y
\frac{i}{p_1^2}\frac{i}{p_2^2}
e^{-i p_1 (x_1-y)} e^{-i p_2 (x_2-y)}
(p_1 - p_2)^\mu
\left[
  \epsilon_\mu e^{-i k y}
  - i\partial_\mu
    \Big(
      \frac{\epsilon\!\cdot\!  u(y)}{k\!\cdot\!  u(y)}\,
      e^{-i k y}
    \Big)
\right] \\
&=ig\int_{p_1, p_2}\int_{y}\,e^{-ip_1 (x_1-y)-ip_2 (x_2-y)-ik y} \frac{1}{p_1^2 p_2^2}\left[(p_2-p_1)\!\cdot\!\epsilon -p_2^2\, \frac{\epsilon\!\cdot\! u(y)}{k\!\cdot\!u(y)}+p_1^2\,\frac{\epsilon\!\cdot\! u(y)}{k\!\cdot\!u(y)}\right],
}{eq:QEDtree_ux}
where we used integration by parts in the final line. 

At first glance this expression is impossible to evaluate since it depends on the whole spacetime profile of $u(y)$. However, we will see momentarily that the terms containing $u(y)$ are much more local than naively expected because they appear in the form $\partial_{\mu}(...)$.  Since the photon couples to a conserved current,  these terms localize to external insertions of charge at $x_1$ and $x_2$. 

To see this explicitly, let us focus on the term in brackets that is proportional to $p_{2}^2$.  The term proportional to $p_{1}^2$ will be treated analogously. The factor of $p_2^2$ pinches the $p_2$ propagator. Diagrammatically, this means that for this term the photon line in the diagram \eqref{diag:QEDtreeUndressed} does not attach to the internal scalar propagator, but to the insertion point $x_2$. Mathematically, we see this by observing that the $p_2$ dependence in this term is trivial and we can immediately integrate $p_2$ to obtain $\delta(x_2-y)$. The upshot is that in the term proportional to $p_2^2$ in \Eq{eq:QEDtree_ux}, we can simply replace the spacetime dependence $u(y)\rightarrow u(x_2)\equiv u_2$, and likewise with $u_1$ in the term proportional to $p_1^2$. This is remarkable, and entirely due to the fact that photons couple to conserved currents. 

Replacing $u(y)$ by $u_2,u_1$ where appropriate, we readily perform the $y$ integral to get the familiar momentum conserving delta function $\delta(p_1+p_2-k)$. Then, as advertised, the $u(x)$-gauge expression in \Eq{eq:QEDtree_ux} is exactly equal to the expression in  \Eq{QED_dressed_tree}  obtained by summing dressed contributions.

\subsection{Loop-Level Correlator}\label{sec:QEDloop}

One emphasis of this work is that \textit{dressing is not on the side}---the behavior of dressed observables can differ significantly from their undressed counterparts. This is first exemplified at tree level in \Eq{QED_dressed_tree}, where there is a new kinematic singularity when the photon momentum $k$ is orthogonal to the worldline velocities $u_{1}, u_{2}$. However, at tree level $k$ is necessarily null and $u$ is timelike, so unless $k$ is soft we always have $k\!\cdot\!u_{i}\neq 0$, and the singularity never appears in physical kinematics.

At loop level however the internal photon momentum is necessarily integrated over all regions, including where $k\!\cdot\!u_{i}=0$, so one expects the effects of dressing to be more profound. To make this intuition concrete, we will now study the one-loop corrections to the two-point vacuum correlation function of dressed scalars. The upshot will be novel nonanalytic features in the dressed correlator which do not occur in conventional calculations of undressed correlators.

\subsubsection{Without Dressing}
To set the stage, let us first calculate the one-loop correction to the \textit{undressed} correlator $\langle \Omega | \phi^\dagger(x_1) \phi(x_2) | \Omega \rangle$.  This is given by the two $\mathcal{O}(g^2)$ diagrams, 

\eq{
\begin{array}{cc}
  \tikzfeynmanset{
    every edge/.style={thick}
    }
\vcenter{\hbox{
\begin{tikzpicture}
    \begin{feynman}
      \vertex[dot, label=90:\(\phi^\dagger(x_1)\)] (x1) {}; 
      \vertex[right=2 of x1, dot, label=90:\(\phi(x_2)\) ] (x2) {};

      \vertex (m1) at ( $(x1)!0.3!(x2)$ );
      \vertex (m2) at ( $(x1)!0.7!(x2)$ );
    
      \diagram*{
        (x1) --[fermion] (x2),
        (m1) --[photon, out=270, in=270, looseness=2.0] (m2),
      };
    \end{feynman}
\end{tikzpicture}
}}
&
\vcenter{\hbox{
\begin{tikzpicture}
  \begin{feynman}
        \vertex[dot, label=90:\(\phi^\dagger(x_1)\)] (x1) {}; 
        \vertex[right=2 of x1, dot, label=90:\(\phi(x_2)\)] (x2) {};

        \vertex (m) at ($(x1)!0.5!(x2)$);
        \diagram*{
          (x1) --[fermion]  (m) --[fermion] (x2),
        };
        \draw[black, thick, photon] (m) arc [start angle=90, end angle=-270, radius=0.4];
    \end{feynman}
\end{tikzpicture}
}}
\\[2em]
\textbf{(a)} & \textbf{(b)}
\end{array}\,.
}{diag:QEDLoopUndressed}
\noindent These corrections organize as 
\eq{
 \langle \Omega| \phi^\dagger(x_1) \phi(x_2)  |\Omega\rangle =  \int_p e^{-ip(x_1-x_2)}\,\left[\frac{i}{p^2} + \left(\frac{i}{p^2}\right)^2 i\Sigma(p)+\cdots\right]\,,
}{eq:selfenergyexpansion}
with the leading self-energy contribution given by diagrams (a) and (b) above. Diagram (b) gives scaleless integrals and thus vanishes in dimensional regularization. Diagram (a) gives
\eq{
    \Sigma(p)=-ig^2 \mu^{4-D}\int_k \frac{N_{\mu\nu}(k)}{k^{2}+i\varepsilon}\frac{(2p+k)^\mu (2p+k)^\nu}{(p+k)^{2}+i\varepsilon}\,,
}{eq:QEDselfenergy_ud}
where $N_{\mu\nu}(k)$ is the numerator of the photon propagator. The correlation function of undressed operators is gauge dependent.  To illustrate this we use a covariant $R_{\xi}$ gauge for the photon, in which the propagator numerator is
\eq{
N_{\mu\nu}^{\xi}(k)=-\eta_{\mu\nu}+(1-\xi)\frac{k_{\mu}k_{\nu}}{k^2}\,.
}{}
Evaluating the self-energy diagram in $R_{\xi}$ gauge,
\eq{
\Sigma^{\xi}(p) = -\frac{g^2 p^2}{4\pi^2}\left(1 + \frac{(\xi-3)}{4}\left(\frac{2}{D-4}+\log\left(\frac{-p^2}{\tilde{\mu}^2}\right)\right)\right)+\OO(D-4)\,,
}{eq:selfenergyrxi}
where we have introduced the modified renormalization scale $\tilde{\mu}^2 = 4\pi e^{-\gamma_{E}}\mu^2$. The presence of the gauge parameter $\xi$ confirms that this quantity is gauge dependent.
 
\subsubsection{With Dressing}
Now we compute the correlator of dressed scalars, $\langle \Omega | \Phi^\dagger(x_1) \Phi(x_2) | \Omega \rangle$. The leading-order contributions to this dressed correlator come from the $\mathcal{O}(g^2)$ one-loop diagrams,

\eq{
\begin{array}{cccc}
\vcenter{\hbox{
\begin{tikzpicture}
\tikzfeynmanset{
every edge/.style={thick}
}
\begin{feynman}
  \vertex[dot, label=90:\(\phi^\dagger(x_1)\)] (x1) {};
  \vertex[right=2 of x1, dot, label=90:\(\phi(x_2)\)] (x2) {};
  \vertex[below=2.4 of x1] (w1) {\(u_1\)};
  \vertex[below=2.4 of x2] (w2) {\(u_2\)};

  \vertex (m1) at ($(x1)!0.3!(x2)$);
  \vertex (m2) at ($(x1)!0.7!(x2)$);

  \diagram*{
    (x1) --[fermion] (x2),
    (x1) --[scalar, blue] (w1),
    (x2) --[scalar, blue] (w2),
    (m1) --[photon, out=270, in=270, looseness=2.0, momentum'={[arrow shorten=0.35]\(k\)}] (m2),
  };
\end{feynman}
\end{tikzpicture}
}}
&
\vcenter{\hbox{
\begin{tikzpicture}
\tikzfeynmanset{
every edge/.style={thick}
}
\begin{feynman}
  \vertex[dot, label=90:\(\phi^\dagger(x_1)\)] (x1) {};
  \vertex[right=2 of x1, dot, label=90:\(\phi(x_2)\)] (x2) {};
  \vertex[below=2.4 of x1] (w1) {\(u_1\)};
  \vertex[below=2.4 of x2] (w2) {\(u_2\)};

  \vertex (m) at ($(x1)!0.5!(x2)$);
  \vertex (g) at ($(x1)!0.5!(w1)$);

  \diagram*{
    (x1) --[fermion] (m) --[fermion] (x2),
    (x1) --[scalar, blue] (w1),
    (x2) --[scalar, blue] (w2),
    (g) --[photon, momentum'={[arrow shorten=0.25]\(k\)}] (m),
  };
\end{feynman}
\end{tikzpicture}
}}
&
\vcenter{\hbox{
\begin{tikzpicture}
\tikzfeynmanset{
every edge/.style={thick}
}
\begin{feynman}
  \vertex[dot, label=90:\(\phi^\dagger(x_1)\)] (x1) {};
  \vertex[right=2 of x1, dot, label=90:\(\phi(x_2)\)] (x2) {};
  \vertex[below=2.4 of x1] (w1) {\(u_1\)};
  \vertex[below=2.4 of x2] (w2) {\(u_2\)};

  \vertex (m) at ($(x1)!0.5!(x2)$);
  \vertex (g) at ($(x2)!0.5!(w2)$);

  \diagram*{
    (x1) --[fermion] (m) --[fermion] (x2),
    (x1) --[scalar, blue] (w1),
    (x2) --[scalar, blue] (w2),
    (m) --[photon, momentum'={[arrow shorten=0.25]\(k\)}] (g),
  };
\end{feynman}
\end{tikzpicture}
}}
&
\vcenter{\hbox{
\begin{tikzpicture}
\tikzfeynmanset{
every edge/.style={thick}
}
\begin{feynman}
  \vertex[dot, label=90:\(\phi^\dagger(x_1)\)] (x1) {};
  \vertex[right=2 of x1, dot, label=90:\(\phi(x_2)\)] (x2) {};
  \vertex[below=2.4 of x1] (w1) {\(u_1\)};
  \vertex[below=2.4 of x2] (w2) {\(u_2\)};
  \vertex (g1) at ($(x1)!0.5!(w1)$);
  \vertex (g2) at ($(x2)!0.5!(w2)$);

  \diagram*{
    (x1) --[fermion] (x2),
    (x1) --[scalar, blue] (w1),
    (x2) --[scalar, blue] (w2),
    (g1) --[photon, momentum={[arrow shorten=0.3]\(k\)}] (g2),
  };
\end{feynman}
\end{tikzpicture}
}}
\\[2em]
\textbf{(I)} & \textbf{(IIa)} & \textbf{(IIb)} & \textbf{(III)}
\end{array}\,.
}{diag:QEDLoopDressed}
\noindent Note that we have dropped additional diagrams with scaleless integrals that trivially vanish in dimensional regularization, so for example 
\newcommand{\diagA}{
    \begin{tikzpicture}
    \tikzfeynmanset{
      every edge/.style={black, thick},
      }
    \begin{feynman}
        \vertex[dot, label=90:\(\phi^\dagger(x_1)\)] (x1) {}; 
        \vertex[right=2 of x1, dot, label=90:\(\phi(x_2)\)] (x2) {};
        \vertex[below=2.4 of x1] (w1) {\(u_1\)};
        \vertex[below=2.4 of x2] (w2) {\(u_2\)};   

        \vertex (m) at ($(x1)!0.5!(x2)$);
        \diagram*{
          (x1) --[fermion] (m) --[fermion] (x2),
          (x1) --[scalar, blue] (w1),
          (x2) --[scalar, blue] (w2),
        };
        \draw[black, thick, photon] (m) arc [start angle=90, end angle=-270, radius=0.4];
    \end{feynman}
    \end{tikzpicture}
}
\newcommand{\diagB}{
    \begin{tikzpicture}
    \tikzfeynmanset{
    every edge/.style={black, thick},
    }
    \begin{feynman}
        \vertex[dot, label=90:\(\phi^\dagger(x_1)\)] (x1) {}; 
        \vertex[right=2 of x1, dot, label=90:\(\phi(x_2)\)] (x2) {};
        \vertex[below=2.4 of x1] (w1) {\(u_1\)};
        \vertex[below=2.4 of x2] (w2) {\(u_2\)};    
        
        \vertex (g) at ( $(x1)!0.435!(w1)$ );
        \vertex[right=0.6 of g] (f);
        \diagram*{
          (x1) --[fermion] (x2),
          (x1) --[scalar, blue] (w1),
          (x2) --[scalar, blue] (w2),
          (g)  --[photon] (f),
        };
        \draw[black, thick] (f) arc [start angle=180, end angle=-180, radius=0.4];
    \end{feynman}
    \end{tikzpicture}
}
\newcommand{\diagC}{
    \begin{tikzpicture}
    \tikzfeynmanset{
    every edge/.style={black, thick},
    }
    \begin{feynman}
        \vertex[dot, label=90:\(\phi^\dagger(x_1)\)] (x1) {}; 
        \vertex[right=2 of x1, dot, label=90:\(\phi(x_2)\)] (x2) {};
        \vertex[below=2.4 of x1] (w1) {\(u_1\)};
        \vertex[below=2.4 of x2] (w2) {\(u_2\)};

        \vertex (g1) at ( $(x1)!0.27!(w1)$ );
        \vertex (g2) at ( $(x1)!0.6!(w1)$ );
        \diagram*{
          (x1) --[fermion] (x2),
          (x1) --[scalar, blue] (w1),
          (x2) --[scalar, blue] (w2),
          (g1)  --[photon, in=0, out=0, looseness=2] (g2),
        };
    \end{feynman}
    \end{tikzpicture}
}
\begin{equation}
\vcenter{\hbox{\diagA}} = 
\vcenter{\hbox{\diagB}} =
\vcenter{\hbox{\diagC}} = 0 .
\label{eq:QED loop other}
\end{equation}

\noindent Next, we calculate the four one-loop diagrams in \Eq{diag:QEDLoopDressed} using the Feynman rules introduced earlier. Diagram (I) is simply the vacuum correlator in the absence of any dressing, namely \Eq{eq:QEDselfenergy_ud}. Since the dressed correlator is gauge invariant, we can choose the most convenient gauge to evaluate these four diagrams, which is $N_{\mu\nu}=-\eta_{\mu\nu}$ corresponding to $\xi=1$. Diagram (I) is then
\eq{
\text{(I)}
& =g^2\int_{p,k} e^{-ip\cdot(x_1-x_2)}\, \frac{1}{(p^2)^2}\frac{(2p+k)^2}{k^2(p+k)^2}.
}{QED_loop_I}
Meanwhile, Diagram (IIa) describes virtual gauge interactions between the dressing and the exchanged scalar field,
\eq{
\text{(IIa)}
& \equiv \big \langle 0 \,\big|\, \phi^\dagger(x_1)\int_{-\infty}^0 ds\, (-ig) u_1^\nu A_\nu(x_1+u_1s)\int_y gA_\mu(y) \bigl[\phi^\dagger(y)\partial^\mu\phi(y)-\mathrm{h.c.}\bigr]\, \phi(x_2)\,\big|\, 0 \big\rangle\\
& = \int_{p_1,p_2,k}e^{-ip_1x_1}e^{-ip_2x_2}e^{ikx_1} \frac{gu_1^\mu}{-k\!\cdot\!u_1}\times ig(p_1-p_2)_\mu\times\frac{i}{p_1^2}\frac{i}{p_2^2}\frac{-i}{k^2}\delta(p_1+p_2-k)\\
& = -g^2 \int_{p,k} e^{-ip\cdot(x_1-x_2)}  \, \frac{1}{p^2}\, \frac{u_1\!\cdot\!(2p+k)}{k^2(p+k)^2} \frac{1}{-k\!\cdot\!u_1},
}{QED_loop_IIa}
where to get the third line we defined $p_{1} = p+k$ and eliminated $p_2$.  Performing a similar calculation for Diagram (IIb), we get
\eq{
\text{(IIb)} 
& = g^2 \int_{p, k} e^{-ip (x_1-x_2)}  \,  \frac{1}{p^2}\, \frac{u_{2}\!\cdot\!(2p+k)}{k^2(p+k)^2} \frac{1}{k\!\cdot\!u_2}.
}{QED_loop_IIb}
Last but not least, we calculate Diagram (III) to be
\eq{
\text{(III)}
& \equiv \big \langle 0 \,\big|\, \phi^\dagger(x_1)\int_{-\infty}^0 ds_1\, (-ig) u_1^\mu A_\mu(x_1+u_1s_1)\,\phi(x_2)\int_{-\infty}^0 ds_2\, (ig) u_2^\nu A_\nu(x_2+u_2s_2)\,\big|\, 0 \big\rangle\\
&= \int_{p_1,p_2,k}e^{-ip_1x_1}e^{-ip_2x_2}e^{ikx_1}e^{-ikx_2}\frac{gu_1^\mu}{-k\!\cdot\!u_1}\frac{-gu_{2\mu}}{k\!\cdot\!u_2}\times\frac{i}{p_1^2}\frac{-i}{k^2}\delta(p_1+p_2)\\
& = -g^2 \int_{p,k} e^{-ip (x_1-x_2)}\, \frac{1}{(p+k)^2 k^2} \frac{u_1\!\cdot\!u_2}{(-k\!\cdot\!u_1)(k\!\cdot\!u_2)}.
}{QED_loop_III}
Assembling these four nontrivial diagrams, we find that the dressed correlator also organizes in terms of self-energies, 
\eq{
 \langle \Omega| \Phi^\dagger(x_1) \Phi(x_2)  |\Omega\rangle =  \int_p e^{-ip(x_1-x_2)}\,\left[\frac{i}{p^2} + \left(\frac{i}{p^2}\right)^2 i\Sigma^{\textrm{dressed}}(p)+\cdots\right]\,,
}{eq:QED_dressed_loop}
with the dressed self-energy
\eq{
\Sigma^{\textrm{dressed}}(p) = ig^{2}\int_{k}\frac{\eta_{\mu\nu}}{(p+k)^2 k^2}\left[(2p+k)^{\mu}-p^2\frac{u_1^{\mu}}{-k\!\cdot\!u_{1}}\right]\left[(2p+k)^{\nu}+p^2\frac{u_2^{\nu}}{k\!\cdot\!u_{2}}\right]\,.
}{eq:QED_dressed_selfenergy}
As with the tree-level expression, we can also write this compactly in terms of the spatial projectors $P^{(i)}_{\mu \nu}$. Notice that terms in the first square bracket can be written as
\eq{
(2p+k)^{\mu}-p^2\frac{u_1^{\mu}}{-k\!\cdot\!u_{1}} = (2p+k)^{\mu}-\frac{(p+k)^2-(2p+k)\!\cdot\!k}{-k\!\cdot\!u_1}u_1^\mu,
}{}
but the term proportional to $(p+k)^2$ will pinch the scalar propagator, leading to a scaleless integral and thus can be dropped. The same treatment can be applied to the other square bracket. So the self-energy can be organized into the following form:
\eq{
\Sigma^{\textrm{dressed}}(p) = ig^{2}\int_{k}\frac{(2p+k)^{\mu}(2p+k)^{\nu}}{(p+k)^2}\frac{\eta^{\alpha\beta}\overline{P^{(1)}_{\mu \alpha}}P^{(2)}_{\nu\beta}}{k^2}\,,
}{}
where $P^{(i)}_{\mu \nu}=\eta_{\mu \nu}-\frac{k_{\mu}u_{i\,\nu}}{k\!\cdot\!u_{i}+i\varepsilon}$, and the overline denotes complex conjugation. In a homogeneous temporal gauge with constant $u(x)=u$, the photon propagator has exactly the form above, but with $u_2=u_1=u.$ Here the effective propagator has the same form as in the homogeneous case, but with different projectors $P^{(i)}_{\mu\nu}$ depending on the dressing of the nearest matter insertion. 
We postpone the evaluation of this Feynman integral until \Sec{sec:QEDloopintegral}.

\subsubsection{Via Gauge Fixing}

We can of course also compute the dressed one-loop correlator $\langle \Omega | \Phi^\dagger(x_1) \Phi(x_2) | \Omega\rangle$ by going to $u(x)$ gauge.  In this case the correlator receives a single contribution from Diagram (I), albeit computed in this gauge. Since the $u(x)$ gauge breaks translation invariance it seems unlikely that this expression would organize into a single Fourier transform of propagators and self-energies, as in \Eq{eq:QED_dressed_loop}. In fact, as we now demonstrate, this is precisely what happens. The mechanism is completely analogous to the tree-level calculation in \Sec{eq:treeQEDviagaugefixing}, but we will also work through it explicitly here at one loop.  The $\OO(g^2)$ contribution is
\eq{
\langle \Omega | \Phi^\dagger(x_1) \Phi(x_2)  | \Omega \rangle\Big|_{g^2} & = g^2 \int_{{y_1},{y_2}} \int_{p_1,p_2,l} e^{-ip_1(x_1-{y_1})-ip_2(x_2-{y_2})-il({y_1}-{y_2})} \frac{i}{p_1^2}\frac{i}{p_2^2}\frac{i}{l^2} (p_1+l)^{\mu_1} (p_2-l)^{\mu_2} \\
&\times \langle 0 | A_{\mu_1}({y_1})A_{\mu_2}({y_2}) | 0 \rangle_{u(x)-\text{gauge}}.
}{}
 Inserting the expression \Eq{eq:dr ph 2pt} for the $u(x)$-gauge photon propagator, we have
\eq{
& \langle \Omega | \Phi^\dagger(x_1) \Phi(x_2)  | \Omega \rangle\Big|_{g^2} \\
& = 
    g^2 \int_{{y_1},{y_2}} \int_{p_1,p_2,l} e^{-ip_1(x_1-{y_1})-ip_2(x_2-{y_2})-il({y_1}-{y_2})} \frac{i}{p_1^2}\frac{i}{p_2^2}\frac{i}{l^2} (p_1+l)^{\mu_1} (p_2-l)^{\mu_2} \\
    & \quad
    \begin{aligned}
    \times \int_k\frac{-i}{k^2}
    \Bigg[
    \eta_{{\mu_1}{\mu_2}}e^{-ik({y_2}-{y_1})}
    -\frac{\partial}{\partial y_1^{\mu_1}}\Big(\frac{iu_{\mu_2}({y_1})}{-k\!\cdot\!u({y_1})}e^{-ik({y_2}-{y_1})}\Big)
    -\frac{\partial}{\partial y_2^{\mu_2}}\Big(\frac{iu_{\mu_1}({y_2})}{k\!\cdot\!u({y_2})}e^{-ik\cdot({y_2}-{y_1})}\Big) \\
    \shoveright{-\frac{\partial}{\partial y_1^{\mu_1}}\frac{\partial}{\partial y_2^{\mu_2}}\Big(\frac{u({y_1})\!\cdot\!u({y_2})}{[-k\!\cdot\!u({y_1})][k\!\cdot\!u({y_2})]}e^{-ik\cdot({y_2}-{y_1})}\Big)\Bigg].}
    \end{aligned}
}{}
After integration by parts for $\partial/\partial y_1^{\mu_1}$ and $\partial/\partial y_2^{\mu_2}$, it becomes 
\eq{
& \langle \Omega | \Phi^\dagger(x_1) \Phi(x_2)  | \Omega \rangle\Big|_{g^2} \\
& = \begin{aligned}[t]
      & g^2 \int_{{y_1},{y_2}} \int_{p_1,p_2,l,k} \frac{e^{-ip_1 x_1} e^{-ip_2 x_2} e^{i(p_1-l+k) {y_1}} e^{i(p_2+l-k) {y_2}}}{p_1^2\,p_2^2\,l^2\,k^2} 
      \times \Bigg[ 
      \underbrace{- (p_1+l)\!\cdot\!(p_2-l)}_{\textrm{I'}} \\
      & + \underbrace{(p_1^2-l^2) \frac{(p_2-l) \!\cdot\! u({y_1})}{-k\!\cdot\!u({y_1})}}_{\textrm{IIa'}} 
      + \underbrace{(p_2^2-l^2) \frac{(p_1+l)\!\cdot\! u({y_2})}{k\!\cdot\!u({y_2})}}_{\textrm{IIb'}}
      - \underbrace{(p_1^2-l^2)(p_2^2-l^2) \frac{u({y_1})\!\cdot\! u({y_2})}{[-k\!\cdot\!u({y_1})][k\!\cdot\!u({y_2})]}}_{\textrm{III'}}
      \Bigg].
    \end{aligned}
}{eq:QED-1loop-integrand}
Now let us  examine this term by term. Term (I') is independent of the internal positions $y_1$ and $y_2$, and so we can readily integrate over these internal positions to obtain momentum conserving delta functions. The other terms (IIa', IIb', and III') depend on the velocity field $u(y_{1,2})$ so we cannot immediately integrate $y_{1,2}$ to enforce momentum conservation---however after some manipulations we will see that the integration can in fact be done.

Next we focus on term (IIa'), as the others are treated completely analogously. It is proportional to $(p_1^2-l^2)$. The $l^2$ part pinches the $l^2$ denominator coming from the internal scalar propagator. This allows us to then shift the integration variable $l\rightarrow l+k$ to remove the $k$ dependence from the phase without changing the denominator. The remaining $k$ integral is scaleless, and thus vanishing. The upshot is that we can drop the $l^2$ from term (IIa'), and in fact, all powers of $l^2$ in the numerator of \Eq{eq:QED-1loop-integrand}. 

Term (IIa') is then proportional to $p_1^2$, and involves a pinch of the $p_1$ propagator. Diagrammatically, this implies that instead of attaching to the internal scalar line, the photon attaches to the insertion point $x_1$. Mathematically we see this by evaluating the $p_1$ integral, obtaining a position-space delta function $\delta(y_1-x_1)$. This allows the $y_1$ integration to be done, trivially, and importantly it fixes the velocity field $u(y_1)$ to $u(x_1)\equiv u_1$. A completely analogous manipulation happens in term (IIb'), where the $p_2$ propagator pinches and the photon effectively attaches to the point $x_2$. In term (III'), both the $p_1$ and $p_2$ propagators pinch and the photon attaches to $x_1$ and $x_2$. 

The upshot of these manipulations is remarkable. In \Eq{eq:QED-1loop-integrand} we have a photon propagator which completely breaks translation invariance via its dependence on a fluid velocity field $u^{\mu}(y)$ that gets doubly integrated over spacetime. For an arbitrary fluid field configuration it appears impossible to evaluate \Eq{eq:QED-1loop-integrand}. However we have just demonstrated that we can simply replace the spacetime dependent fields by constants $u(y_1) \rightarrow u(x_1)\equiv u_1$ and $u(y_2) \rightarrow u(x_2)\equiv u_2$. The only value of the fluid velocity that matters is the value at the position of the $\phi$ operator insertions. 

If we use this simplification, and drop the powers of $l^2$ in the numerator as discussed, we can trivially evaluate the $y_1, y_2$ integrals to obtain two momentum conserving delta functions $\delta(p_1-l+k)\delta(p_2+l-k)$. Using these to eliminate the $p_2$ and $l$ integrals, we obtain
\eq{
&\langle \Omega | \Phi^\dagger(x_1) \Phi(x_2)  | \Omega \rangle\Big|_{g^2}  =g^2  \int_{p} \frac{e^{-ip (x_1-x_2)}}{(p^2)^2}\int_{k} \frac{1}{(p+k)^2\,k^2} 
      \times \Bigg[ 
      \underbrace{(2p+k)^2}_{\textrm{I'}} \\
      &- \underbrace{p^2 \frac{(2p+k) \!\cdot\! u_1}{-k\!\cdot\!u_1}}_{\textrm{IIa'}} 
      + \underbrace{p^2 \frac{(2p+k)\!\cdot\! u_2}{k\!\cdot\!u_2}}_{\textrm{IIb'}}
      - \underbrace{(p^2)^2 \frac{u_1\!\cdot\! u_2}{(-k\!\cdot\!u_1)(k\!\cdot\!u_2)}}_{\textrm{III'}}
      \Bigg].
}{eq:QED-1loop-integrand2}
After this simplification we can see explicitly that each of the terms (I'), (IIa'), (IIb'), and (III') in \Eq{eq:QED-1loop-integrand2} coming from the $u(x)$-gauge calculation agrees exactly with the corresponding  terms of (I), (IIa), (IIb), and (III) in the dressing calculations in Eqs. (\ref{QED_loop_I})-(\ref{QED_loop_III}). Therefore we have shown, at loop level, that all dressing effects are encoded into the $u(x)$-gauge photon propagator!

\subsubsection{Loop Integration}\label{sec:QEDloopintegral}

Now that we have demonstrated the loop-level equivalence between dressing and using $u(x)$ gauge, we will evaluate the loop integral to understand the physical effects of dressing. The dressed correlation function can be written as a self-energy expansion, \Eq{eq:QED_dressed_loop}, with the leading contribution given by
\eq{
    \Sigma^{\textrm{dressed}}(p)=-ig^2 \mu^{4-D}\int_k \frac{N_{\mu\nu}(k,u)}{k^{2}+i\varepsilon}\frac{(2p-k)^\mu (2p-k)^\nu}{(p-k)^{2}+i\varepsilon}\,,
}{}
where the relevant photon numerator is
\eq{
    N_{\mu\nu}(k,u)=-\eta^{\alpha\beta}P^{(1)}_{\mu \alpha}\overline{P^{(2)}_{\nu \beta}}\,.
}{eq:photonnumerator}
We again have the spatial projector \Eq{eq:spatialprojector} and its complex conjugate, and we have rerouted the direction of the loop momentum $k$ to conform with conventional Feynman integral tables~\cite{Smirnov2012AnalyticTools}. In anticipation of the $D=4$ limit we have defined a renormalization scale $\mu$.

Due to the projectors in \Eq{eq:photonnumerator}, the dressed self-energy involves integrals of the form
\eq{
G(\lambda_{1},\lambda_{2},\lambda_{3},\lambda_{4})=\int_k \frac{1}{(k^{2}+i\varepsilon)^{\lambda_{1}}[(p-k)^{2}+i\varepsilon]^{\lambda_{2}}(2k\!\cdot\!u_1+i\varepsilon)^{\lambda_{3}}(-2k\!\cdot\!u_2+i\varepsilon)^{\lambda_{4}}}\,.
}{eq:int_family}
Generally, these integrals depend on six independent Lorentz scalars and so their evaluation is quite challenging. 

With the simplifying choice $u_2=-u_1$ (discussed in the subsequent section), the integrals depend on only three Lorentz scalars, $(u_1^2, p^2, p \!\cdot\! u_1)$, and each integral has $\lambda_4=0$. For this choice we have evaluated the loop integral, yielding the dressed scalar QED self-energy,
\eq{
&\Sigma^{\textrm{dressed}}(p)\\
&= \frac{g^2}{8\pi^2}\bigg[
\frac{6 p^2}{D-4}+2p^2\left(2\log\left(\frac{-p^2}{\tilde{\mu}^2}\right)-\log\left(\frac{-\omega}{\tilde{\mu}}\right)\right)-4p^2+\frac{\omega^2\sin^{-1}(\sqrt{y})}{\sqrt{y(1-y)}}\left(2+\log\left(\frac{-p^2}{\omega^2}\right)\right) \\
&\hspace{35pt}
-\omega^2\,{}_2F_1^{(0,1,0,0)}(1,1,\tfrac{3}{2},y)
-\omega^2\,{}_2F_1^{(0,0,1,0)}(1,1,\tfrac{3}{2},y)+\OO(D-4)
\bigg]\,,
}{eq:QED1loopselfenergy}
where we have introduced the shorthand variables $\omega=2p \cdot u_1+i\varepsilon$ and  $y=(p \cdot u_1+i\varepsilon)^{2}/(p^2+i\varepsilon)$,  and used $u_1^2=1$. The branch cut prescription is determined by the replacement rule $-p^2\rightarrow -(p^2+i\varepsilon)$. Details of the integration are given in \App{app:one_loop}. As with the undressed $R_{\xi}$-gauge result, \Eq{eq:selfenergyrxi}, there is a UV divergence which can be absorbed by a wavefunction renormalization counterterm, and it is accompanied by a log describing the running of this counterterm. 

Relative to the undressed result computed in $R_{\xi}$ gauge, however, the presence of $\log(-\omega)$,  $\sin^{-1}$ and ${}_2F_{1}$ functions in Eq. \eqref{eq:QED1loopselfenergy} introduces a considerably more interesting analytic structure. This suggests nontrivial long-distance modifications to the correlation function in position space, but we leave a thorough analysis to future work.

\subsection{Dressing Induces Kinematic Singularities}

A central observation of this work is that dressing actually generates new analytic structures in the kinematic variables of local operators.  We saw this in both the tree- and loop-level calculations.  Here we take stock of these results.

\subsubsection{Tree-Level Singularities}

Our expression for the dressed tree-level matrix element $\langle \Omega | \Phi^\dagger(x_1) \Phi(x_2) | A(k) \rangle$ is summarized in \Eq{QED_dressed_tree}.  This formula exhibits poles of the form $1/u_1\cdot k$ and $1/u_2\cdot k$.  So if $k$ is orthogonal to the fluid velocity at $x_1$ or $x_2$, then the correlator actually diverges. This has an obvious physical interpretation.  Since the worldlines accommodating the dressing are dynamical fields, they exhibit singularities when they go on shell. These nonanalyticities are precisely activated when the gauge boson has zero frequency, as viewed from the rest frame of the fluid.  Given the interpretation of $u^{\mu}$ as the timelike four-velocity field of a fluid, if $k$ is the momentum of an on-shell gauge boson then $u\!\cdot\!k=0$ only in the soft limit. 

This implies that the operator product $\Phi^\dagger(x_1) \Phi(x_2)$ has singular matrix elements.  An important corollary of this is that dressing is not an innocuous correction to the undressed product.  Rather, it has an enormous, infinite correction, at least when probing the matrix elements near the soft limit.  Consequently, dressing is always detectable by appropriately soft on-shell gauge bosons.

This soft pole is precisely the familiar soft pole in scattering amplitudes with external charged particles, which is described by soft theorems. One does not typically see such soft poles in correlation functions, which usually describe processes in a compact region. The appearance of the soft pole in the dressed correlator is actually quite intuitive. Gauge invariance demands that charge cannot be created or destroyed, so the charged particles created and destroyed by the operators $\phi^{\dagger}(x_1), \phi(x_2)$ are necessarily accompanied by Wilson lines that carry the charge entering $x_1$ and exiting $x_2$. The Wilson lines then play the role of hard ``external'' charged particles propagating to infinity. The soft photon then sees these ``external legs'' as it typically does in scattering amplitudes and produces a soft pole for the usual reasons.  In particular, processes with a finite number of soft photons are unphysical, requiring instead inclusive quantities or dressed features like soft photon clouds.

When $u_{1}=u_{2}$ the soft pole in \Eq{eq:QEDtreedressedprojectors} vanishes. This too has an explanation in the language of scattering amplitudes. In the $u_1=u_2$ configuration the Wilson lines emerge from past infinity along the same point on the celestial sphere, implying net zero charge coming from that direction. So soft photons see this pair of parallel Wilson lines as inert and there is no soft pole from coupling to them.  In other words, there is a vanishing hard contribution to the large gauge charge at the point on the celestial sphere from which the Wilson lines emerge.  Hence no soft radiation is required to dress the state---in the Faddeev-Kulish sense of dressing to neutralize large gauge charge---nor is soft radiation produced by the abrupt termination of this pair of parallel Wilson lines when they land on the operator $\phi^{\dagger}(x_1)\phi(x_2)$.

\subsubsection{Loop-Level Singularities}
\label{sec:poleprescription}

Our results lean heavily on loop integration methods, which depend in general on prescriptions for integrating around poles.  Remarkably, this is an instance where physical insight into these dressing fields offers some guidance on the pole prescription.

Historically, past literature on temporal or axial gauge calculations has discussed at length ambiguities in the pole  prescription for the $k\!\cdot\!u$ denominator. A variety of mathematical methods were proposed for treating this pole, but a clear physical prescription remained lacking~
\cite{RevModPhys.59.1067, Alekseev1991}. For example, a principal value prescription was popular for its convenience at one loop, despite its failure to properly describe the exponentiation of Wilson loops \cite{CARACCIOLO1982311}.

In the present work, the mechanics of dressing provides a physical interpretation of this pole, as well as a motivated prescription for continuation around the pole. Our local operators are tethered to past infinity by straight Wilson lines with future-directed four-velocities $u_1, u_2$. It is precisely these infinitely-long Wilson lines which produce the $k\!\cdot\!u_i$ denominators. For example, with a Wilson line along $u_1^{\mu}$, writing \Eq{Wilson_line} in momentum space gives
\eq{
W_1(x)&= \exp\left[ig\int_k e^{-ikx}u_{1}^{\mu}A_{\mu}(k) \int_{-\infty}^{0}ds\, e^{-is (ku_{1}+i\varepsilon)}\right] \\
&=\exp\left[ig\int_k e^{-ikx}u_{1}^{\mu}A_{\mu}(k) \frac{i}{k\!\cdot\!u_1+i\varepsilon}\right].\,
}{}
This is an important point that we would like to reiterate. In gauge invariant calculations, for example in on-shell scattering amplitudes, the unphysical poles that arise from the gauge fixed photon propagator are never activated and one need not worry about the pole prescription.  When calculating dressed observables, however, the gauge fixing becomes physical in that it encodes the dynamics of a physical object that tethers our system to infinity. The orientation of the tethering decides the pole prescription. 

The $i\varepsilon$ prescription is a direct consequence of our choice to tether to past infinity. One can just as well anchor to the future by choosing the four-velocity to be past-directed, or anchor to spatial or null infinity by choosing $u$ to be spacelike or null. In this work we restrict to timelike $u$.

The integrals in \Eq{eq:int_family} are difficult to evaluate because they depend on multiple independent Lorentz scalars. To proceed we found it necessary to simplify the problem and restrict to a particular orientation of $u_{1}$ and $ u_{2}$. A natural first choice would be $u_2=u_1$, corresponding to parallel Wilson lines. However one can readily see from \Eq{eq:int_family} that when $u_2=u_1$ the integral has a pinch singularity at $k \cdot u_1=0$. Such a singularity is not regulated by dimensional regularization and actually describes a genuine physical feature. 

The pinch singularities come only from Diagram (III), where the two Wilson lines interact directly with each other. To understand the singularity it can be helpful to work in position space. To focus on the subdivergence we will ignore the $\phi^{\dagger}\phi$ insertions and look directly at the correlator of Wilson lines. Since the time evolution operator is $T\{\exp(-i\int dt H(t))\}$, and these states have been evolved for an infinite amount of time, the imaginary part of the argument above is the time-integrated energy, 
\eq{
 \int_{-\infty}^{0} dt E(t
) = -\text{Im}\log\langle \Omega |W^{\dagger}_{1}(x)W_{2}(0)|\Omega\rangle\,.
}{eq:phase}
When the energy is constant this factor is proportional to the total time of evolution, and when the states are evolved for an infinite amount of time we expect a divergence in any dimension $D$. This is precisely what happens when the Wilson lines are taken parallel, as we demonstrate below.

Ignoring loops of charged particles, which will not qualitatively change our conclusions, we can evaluate this Wilson line correlation function exactly in free Maxwell theory. Defining the conserved current,
\eq{
J^{\mu}(z)=-g\int_{-\infty}^{0}ds\,u_{1}^{\mu} \delta^{(D)}(z-x-u_1 s)+g\int_{-\infty}^{0}ds\,u_{2}^{\mu} \delta^{(D)}(z-u_2 s)\,,}{}
we find that
\eq{
\langle 0 |W^{\dagger}_{1}(x)W_{2}(0)|0\rangle &= \big\langle 0\big|e^{i\int_y\,J^{\mu}(y)A_{\mu}(y)} \big| 0 \big\rangle  \\
&= \exp\left[-\frac{1}{2} \int_{y,\,y'}  J^{\mu}(y)J^{\nu}(y') \langle A_{\mu}(y) A_{\nu}(y')\rangle\right]  \\
&=\exp\left[\frac{1}{2}\frac{\Gamma(D/2-1)}{4\pi^{D/2}} \int_{y,\,y'}\,\frac{J(y) \!\cdot\! J(y')}{(-(y-y')^2+i\varepsilon)^{D/2-1}}\right]\,.
}{eq:wilsonwilson}
In the product of currents, the terms where dressings self-interact give scaleless integrals which are zero in dimensional regularization. The only remaining terms are those for which the Wilson line dressings interact with each other,
\eq{
\langle 0 |W^{\dagger}_{1}(x)W_{2}(0)|0\rangle = \exp\left[-g^2\frac{\Gamma(D/2-1)}{4\pi^{D/2}} u_{1} \!\cdot\! u_{2}\int_{-\infty}^{0} ds ds' \,\frac{1}{(-(x+u_1 s- u_2 s')^2+i\varepsilon)^{D/2-1}}\right]\,.
}{eq:photonpotential}
For generic $u_1, u_2$ the integral scales as $(-x^2+i\varepsilon)^{2-D/2}$. In $D>4$ it is finite, and in $D=4$ it features the well-known Coulomb logarithm describing the IR divergent phase accumulated between asymptotic particles. This IR divergence is clearly dimension-dependent, occurs for generic $u_1, u_2$, and is already well understood as originating from the $1/r$ potential not falling off fast enough to the asymptotic states to be truly free. 

If we take $u_2=u_1$ in \Eq{eq:photonpotential} the integrand depends only on the relative time $s-s'$, and not the total time $T=-(s+s')/2$, which runs from $0$ to infinity. We can evaluate the relative time integral in general $D$ in terms of hypergeometric functions, although its precise form is not important here. What matters is that when we expand in the long time limit there is a constant term which accumulates over the total $T$ integration, 
\eq{
\langle 0 |W^{\dagger}_{1}(x)W_{1}(0)|0\rangle &= \lim_{T_{\max}\rightarrow \infty}\exp\left[i\frac{g^2\,\Gamma((D-3)/2)}{4\pi^{\frac{D-1}{2}}|\vec{x}|^{D-3}}\int^{T_{\max}}_{0} dT \left(1+\OO(T^{3-D})\right)\right] \\ 
&\underset{D=4}{=}\lim_{T_{\max}\rightarrow \infty}\exp\left[\frac{ig^2}{4\pi |\vec{x}|}\left(T_{\max}+\OO(\log(T_{\max}))\right)\right]\,,
}{}
and we recognize the Coulomb energy of a pair of oppositely charged Wilson lines multiplied by a divergent total time integral. This is the total time divergence in \Eq{eq:phase}. Had we integrated over $s,s'$ while still in momentum space, this divergence would instead manifest as a pinch singularity at zero frequency, $k\!\cdot\!u=0$.

To avoid such a divergence at $u_1 = u_2$, we instead made the next simplest choice, $u_2 = -u_1$, so that $u^{\mu}_2$ is a past-directed timelike vector.  Mathematically, this choice eliminates the pinch singularity by ensuring that all integrals are in the simpler family $G(\lambda_1,\lambda_2,\lambda_3,0)$. Physically, in the two-point correlator of dressed operators this implies that the operator which creates the charged particle is dressed to past infinity whereas the operator which annihilates the charged particle is dressed to future infinity. 

One expects that with a suitable regularization method and renormalization of the ground state energy, reliable calculations could be performed for $u_1 = u_2$. This orientation would be particularly interesting to study because it aligns better with the physical intuition that the dressing should describe the state preparation conducted by an experimentalist. However we leave this more involved calculation to future work.

\section{Gravity}\label{sec:gravity}

With the case of QED as a guide, we now turn to perturbative gravity. Much of the construction will parallel gauge theory, but with important technical and conceptual differences. We first define gauge-invariant, dressed bulk observables in gravity and relate them to a convenient gauge choice. We then perturbatively compute the corresponding undressed bulk correlators before adding dressing to restore diffeomorphism invariance.

Our system is a gravitationally interacting massless scalar whose Lagrangian is
\eq{
  S = \int d^D\z \sqrt{-g}\left(-\frac{2}{\kappa^2}R+\frac{1}{2}g^{\mu\nu}\nabla_\mu\phi\nabla_\nu\phi\right)\,,
}{eq:gravityaction}
where the gravitational coupling constant is defined by $\kappa^2=32\pi G$. Dressed observables can of course be computed in any curved background but for ease of calculation we will focus on perturbations about flat spacetime, $\kappa h_{\mu\nu} = g_{\mu\nu}-\eta_{\mu\nu}$. At $\OO(\kappa)$, the action for this setup is
\eq{
S=\int d^Dx \left(\frac{1}{2}P_4^{\alpha\beta\gamma\delta}\partial_{\mu}h_{\alpha\beta}\partial^{\mu}h_{\gamma\delta} + \frac{1}{2}\partial_{\mu}\phi\partial^{\mu}\phi-\frac{\kappa}{2}h_{\mu\nu}T^{\mu\nu}\right)\,,
}{}
where we have defined the tensor
\eq{
P_{D}^{\alpha\beta\gamma\delta}=\frac{1}{2}\left(\eta^{\alpha\gamma}\eta^{\beta\delta}+\eta^{\alpha\delta}\eta^{\beta\gamma}-\frac{2}{D-2}\eta^{\alpha\beta}\eta^{\gamma\delta}\right)\,,
}{}
in terms of which the stress tensor is\footnote{While the appearance of $P_{4}$ in the $D$-dimensional action may seem peculiar, it arises directly from the expansion of $\sqrt{-g}g^{\mu\nu}$. In general $D$ spacetime dimensions, $P_D$ is the inverse of $P_4$, which explains its appearance in the graviton propagator.}
\eq{
T^{\mu\nu} = P_4^{\mu\nu\alpha\beta}\partial_{\alpha}\phi\partial_{\beta}\phi\,.
}{}
Here we have used de Donder gauge fixing in the action.  Since all dressed observables are gauge invariant, this choice is simply a matter of convenience. In terms of Feynman rules, we have the de Donder gauge graviton propagator $iP_{D}^{\alpha\beta\mu\nu}/k^2$, scalar propagator $i/p^2$, and the three-point $h\phi\phi$ interaction vertex,
\begin{equation}
    \vcenter{\hbox{
    \begin{tikzpicture}
    \begin{feynman}
      \vertex[] (x1) {}; 
      \vertex[right=2.6 of x1] (x2) {};
      \vertex (m) at ( $(x1)!0.5!(x2)$ );
      \vertex[below=1 of m] (f) {\(\mu\nu\)};

      \diagram*{
        (x1) --[plain, reversed momentum={[arrow shorten=0.2]\(p_1\)}] (m) --[plain, momentum={[arrow shorten=0.2]\(p_2\)}](x2),
        (f) --[graviton] (m),
      };
    \end{feynman}
    \end{tikzpicture}}} = i\kappa p_{1}^{\alpha}p_{2}^{\beta}P_{4\,\,\alpha\beta\mu\nu}\,.
\label{eq:grav-vertices}
\end{equation}
In this section, coordinates on the curved spacetime manifold will be denoted by $\z$, whereas  $x$  will describe flat spacetime, matching the notation of \Sec{sec:gaugetheory}. Additionally, the four-velocity vector field in curved spacetime is denoted $\UU$, whereas $u$ is used in flat spacetime. The distinction between these notations will be important in subsequent sections when we work perturbatively about flat spacetime. 

\subsection{Dressed Operators}

In gauge theory, we reviewed in detail how a local operator $\phi(x)$ with charge $g$ can be made gauge invariant by attaching a Wilson line $W(x)=\exp\left[ig\int_{\mathcal{C}}A\right]$, where $\mathcal{C}$ is any curve between the bulk point $x$ and infinity. In \Sec{sec:gaugetheory} we made the choice that all Wilson lines are straight lines, or equivalently, integral curves of a velocity field satisfying $u^{\nu}\partial_{\nu}u^{\mu}=0$. For the case of gauge theory, we made this choice as a matter of technical convenience, but in principle arbitrary curves could have instead been chosen.

In gravitational systems, the situation is far more restrictive. We cannot specify arbitrary curves in spacetime in a diffeomorphism invariant manner. There is, however, a physically motivated set of curves that we can use, which are geodesic congruences. Given a space-filling, pressureless dust, each dust element follows a geodesic in the spacetime for which  the tangent to each curve is a timelike vector field satisfying $\UU^{\nu}\nabla_{\nu}\UU^{\mu}=0$. 

Each dust element is a particle following the geodesic $\z^{\mu}(Y^{A})$, where $\z^{\mu}$ are coordinates on spacetime.  Meanwhile, $Y^{A}=\{y^{a}, \tau\}$ consists of the point $y^a$ on the initial data surface from which the geodesic originates, and the proper time $\tau$ elapsed along that geodesic. The geodesic $\z^\mu$ also has an implicit dependence on the initial conditions of the dust field, $u(y^a)=U|_{\Sigma}(y^a)$, which are specified on the initial data surface $\Sigma$. Since this surface is in the asymptotic past, both its coordinates as well as the initial data specified on the surface are physical data in our theory. Consequently, they are \textit{not} acted upon by diffeomorphisms. Given a set of initial conditions for the dust field, one solves for $\UU^{\mu}(x)$ throughout spacetime, which then defines $\z^{\mu}(Y^{A})$ in the bulk.\footnote{For generic spacetimes, this map may fail to be bijective due to caustics, orbits, singularities, etc., and one may need an atlas of different maps (see \cite{Goeller:2022rsx}). Here we will work perturbatively in a weak field limit so no such issues arise. Caustics can form, in principle, after a proper time of order $\tau\sim 1/\sqrt{\textrm{curvature}}\sim 1/\sqrt{\kappa\partial\partial h}$, which is long when $h$ is perturbatively small. For oscillating or fluctuating backgrounds, this formation time may be parametrically increased further in $h$ if the average shear is vanishing, because then only its variance will drive focusing. To linear order in $h$, which we restrict ourselves to in this work, focusing can be neglected.} In this work we assume the dust field to be sufficiently light that it does not backreact, and indeed follows geodesic motion. We discuss the validity of this assumption in \Sec{sec:gravity_physical_interpetation}.

\subsubsection{Dressed Scalars}

Under a diffeomorphism $\z\rightarrow \z^\prime = f(\z)$, dynamical fields transform via the pushforward 
\eq{
\phi(\z) \rightarrow \phi'(\z^\prime) = \phi(f^{-1}(\z^\prime))\,.
}{}
The coordinates along the geodesics transform as
\eq{
\z(Y^{A})\rightarrow \z^\prime(Y^{A}) = f(\z(Y^A))\,.
}{}
We can then define the manifestly diffeomorphism invariant scalar field operator
\eq{
\Phi(Y^A)\equiv \phi(\z(Y^A))\,.
}{eq:relationalscalar}
The physical interpretation of $\Phi(Y^A)$ is straightforward.  Starting from a point $y^a$ in the asymptotic past, take a clock and release it to free fall in unison with a dust field.  They then evolve dynamically together under the influence of the metric.  After a time $\tau$ elapses on the clock, we then insert the $\phi$ operator.  Note that this picture is manifestly independent of how one chooses to coordinatize the bulk spacetime.

To perform explicit calculations involving these operators in quantum gravity, we will work to leading order in perturbations about flat spacetime. In this case the diffeomorphisms can be linearized, $\z^{\mu} \rightarrow \z^{\mu} - \kappa \xi^{\mu}(\z)$, and act on our scalar field as
\eq{
\delta_{\xi}\phi(\z) = \phi(f^{-1}(\z)) - \phi(\z) = \kappa\xi^\mu(\z)\partial_\mu\phi(\z) + \mathcal{O}(\kappa^2) \, ,
}{eq:diffscalar}
and on the metric perturbation and the Levi-Civita connection as
\eq{
\delta_{\xi} h_{\mu\nu} (\z) &= \partial_\mu\xi_\nu(\z) + \partial_\nu\xi_\mu(\z) + \mathcal{O}(\kappa), \\
\delta_{\xi} \Gamma^{\mu}_{\ \alpha\beta}(\z) &= \kappa\partial_{\alpha}\partial_{\beta}\xi^{\mu}(\z) + \mathcal{O}(\kappa^2).
}{eq:diffmetric}

To calculate $\Phi(Y^A)$ perturbatively requires an expression for the geodesic. To obtain this we compute the geodesics as perturbations about a straight line\footnote{The coordinate $y^a$ lives on the ($D-1$)-dimensional boundary surface $\Sigma$ where we specify initial data, while the corresponding location in the $D$-dimensional spacetime is denoted by $y^\mu$. In our setup $\Sigma$ lies at the past asymptotic boundary. See, however, \Sec{sec:gravity_physical_interpetation} for comments on this setup when quantum fluctuations of the worldline are considered.
},
\eq{
\z^{\mu}(Y^A) = y^{\mu} + u^{\mu}(y)(\tau-\tau_0)+w^{\mu}(Y^A)\,,
}{}
where $u^\mu(y)=U^\mu|_{\Sigma}(y)$ is the initial velocity specified on the boundary, and the deflection $w^{\mu}$ is an $\OO(\kappa)$ function that vanishes at the initial proper time $\tau=\tau_0$. At zeroth order, the bulk point evaluated at $\tau=0$ would be $x^\mu\equiv y^\mu-u^\mu(y)\tau_0$, and the velocity would be a constant along the line, $u^\mu(x)=u^\mu(y)$. Treating $x$ as a reference point, we can just as well take the invariant data that specifies bulk points to be $Y^A=\{x^{\mu}\equiv y^{\mu}-u^{\mu}\tau_0, \tau\}$, and write\footnote{We reiterate that $x^{\mu}$ is not a bulk coordinate on the curved spacetime manifold, and is not acted upon by diffeomorphisms. Rather it is a particular linear combination of boundary data. Its utility comes from the fact that it can be thought of as a bulk coordinate on an auxiliary flat spacetime in which the same $\{y^{a}, u^{\mu}, \tau\}$ are specified. As a reminder, $\z^{\mu}$ denotes the bulk coordinates on the curved manifold. The four-velocity vector field in curved spacetime is $\UU$, while $u$ is used in the auxiliary flat spacetime and specified by the initial data.} 
\eq{
\z^{\mu}(Y^A) = x^{\mu} + u^{\mu}(x)\tau+w^{\mu}(Y^A)\,.
}{}  
This choice is convenient for two reasons. First,  $x^{\mu}$ remains finite as we simultaneously take the initial point $y^{\mu}$ and the initial time $\tau_0$ to past infinity. Second, at time $\tau=0$ the difference $w^{\mu}(x,\tau) = \z^{\mu}(x,\tau)-x^{\mu}$ between the true position of the geodesic and the reference point is $\OO(\kappa)$.

Next, we consider the geodesic equation,
\eq{
\frac{d^2\z^{\mu}(Y^A)}{d\tau^{2}}+\Gamma^{\mu}_{\alpha\beta}(\z(Y^A))\frac{d\z^{\alpha}(Y^A)}{d \tau}\frac{d\z^{\beta}(Y^A)}{d \tau}=0\,,
}{}
which reduces at $\OO(\kappa^0)$ to
\eq{
u^{\nu}(x)\frac{\partial}{\partial x^{\nu}}u^{\mu}(x)=0\,,
}{}
and at \(\mathcal{O}(\kappa^1)\) gives the equation for the position deflection,
\eq{
\frac{d^2w^{\mu}(x,\tau)}{d\tau^{2}}+\Gamma^{\mu}_{\alpha\beta}(x+u(x)\tau)u^\alpha(x) u^\beta(x)=0\,.
}{}
The field $w^{\mu}(x,\tau)$ will depend on $x$ only in the combination $x^{\mu}+u^{\mu}(x)\tau$. Without loss of generality we can henceforth set $\tau=0$, and we can restore finite $\tau$ at the end of calculations by simply shifting $x^{\mu}\rightarrow x^{\mu}+u^{\mu}(x)\tau$.

Putting this all together, we find that the dressed scalar field is
\eq{
	\Phi(x) = \phi(x)+w^{\mu}(x)\partial_{\mu}\phi(x)+\OO(\kappa^2)\,,
}{}
where the solution for the deflection is the double line integral\footnote{It is also equivalent to write the solution as a single integral that convolves the right-hand side with the worldline propagator, which inverts the $d^2/d\tau^2$ operator.}
\eq{
	w^\mu(x) = - \int_{-\infty}^0 ds \int_{-\infty}^{s} ds^{\prime} \,  \Gamma^\mu_{\ \alpha\beta} (x+u(x)s^{\prime}) u^\alpha(x) u^\beta(x)\,.
}{}
Using \Eq{eq:diffmetric} we can verify how $w^{\mu}$ transforms under linearized diffeomorphisms, 
\eq{
	\delta_{\xi} w^\mu(x)
	& = 	- \kappa \int_{-\infty}^0 ds \int_{-\infty}^{s} ds^{\prime} \,  \partial_\alpha\partial_\beta\xi^\mu(x+u(x)s^{\prime}) u^\alpha(x) u^\beta(x) \\
	& = - \kappa \int_{-\infty}^0 ds \int_{-\infty}^{s} ds^{\prime} \, \frac{d^2 \xi^\mu(x+u(x)s^{\prime})}{ds^{\prime}{}^2} \\
	& = -\kappa \xi^\mu(x),
}{}
where we assumed that $\xi^\mu(x)$ vanishes at infinity. Together with \Eq{eq:diffscalar}, we then confirm linearized diffeomorphism invariance of the dressed scalar,
\eq{
\delta_{\xi}\Phi(x)
& = \delta_{\xi}\phi(x)+\delta_{\xi} w^\mu(x)\partial_\mu\phi(x)\\
& = \kappa \xi^\mu(x) \partial_\mu\phi(x) - \kappa\xi^\mu(x)\partial_\mu\phi(x) = 0\,.
}{}
For perturbative calculations it will be useful to Fourier transform the deflection,
\eq{
w^\mu(x)
&= - \int_{-\infty}^0 ds \int_{-\infty}^{s} ds^{\prime} \int_{k} e^{-ik(x+u(x)s^\prime)} \,  \Gamma^\mu_{\ \alpha\beta}(k) u^\alpha(x) u^\beta(x)\\
&= \int_k e^{-ikx} \frac{\Gamma^\mu_{\ \alpha\beta}(k) u^\alpha(x) u^\beta(x)}{[k\!\cdot\!u(x)]^2}\\
& = -i\kappa\int_k e^{-ikx} \frac{k_\beta h^\mu_\alpha(k)+k_\alpha h^\mu_\beta(k)-k^\mu h_{\alpha\beta}(k)}{2[k\!\cdot\!u(x)]^2} u^\alpha(x) u^\beta(x)\\
& = -i\kappa\int_k e^{-ikx} \frac{\OO^{\mu\alpha\beta}(k,u(x)) h_{\alpha\beta}(k)}{[k\!\cdot\!u(x)]^2}\,,
}{eq:w-fourier}
where in the numerator we have defined
\eq{
\OO^{\mu\alpha\beta}(k,u(x)) = \frac{1}{2}\big([k\!\cdot\!u(x)]u^{\alpha}(x)\eta^{\beta\mu}+[k\!\cdot\!u(x)]u^{\beta}(x)\eta^{\alpha\mu}-k^{\mu}u^{\alpha}(x)u^{\beta}(x)\big)\,.
}{eq:OO}
Then the dressed scalar $\Phi(x)$ can be expressed as follows, with the leading-order dressing displayed explicitly,
\eq{
\Phi(x) &= \phi(x)+w^\mu(x)\partial_\mu\phi(x)+\cdots \\
&= \phi(x) + \int_{p} e^{-ipx} \phi(p) \int_{k} e^{-ikx} \kappa h_{\alpha\beta}(k) \frac{-p_{\mu}\OO^{\mu\alpha\beta}(k,u(x))}{ [k\!\cdot\!u(x)]^2} + \cdots,
}{}
which gives the following Feynman vertex for worldline single-graviton emission/absorption:
\begin{equation}
\tikzfeynmanset{
every edge/.style={thick},
}
\vcenter{\hbox{
    \begin{tikzpicture}
    \begin{feynman}
      \vertex[dot, label=90:\(x\)] (x1) {}; 
      \vertex[right=1.3 of x1] (x2) {};
      \vertex[below=2.6 of x1] (w1) {\(u\)};
      \vertex[below=2.6 of x2] (w2) {};
      \vertex (g1) at ( $(x1)!0.5!(w1)$ );
      \vertex (g2) at ( $(x2)!0.5!(w2)$ );
    
      \diagram*{
        (x2) --[plain, momentum={[arrow shorten=0.3] \(p\)}] (x1),
        (x1) --[scalar, blue] (w1),
        (g2) --[graviton] (g1),
        (g2) --[draw=none, momentum={[arrow shorten=0.3] \(k,\alpha\beta\)}] (g1),
      };
    \end{feynman}
    \end{tikzpicture}}} = \int_{p,k}e^{-ipx-ikx}\,\frac{-\kappa p^{\mu}\OO_{\mu\alpha\beta}(k,u)}{(k\!\cdot\!u)^2}\,.
\label{eq:WL-graviton-feynman}
\end{equation}
Note that the reversal of any momentum arrow introduces a minus sign to the corresponding momentum, according to the conventions specified in Eq.~\eqref{eq:momentum-arrow}.

\subsubsection{Dressed Gravitons}

We are now equipped to define the dressed graviton field and Levi-Civita connection
\eq{
\hh_{\mu\nu}(x) &= h_{\mu\nu}(x) + \kappa^{-1}\partial_\mu w_\nu(x) + \kappa^{-1}\partial_\nu w_\mu(x), \\
\GG^\mu_{\ \alpha\beta}(x) &= \Gamma^\mu_{\alpha\beta}(x) + \partial_\alpha\partial_\beta w^\mu(x).
}{}
Under a diffeomorphism, we find that $\delta_{\xi} \hh_{\mu\nu}(x)=0$ and $\delta_{\xi} \GG^\mu_{\ \alpha\beta}(x)=0$, so the dressed metric perturbation and Levi-Civita connection are invariant. Also, we can show that they are orthogonal to the velocity field. First, for $\GG^{\mu}_{\ \alpha\beta}$,
\eq{
	\GG^{\mu}_{\ \alpha\beta}(x) u^\alpha(x) u^\beta(x)
	& = \Gamma^{\mu}_{\ \alpha\beta}(x) u^\alpha(x) u^\beta(x) \\
    & -\int_{-\infty}^0 ds \int_{-\infty}^{s} ds^\prime \,u^\alpha(x)u^\beta(x)\partial_{\alpha} \partial_{\beta} \left( \Gamma^\mu_{\nu\rho} (x+u(x)s^\prime)u^{\nu}(x)u^{\rho}(x)\right) = 0,
}{}
where the integral gives only boundary terms, which then cancel the previous term. Here we used the condition $u^\nu\partial_\nu u^\mu(x) = 0$ and assumed that $\Gamma^\mu_{\ \nu\rho}$ vanishes at infinity, in other words flat boundary conditions. For $\hh_{\mu\nu}$ the expression is lengthier but it is most conveniently presented, and without loss of generality, for a single Fourier mode, $h_{\mu\nu}(x)=h_{\mu\nu}(k)e^{-ikx}$. For this we have three contributions,
\eq{
u^{\mu}h_{\mu\nu}(x) &= e^{-ikx}(u\!\cdot\! h)_{\nu}\,, \\
\kappa^{-1}u^{\mu}\partial_{\mu}w_{\nu}(x) &= e^{-ikx} \frac{-i}{(u\!\cdot\! k)}\left[-i(u\!\cdot\! k)(u\!\cdot\! h)_{\nu}+\frac{i}{2}k_{\nu}(u\!\cdot\! h\!\cdot\! u)\right]\,,\\
\kappa^{-1}u^{\mu}\partial_{\nu}w_{\mu}(x) &=-\frac{k_{\nu}}{2}e^{-ikx}\frac{(u\!\cdot\!h\!\cdot\!u)}{(u\!\cdot\!k)}+ie^{-ikx}\frac{(u\!\cdot\!h\!\cdot\!u)}{(u\!\cdot\!k)^2}k_{\alpha}\partial_{\nu}u^{\alpha}\\
&-ie^{-ikx}\frac{u^{\alpha}\partial_{\nu}u^{\beta}}{(u\!\cdot\!k)^2}\left[k_{\alpha}(u\!\cdot\!h)_{\beta}+k_{\beta}(u\!\cdot\!h)_{\alpha}-(u\!\cdot\!k)h_{\alpha\beta}\right]\,,
}{}
which nicely sum to zero,
\eq{
u^\mu(x)\hh_{\mu\nu}(x) = 0\,.
}{}
Mirroring the QED case, this again suggests that we define an inhomogeneous gauge condition, here also called $u(x)$ gauge,
\eq{
	u^\mu(x)h_{\mu\nu}(x) &= 0 \,,\\
	\Gamma^{\mu}_{\ \alpha\beta}(x) u^\alpha(x) u^\beta(x) &= 0\,.
}{}
For general configurations $u^{\mu}(x)$, the conditions $u^{\mu}(x)h_{\mu\nu}(x)=0$ and $\Gamma^{\mu}_{\ \alpha\beta}(x) u^\alpha(x) u^\beta(x)=0$ define different gauges. However in the case of interest, with $u^{\nu}\partial_{\nu} u^{\mu}=0$ and vanishing boundary conditions for $h_{\mu\nu}$, the two conditions are in fact equivalent. That is why both are satisfied here. For further discussion, see \App{app:Gammauu_vs_uh}.

In $u(x)$ gauge we then obtain the one-point matrix element, or external wavefunction,
\eq{
\big\langle 0 \big| h_{\mu\nu}(x) \big| h(k) \big\rangle_{u(x)\textrm{-gauge}} &= \big\langle 0 \big| \hh_{\mu\nu}(x) \big| h(k) \big\rangle \\
& = \big \langle 0 \big | h_{\mu\nu}(x) + \kappa^{-1}\partial_\mu w_\nu(x) + \kappa^{-1}\partial_\nu w_\mu(x) \big | h(k) \big\rangle\\ 
& \hspace{-30pt}= \epsilon_{\mu\nu}e^{-ik x}-i\partial_{\mu}\left(\frac{\epsilon^{\alpha\beta}\OO_{\nu\alpha\beta}(k,u(x))}{[k\!\cdot\!u(x)]^2}e^{-ikx}\right)-i\partial_{\nu}\left(\frac{\epsilon^{\alpha\beta}\OO_{\mu\alpha\beta}(k,u(x))}{[k\!\cdot\!u(x)]^2}e^{-ikx}\right)\,,
}{eq:uxgaugegravitonwf}
where $\OO_{\mu\alpha\beta}$ is defined in Eq.~\eqref{eq:OO}. Meanwhile, we can also evaluate the two-point vacuum correlator of dressed gravitons, 
\eq{
& \big \langle 0 \big| h_{\mu_1\nu_1}(x_1) h_{\mu_2\nu_2}(x_2) \big| 0 \big\rangle_{u(x)\textrm{-gauge}} = \big\langle 0 \big| \hh_{\mu_1\nu_1}(x_1)\hh_{\mu_2\nu_2}(x_2) \big| 0 \big\rangle \\
& = \big \langle 0 \big|
	\left[h_{\mu_1\nu_1}(x_1) + \kappa^{-1}\partial_{\mu_1} w_{\nu_1}(x_1) + \kappa^{-1}\partial_{\nu_1} w_{\mu_1}(x_1)\right] 
	\left[h_{\mu_2\nu_2}(x_2) + \kappa^{-1}\partial_{\mu_2} w_{\nu_2}(x_2) + \kappa^{-1}\partial_{\nu_2} w_{\mu_2}(x_2)\right]
	\big| 0 \big\rangle \\
& = \int_k \frac{i}{k^2} \mathcal{I}_{\mu_1\nu_1\mu_2\nu_2}(x_1,x_2,k)\, ,
}{eq:gravpropuxgauge}
where the numerator tensor structure is
\eq{
\mathcal{I}_{\mu_1\nu_1\mu_2\nu_2}(x_1,x_2,k)=P_{D}^{\alpha_1\beta_1\alpha_2\beta_2}&\left[e^{-ikx_1}\eta_{\alpha_1\mu_1}\eta_{\beta_1\nu_1}-2i\partial_{(\mu_1}\left(\frac{e^{-ikx_1} \OO_{\nu_1)\alpha_1\beta_1}(k,u(x_1))}{[k\!\cdot\!u(x_1)]^2}\right)\right] \\
\times &\left[e^{ikx_2}\eta_{\alpha_2\mu_2}\eta_{\beta_2\nu_2}+2i\partial_{(\mu_2}\left(\frac{e^{ikx_2} \OO_{\nu_2)\alpha_2\beta_2}(k,u(x_2))}{[-k\!\cdot\!u(x_2)]^2}\right)\right]\, ,
}{eq:gravpropuxgauge2}
and the brackets denote normalized symmetrization, $T_{(\mu\nu)}=\tfrac{1}{2}\left(T_{\mu\nu}+T_{\nu\mu}\right)$.

\subsubsection{Dressing as Gauge Fixing}

Just as in the case of gauge theory, in gravity we can also make the mathematical equivalence between dressing and gauge fixing manifest at the level of the action. To achieve this, we again observe that the field $\phi(z)$ and its dressed counterpart $\Phi(Y(\z))$ are related by a coordinate transformation.  Applying this to the action in \Eq{eq:gravityaction}, we obtain
\eq{
  S=  \int d^D\z \sqrt{-g(\z)}\left(-\frac{2}{\kappa^2}R(g_{\mu\nu}(\z))+\frac{1}{2}g^{\mu\nu}(\z) \frac{\partial \Phi(Y(\z))}{\partial \z^{\mu}}\frac{\partial\Phi(Y(\z))}{\partial \z^{\nu}}\right)\,,
}{}
where we have restored the $z$ arguments to clarify the subsequent manipulations. In the present form it does not appear very useful, since the explicit map $Y^{A}(z)$ will be very complicated, and certainly dependent on the metric.

We are, however, free to change integration variables from $\z^{\mu}$ to  $\z^{\prime\,\mu'}(\z)$. A convenient choice of variables is $z^{\prime\,\mu'}(\z) = Y^A(\z)$, in which case $\z(\z') = Y^{-1}(\z')$. The new integration variables trivialize the map $Y$, in the sense that $Y$ acts on them as the identity $Y(\z(\z'))=\z'$.  Physically, this selects  $Y^{A}$ to be the coordinates of the spacetime manifold.  Provided we also perform a coordinate transformation of the metric,
\eq{
\dressedmetric^{AB}(Y)=g^{\mu\nu}(\z(Y))\frac{\partial Y^{A}}{\partial \z^{\mu}}\frac{\partial Y^{B}}{\partial \z^{\nu}}  \,,
}{}
we can use the fact that the Einstein-Hilbert action is invariant under coordinate transformations to write the action as
\eq{
  S = \int d^D Y \sqrt{-\dressedmetric(Y)}\left(-\frac{2}{\kappa^2}R(\dressedmetric_{AB}(Y))+\frac{1}{2}\dressedmetric^{AB}(Y) \frac{\partial \Phi(Y)}{\partial Y^{A}}\frac{\partial\Phi(Y)}{\partial Y^{B}}\right)\,,
}{}
which is simply a change of coordinates.
Recall that these coordinates are $Y^{A}=\{y^{a}, \tau\}$ where $y^a$ are spatial coordinates on an initial spacelike surface and $\tau$ is the elapsed proper time along a timelike curve with constant $y^{a}$. In these coordinates, the dust field is
\eq{
U^{A}=\frac{dY^{A}}{d\tau}\,,
}{}
and the definition of the coordinates implies 
\eq{
U^{\tau}(\tau,y^{a}) = 1 \qquad \textrm{and} \qquad U^{a}(\tau,y^{a}) =0 \,,
}{}
while the geodesic condition implies
\eq{
\Gamma^{A}_{\,\tau\tau}(Y)=0\,.
}{}
This mirrors the usual construction of synchronous coordinates in general relativity. 

The above procedure transforms to coordinates where rather than dressing the field $\phi$, we simply introduce a bare insertion of $\Phi$, at the expense of transforming the original metric $g$ to a new metric $\dressedmetric$ that is in $u(x)$ gauge.  So again, we learn that dressing is gauge fixing.  While the general arguments here parallel the construction of synchronous coordinates, and synchronous gauge is a special case of our construction, we allow for general inhomogeneous initial conditions for the dust field, so that perturbatively we are imposing $u^{\mu}(x)h_{\mu\nu}=0$ with $u^{\mu}(x)$ not necessarily being a constant vector. This generalization is what allows us to avoid technical obstructions that typically plague loop calculations in gauges that entail a fixed reference vector, such as temporal or axial gauges (see \Sec{sec:poleprescription}).\footnote{For constant $u^{\mu}$, this gauge choice is typically referred to in the field theory literature as temporal or axial gauge depending on whether $u^{\mu}$ is timelike or spacelike. In the relativity literature it is more commonly referred to as synchronous gauge.} 

The equivalence between dressing and gauge fixing in classical gravity was in fact pointed out long ago by Komar~\cite{Komar:1958ymq}. In applying this idea in quantum gravity, we find it works formally and also in practice at loop level. 
\subsection{Tree-Level Correlator}

In analogy with \Sec{sec:QEDtreecorrelator}, we  now present various perturbative calculations involving gravitationally dressed operators. Specifically, we compute matrix elements of the time-ordered product $\Phi(x_1)\Phi(x_2)$. The vacuum expectation value of this operator is first sensitive to the effects of dressing at $\OO(\kappa^2)$, which corresponds to the one-loop calculation in \Sec{sec:gravityloopcorrelator}. In this section we instead focus on the simplest tree-level matrix element that is sensitive to the dressing, $\langle \Omega | \Phi(x_1)\, \Phi(x_2) | h(k)\rangle$.  This expression involves an asymptotic, on-shell external graviton, and is closely related to an expectation value on a plane wave graviton coherent state, as discussed in \Sec{sec:causalitygravity}. 

\subsubsection{Without Dressing}
To begin, let us compute the correlator $\langle \Omega | \phi(x_1) \phi(x_2) | h(k)\rangle$, without any dressing.  This quantity is calculated from the insertion of the three-point interaction vertex, and has the same diagrammatic topology as in \Eq{diag:QEDtreeUndressed},
\eq{
  \big\langle \Omega \big| \phi(x_1) \phi(x_2)\big| h(k) \big\rangle
  &= -i\kappa \int_{p_1,p_2}\, e^{-ip_1x_1}e^{-ip_2x_2}\, \frac{(p_1\!\cdot\!\epsilon \!\cdot\!p_2)}{p_1^2p_2^2} \, \delta(p_1+p_2-k),
}{eq:GR undressed}
where we have used the external contraction rule $\langle 0|h_{\mu\nu}(x)|h(k)\rangle = \epsilon_{\mu\nu}(k)e^{-ikx}$ and tracelessness of gravitons $\epsilon^{\mu}_{\ \mu}=0$, and defined the shorthand notation $(p_1\!\cdot\!\epsilon \cdot p_2)\equiv p_1^\mu\epsilon_{\mu\nu}p_2^\nu$. Computing the integral, we obtain
\eq{
  \big\langle \Omega \big| \phi(x_1) \phi(x_2) \big| h(k) \big\rangle =\frac{i\kappa\Gamma\left(\tfrac{D}{2}\right)}{4\pi^{D/2}}\frac{(e^{-ikx_1}-e^{-ikx_2})}{(-x_{12}^{2}+i\varepsilon)^{\frac{D}{2}}}\frac{x_{12}\!\cdot\!\epsilon\!\cdot\!x_{12}}{k\!\cdot\!x_{12}},
}{}
where $x_{12}^\mu\equiv x_1^\mu - x_2^\mu$.

Under diffeomorphism, $\delta_{\xi} h_{\mu\nu} = \partial_\mu\xi_\nu+\partial_\nu\xi_\mu$, the external graviton polarization changes by $\delta\epsilon_{\mu\nu} = -i k_\mu\tilde{\xi}_\nu -i k_\nu\tilde{\xi}_\mu$, and therefore the momentum space correlator
\eq{
	\big\langle \Omega \big| \tilde{\phi}(p_1) \tilde{\phi}(p_2) \big| h(k) \big\rangle = \int_{x_1,x_2} e^{ip_1x_1}e^{ip_2x_2}\big\langle \Omega \big| \phi(x_1) \phi(x_2) \big| h(k) \big\rangle
}{}
has the variation
\eq{
  \delta_{\xi} \big\langle \Omega \big| \tilde{\phi}(p_1) \tilde{\phi}(p_2) \big| h(k) \big\rangle  
  = -\kappa\, (p_1 \!\cdot\! \tilde{\xi})\left(\frac{1}{p_1^2}-\frac{1}{p_2^2}\right)\delta(p_1+p_2-k),
}{}
which is nonzero for general $p_1$ and $p_2$, indicating that the undressed correlator is not gauge invariant. In position space, the diffeomorphism variation is
\eq{
  \delta_{\xi} \big\langle \Omega \big| \phi(x_1) \phi(x_2)\big| h(k) \big\rangle = \frac{\kappa\Gamma\left(\tfrac{D}{2}\right)}{2\pi^{D/2}}\frac{(e^{-ikx_1}-e^{-ikx_2})}{(-x_{12}^{2}+i\varepsilon)^{\frac{D}{2}}}(x_{12}\!\cdot\!\tilde{\xi})\,,
}{}
which is generally not zero unless $k\!\cdot\! x_{12}=0$ or $x_{12}\!\cdot\!\tilde{\xi}=0$.

\subsubsection{With Dressing}\label{subsec:GR_dressed_tree}

Now we calculate the dressed correlator $\langle \Omega |\Phi(x_1) \Phi(x_2) | h(k)\rangle$.
\eq{
  &\quad \big\langle \Omega \big| \Phi(x_1) \Phi(x_2) \big| h(k) \big\rangle \\
  & = \underbrace{\big\langle 0  \big| \phi(x_1) \phi(x_2) \big| h(k)\big\rangle}_{\mathrm{I}}
  + \underbrace{\int_{-\infty}^0 ds\int_{-\infty}^{s} ds^\prime\,
  \big\langle 0\, \big| - \Gamma^\mu_{\alpha\beta}(x_1+u_1s^\prime)\, u_1^\alpha u_1^\beta\, \partial_\mu\phi(x_1)\, \phi(x_2) \,\big|\, h(k)\big\rangle}_{\mathrm{II}} \\
  &\quad + \underbrace{\int_{-\infty}^0 ds\int_{-\infty}^{s} ds^\prime\,
  \big\langle 0\, \big| - \Gamma^\mu_{\alpha\beta}(x_2+u_2s^\prime)\, u_2^\alpha u_2^\beta\, \phi(x_1)\, \partial_\mu\phi(x_2) \,\big|\, h(k)\big\rangle}_{\mathrm{III}} \,+\, \mathcal{O}(\kappa^2).
}{}
These three terms correspond to the same diagram topologies as in QED, see \Eq{diag:QEDtreeDressed}, but with the photon lines replaced by graviton lines, and complex scalars replaced by real scalars,

\eq{
\begin{array}{ccc}
\vcenter{\hbox{
\begin{tikzpicture}
\tikzfeynmanset{
every edge/.style={thick}
}
\begin{feynman}
  \vertex[dot, label=90:\(\phi(x_1)\)] (x1) {};
  \vertex[right=2 of x1, dot, label=90:\(\phi(x_2)\)] (x2) {};
  \vertex[below=2.4 of x1] (w1) {\(u_1\)};
  \vertex[below=2.4 of x2] (w2) {\(u_2\)};

  \vertex (m) at ($(x1)!0.5!(x2)$);
  \vertex[below=1.6 of m] (f);

  \diagram*{
    (x1) --[plain] (m) --[plain] (x2),
    (x1) --[scalar, blue] (w1),
    (x2) --[scalar, blue] (w2),
    (f) --[graviton] (m),
    (f) --[draw=none, momentum={[arrow shorten=0.2]\(k\)}] (m)
  };
\end{feynman}
\end{tikzpicture}
}}
&
\vcenter{\hbox{
\begin{tikzpicture}
\tikzfeynmanset{
every edge/.style={thick}
}
\begin{feynman}
  \vertex[dot, label=90:\(\phi(x_1)\)] (x1) {};
  \vertex[right=2 of x1, dot, label=90:\(\phi(x_2)\)] (x2) {};
  \vertex[below=2.4 of x1] (w1) {\(u_1\)};
  \vertex[below=2.4 of x2] (w2) {\(u_2\)};

  \vertex (g) at ($(x1)!0.5!(w1)$);
  \vertex[left=1 of g] (f);

  \diagram*{
    (x1) --[plain] (x2),
    (x1) --[scalar, blue] (w1),
    (x2) --[scalar, blue] (w2),
    (f) --[graviton] (g),
    (f) --[draw=none, momentum'={[arrow shorten=0.2] \(k\)}] (g),
  };
\end{feynman}
\end{tikzpicture}
}}
&
\vcenter{\hbox{
\begin{tikzpicture}
\tikzfeynmanset{
every edge/.style={thick}
}
\begin{feynman}
  \vertex[dot, label=90:\(\phi(x_1)\)] (x1) {};
  \vertex[right=2 of x1, dot, label=90:\(\phi(x_2)\)] (x2) {};
  \vertex[below=2.4 of x1] (w1) {\(u_1\)};
  \vertex[below=2.4 of x2] (w2) {\(u_2\)};

  \vertex (g) at ($(x2)!0.5!(w2)$);
  \vertex[right=1 of g] (f);

  \diagram*{
    (x1) --[plain] (x2),
    (x1) --[scalar, blue] (w1),
    (x2) --[scalar, blue] (w2),
    (f) --[graviton] (g),
    (f) --[draw=none, momentum={[arrow shorten=0.2] \(k\)}] (g),
  };
\end{feynman}
\end{tikzpicture}
}}
\\[2em]
\textbf{(I)} & \textbf{(II)} & \textbf{(III)}
\end{array}\,,
}{diag:GRtree}
where, as before, the scalar fields $\phi(x_1)$ and $\phi(x_2)$ carry momenta $p_1$ and $p_2$ that are implicitly defined to be flowing towards them, according to our convention in Eq.~\eqref{eq:momentum-arrow}.

Term I is just the undressed correlator calculated in Eq.~\eqref{eq:GR undressed}. Term II is calculated using the rules Eq.~\eqref{eq:WL-graviton-feynman}, Eq.~\eqref{eq:X-X-contraction}, and Eq.~\eqref{eq:photon-plr},
\eq{
 \mathrm{II} 
 &=\int_{p_1,p_2}e^{-ip_1x_1}e^{-ip_2x_2}\underbrace{\left\langle\tilde{\phi}(p_1)\tilde{\phi}(p_2)\right\rangle}_{i\delta(p_1+p_2)/p_2^2}\int_{q}e^{-iqx_1}\frac{-\kappa p_{1\mu}\OO^{\mu}_{\alpha\beta}(q,u_1)}{(q \!\cdot\! u_1)^2}\underbrace{\left\langle 0 \middle| \tilde{h}^{\alpha\beta}(q) \middle| h(k) \right\rangle}_{\epsilon^{\alpha\beta}(k)\delta(q-k)} \\
 &= -i\kappa\int_{p_1,p_2}\,e^{-ip_1 x_1-ip_2x_2-ikx_1}\times \delta(p_1+p_2)\,\frac{1}{p_2^2}\times\frac{p_{1\mu}\OO^{\mu}_{\alpha\beta}(k,u_1)\epsilon^{\alpha\beta}}{(k\!\cdot\!u_1)^2} \\
 &= -i\kappa\int_{p_1,p_2}\,e^{-ip_1 x_1-ip_2x_2}\times \delta(p_1+p_2-k)\,\frac{1}{p_2^2}\times\frac{-p_{2\mu}\OO^{\mu}_{\alpha\beta}(k,u_1)\epsilon^{\alpha\beta}}{(k\!\cdot\!u_1)^2} \\
 &= -i\kappa\int_{p_1,p_2}\,e^{-ip_1 x_1-ip_2 x_2}\times \delta(p_1+p_2-k)\,\frac{1}{p_2^2}\times
 \frac{-2(p_2\!\cdot\!\epsilon\!\cdot\! u_1)(k\!\cdot\!u_1)+(p_2\!\cdot\! k)(u_1\!\cdot\!\epsilon\!\cdot\! u_1)}{2(k\!\cdot\!u_1)^2},
}{}
where in the third equality we shifted the integration variable as $p_1\to p_1-k$ to get the standard two-momenta Fourier factor $e^{-ip_1x_1-ip_2x_2}$ and the momentum conserving delta function $\delta(p_1+p_2-k)$.  Term III can be obtained by swapping $1\leftrightarrow 2$ in term II.

After adding up the three terms I, II and III, the final result is\footnote{
We remind the reader that denominators have implicit $+i\varepsilon$ prescriptions unless otherwise specified, e.g. $p^2\rightarrow p^2+i\varepsilon$ and $k\!\cdot\!u\rightarrow k\!\cdot\!u+i\varepsilon$.}
\eq{
  &\big\langle 0 \big| \Phi(x_1) \Phi(x_2) \big| h(k) \big\rangle 
  = -i\kappa \int_{p_1,p_2}\, e^{-ip_1x_1-ip_2x_2}\times {\delta(p_1+p_2-k)}\\
  &\times \bigg[
  \frac{(p_1 \!\cdot\! \epsilon \!\cdot\!  p_2)}{p_1^2p_2^2}
  +
  \frac{1}{p_2^2}
  \frac{-2(p_2 \!\cdot\! \epsilon \!\cdot\!  u_1)(k\!\cdot\! u_1)+(p_2\!\cdot\! k)(u_1 \!\cdot\! \epsilon  \!\cdot\! u_1)}{2(k\!\cdot\! u_1)^2}
  +
  \frac{1}{p_1^2}
  \frac{-2(p_1 \!\cdot\! \epsilon  \!\cdot\! u_2)(k\!\cdot\! u_2)+(p_1\!\cdot\! k)(u_2 \!\cdot\! \epsilon \!\cdot\!  u_2)}{2(k\!\cdot\!u_2)^2}
  \bigg].
}{}
Under a diffeomorphism, $\epsilon_{\alpha\beta}\to\epsilon_{\alpha\beta} + k_\alpha\tilde{\xi}_\beta + k_\beta\tilde{\xi}_\alpha$, the quantity in brackets transforms as
\eq{
  \delta_{\xi}[\cdots]=& \quad \frac{(p_1 \!\cdot\! k)(p_2  \!\cdot\! \tilde{\xi})+(p_2  \!\cdot\! k)(p_1  \!\cdot\! \tilde{\xi})}{p_1^2 p_2^2}\\
  &+
  \bigg[\frac{\cancel{-2(p_2  \!\cdot\! k)(\tilde{\xi}  \!\cdot\! u_1)(k  \!\cdot\! u_1)} - 2(p_2  \!\cdot\! \tilde{\xi})(k  \!\cdot\! u_1)^{2} + \cancel{2(p_2  \!\cdot\! k)(k  \!\cdot\! u_1)(\tilde{\xi}  \!\cdot\! u_1)} }{2p_2^2(k  \!\cdot\! u_1)^2} 
  +(1\leftrightarrow 2)\bigg]\\
  & = \frac{p_1^2(p_2  \!\cdot\! \tilde{\xi}) + p_2^2(p_1  \!\cdot\! \tilde{\xi})+(p_1  \!\cdot\! p_2)\cancel{(p_1  \!\cdot\! \tilde{\xi} + p_2  \!\cdot\! \tilde{\xi})}}{p_1^2 p_2^2} + \frac{-(p_2  \!\cdot\! \tilde{\xi})}{p_2^2} + \frac{-(p_1  \!\cdot\! \tilde{\xi})}{p_1^2} = 0,
}{}
where we used $k=p_1+p_2$ and $k \!\cdot\!  \tilde{\xi}=0$. For efficient comparison with later sections, we'll repackage the result in terms of the $\OO^{\alpha\mu\nu}$ tensor appearing in the Feynman rules,
\eq{
  &\big\langle 0 \big| \Phi(x_1) \Phi(x_2) \big| h(k) \big\rangle\\ 
  &= -i\kappa \int_{p_1,p_2}\, \frac{e^{-ip_1x_1-ip_2x_2}}{p_1^2 p_2^2}\delta(p_1+p_2-k)
  \bigg[
  (p_1\!\cdot\!\epsilon\!\cdot\! p_2)
  -p_1^2 p_2^a\frac{\OO_{a\mu\nu}(k,u_1)\epsilon^{\mu\nu}}{(k\!\cdot\! u_1)^2}
  -p_2^2 p_1^a\frac{\OO_{a\mu\nu}(k,u_2)\epsilon^{\mu\nu}}{(k\!\cdot\! u_2)^2}\bigg]\,.
}{eq:treegravitydressing}
Evaluating the integrals, see \App{app:tree_fourier}, the result in position space is
\eq{
&\big\langle 0 \big|  \Phi(x_1) \Phi(x_2)\big| h(k) \big\rangle \\
&=
\begin{aligned}[t]
\frac{i\kappa \Gamma(D/2)}{4\pi^{D/2}\,(-x_{12}^{2}+i\varepsilon)^{D/2}}
\Bigg[\,
&
\left(
 \frac{(x_{12}\!\cdot\!\epsilon\!\cdot\!x_{12})}{k\!\cdot\! x_{12}}
-\frac{2(u_{1}\!\cdot\!\epsilon\!\cdot\!x_{12})}{u_{1}\!\cdot\! k}
+\frac{(k\!\cdot\! x_{12})\,(u_{1}\!\cdot\!\epsilon\!\cdot\! u_{1})}{(u_{1}\!\cdot\! k)^{2}}
\right)e^{-i k x_{1}}\\
-\,
&
\left(
 \frac{(x_{12}\!\cdot\!\epsilon\!\cdot\!x_{12})}{k\!\cdot\! x_{12}}
-\frac{2\,(u_{2}\!\cdot\!\epsilon\!\cdot\!x_{12})}{u_{2}\!\cdot\! k}
+\frac{(k\!\cdot\! x_{12})\,(u_{2}\!\cdot\!\epsilon \!\cdot\!u_{2})}{(u_{2}\!\cdot\! k)^{2}}
\right)e^{-i k x_{2}}\,
\Bigg]
\end{aligned}\\
&=\frac{i\kappa \Gamma(D/2)}{4\pi^{D/2}\,(-x_{12}^{2}+i\varepsilon)^{D/2}}
\frac{1}{x_{12}\!\cdot\!k}
\Bigg(\left(x_{12}\!\cdot\! P^{(1)}\!\cdot\!\epsilon\right)^2 e^{-ik x_1}
-\left(x_{12}\!\cdot\! P^{(2)}\!\cdot\!\epsilon\right)^2e^{-ik x_2}\Bigg)\,,
}{eq:dressed GR 3pt position}
where in the second line we wrote the graviton polarization tensor in terms of polarization vectors $\epsilon_{\mu\nu}=\epsilon_\mu\epsilon_\nu$. This result is manifestly diffeomorphism invariant because  $P^{(i)}\cdot k=0$.

This expression has a nearly identical structure to the corresponding scalar QED result, \Eq{eq:QEDtreedressedprojectors}, except for some features which exhibit a double copy structure. Since the only interaction in play is the three-point vertex, this is perhaps unsurprising \cite{Goldberger:2016iau, Shi:2021qsb}.  Nevertheless, the appearance of such a double copy structure at the level of dressed correlators, rather than on-shell amplitudes, is still suggestive.  In future work it would be interesting to investigate double copy structure for dressed correlators, specifically to see if it persists beyond three points. 

\subsubsection{Via Gauge Fixing}
We will now compute the dressed correlator using $u(x)$ gauge rather than summing dressing diagrams. The only diagram to compute is the standard three-point correlator,  \Eq{diag:QEDtreeUndressed}, but with the external graviton wavefunction now given by \Eq{eq:uxgaugegravitonwf}. 
\eq{
&\big\langle \Omega\big|\Phi(x_1) \Phi(x_2)\big|h(k) \big\rangle
=\big\langle \Omega\big|\phi(x_1)\phi(x_2)\big|h(k) \big\rangle_{u(x)\textrm{-gauge}}\\
&=i\kappa\int_{p_1,p_2}\int_y \frac{i}{p_1^2}\frac{i}{p_2^2}e^{-ip_1 (x_1-y)}e^{-ip_2 (x_2-y)}p_{1\,a}p_{2\,b}P_{4}^{ab\mu\nu}\bigg[\epsilon_{\mu\nu}e^{-ik y}-2i\partial_{\mu}\left(\frac{\OO_{\nu\alpha\beta}(k,u(y))\epsilon^{\alpha\beta}e^{-ik y}}{[k\!\cdot\!u(y)]^2}\right)\bigg] \\
&=i\kappa\int_{p_1,p_2}\int_y \frac{i}{p_1^2}\frac{i}{p_2^2}e^{-ip_1 (x_1-y)}e^{-ip_2 (x_2-y)} \\
&\hspace{80pt}\times p_{1\,a}p_{2\,b}P_{4}^{ab\mu\nu}\bigg[\epsilon_{\mu\nu}e^{-ik y}-2(p_1+p_2)_{\mu}\left(\frac{\OO_{\nu\alpha\beta}(k,u(y))\epsilon^{\alpha\beta}e^{-ik y}}{[k\!\cdot\!u(y)]^2}\right)\bigg]\,,
}{eq:gravitytreeux}
where we integrated by parts to obtain the last line.

The graviton wavefunction has nontrivial position dependence via $u(y)$. It initially appears impossible to integrate over the position $y$ of the internal vertex for general $u(y)$. However, a simplification occurs because all nontrivial $y$ dependence is inside a derivative, in the form of a gauge transformation. To see this, we evaluate the contraction between this total derivative, $(p_1+p_2)_{\mu}$, and the stress tensor $p_{1\,a}p_{2\,b}P_{4}^{ab\mu\nu}$, 
\eq{
p_{1\,a}p_{2\,b}P_{4}^{ab\mu\nu}(p_1+p_2)_{\mu} = \frac{1}{2}(p_1^2 p_2^{\nu}+p_2^2 p_1^\nu)\,.
}{}
Inserting this into the correlator, we have
\eq{
&\big\langle \Omega\big|\phi(x_1)\phi(x_2)\big|h(k) \big\rangle_{u(x)\textrm{-gauge}}=i\kappa\int_{p_1,p_2}\int_y \frac{i}{p_1^2}\frac{i}{p_2^2}e^{-ip_1 (x_1-y)}e^{-ip_2 (x_2-y)}e^{-ik y} \\
&\hspace{80pt}\times \bigg[(p_1\!\cdot\!\epsilon\!\cdot\!p_2)-(p_1^2 p_2^{\nu}+p_2^2 p_1^{\nu})\left(\frac{\OO_{\nu\alpha\beta}(k,u(y))\epsilon^{\alpha\beta}}{[k\!\cdot\!u(y)]^2}\right)\bigg]\,.
}{}
The result involves two terms, each proportional to $p_i^{2}$.  These pinch the $p_i$ scalar propagator in \Eq{eq:gravitytreeux} so that the first term localizes to $x_1$ and the second term localizes to $x_2$. 

Physically, this is the momentum space manifestation of the fact that stress tensors are conserved unless an operator is inserted, and so inserting $\partial_\mu T^{\mu\nu}$ localizes a correlator to the locations $x_1,x_2$ of the other operators. Mathematically, in the term where $p_i^2$ pinches the $p_i$ propagator, we can trivially perform the $p_i$ integral to obtain $\delta(x_i-y)$. The upshot is that in the term proportional to $p_i^2$ we can replace $u(y)\rightarrow u_i$, and all nontrivial $y$ dependence is gone,
\eq{
&\big\langle \Omega\big|\phi(x_1)\phi(x_2)\big|h(k) \big\rangle_{u(x)\textrm{-gauge}}=i\kappa\int_{p_1,p_2}\int_y \frac{i}{p_1^2}\frac{i}{p_2^2}e^{-ip_1 (x_1-y)}e^{-ip_2 (x_2-y)}e^{-ik y} \\
&\hspace{80pt}\times \bigg[(p_1\!\cdot\!\epsilon\!\cdot\!p_2)-p_1^2 p_2^{\nu}\left(\frac{\OO_{\nu\alpha\beta}(k,u_1)\epsilon^{\alpha\beta}}{(k\!\cdot\!u_1)^2}\right)-p_2^2 p_1^{\nu}\left(\frac{\OO_{\nu\alpha\beta}(k,u_2)\epsilon^{\alpha\beta}}{(k\!\cdot\!u_2)^2}\right)\bigg]\,.
}{}
We can now freely integrate over $y$ to obtain the usual momentum conserving delta function, obtaining exactly \Eq{eq:treegravitydressing}, and again demonstrating the equivalence between the dressing and $u(x)$-gauge calculations.

\subsubsection{Causality with External Gravitons}\label{sec:causalitygravity}

The commutator of the dressed operators we study gives a diffeomorphism-invariant diagnostic of causality. To illustrate this, we will compute $\braket{\Psi|[\Phi(x_1), \Phi(x_2)]|\Psi}$ for a nonvacuum state $\ket{\Psi}$. Recall that in quantum field theory, the vanishing of $[\phi(x_1),\phi(x_2)]$ outside the light cone is an operator statement, which means causality requires that any matrix element of the commutator is zero outside the light cone. We will choose the state $|\Psi\rangle$ to be a coherent state of plane wave gravitons, $\ket{\gamma}$, and we will calculate the associated matrix elements to leading order in $\kappa$.

In this section only, we consider correlation functions with a variety of operator orderings, and so we no longer implicitly treat operator products as time ordered. The symbol $T\{...\}$ will be used when time ordering is intended.


A coherent state of gravitons can be written as
\eq{
|\gamma\rangle = 
\exp\bigg[\sum_{\lambda}\int\frac{d^{d}k}{(2\pi)^d}\frac{1}{\sqrt{2|k|}}\big(\gamma^{\lambda}(k)a^{\dagger}_{\lambda,k}-\gamma^{\lambda\,*}(k)a_{\lambda,k}\big)\bigg]|0\rangle\,,
}{}
where $d=D-1$, the helicity states are labeled by $\lambda$, and $\gamma(k)$ is a wavepacket in each helicity sector. This state is the semiclassical state of a classical metric $g_{\mu\nu}(x)=\eta_{\mu\nu}+\kappa \gamma_{\mu\nu}(x)$. That is, it is the minimal uncertainty state with expectation value
\eq{
\langle \gamma | h_{\mu\nu}(x) |\gamma\rangle = \gamma_{\mu\nu}(x) = \sum_{\lambda}\int\frac{d^{d}k}{(2\pi)^d}\frac{1}{2|k|}\big(\gamma^{\lambda}(k)\epsilon^{\lambda}_{\mu\nu}(k) e^{-ikx}+\gamma^{\lambda\,*}(k)\epsilon^{\lambda\,*}_{\mu\nu}(k)e^{ikx}\big)\,.
}{}
We will need only the one-particle content of this state, as we'll be calculating to $\OO(\kappa)$,
\eq{
|\gamma\rangle = |0\rangle +\sum_{\lambda}\int\frac{d^{d}k}{(2\pi)^d}\frac{1}{2|k|}\gamma^{\lambda}(k)|h^{\lambda}(k)\rangle+\cdots\,.
}{}
The expectation value of an operator $O$ in this coherent state is then, to linear order,
\eq{
\langle \gamma | O |\gamma\rangle = \langle 0| O |0\rangle  + \sum_{\lambda}\int\frac{d^{d}k}{(2\pi)^d}\frac{1}{2|k|}\bigg[\gamma^{\lambda}(k) \langle 0|O|h^{\lambda}(k)\rangle + \gamma^{\lambda\,*}(k)\langle h^{\lambda}(k)|O|0\rangle\bigg]\,.
}{}

We are ultimately interested in matrix elements for which the operator $O$ is the commutator of dressed fields. We have not yet calculated these matrix elements, but we have already calculated analogous matrix elements of the time-ordered product of fields. Fortunately, the former can be obtained from the latter via analytic continuation. We review this here.

Lorentzian correlation functions are defined by analytic continuation from their Euclidean counterpart, where operators are necessarily time ordered in Euclidean time. Which Lorentzian correlator we arrive at depends on how we analytically continue the Euclidean time $x^{D}_{i}$ for each of the $i$ operator insertions. A uniform rotation $x^{D}_{i}\rightarrow x^{0}_{i}e^{i(\pi/2-\varepsilon)}$ produces Lorentzian time-ordered correlators. For example, denoting the Lorentzian spatial coordinates as $\mathbf{x} = (x^1, \cdots, x^{D-1})$, the rotation needed for a two-point function is 
\eq{    
(x_1^D-x_2^D)^2+\mathbf{x}_{12}^2 &\rightarrow -(x_1^0-x_2^0-i\varepsilon(x_1^0-x_2^0))^2+\mathbf{x}_{12}^2 \\
&=-(x_1^0-x_2^0)^2+\mathbf{x}_{12}^2 +i\varepsilon \\
&=-x_{12}^2 +i\varepsilon\,.
}{}
We could instead analytically continue the $x_i^{D}$ independently, $x_i^{D}\rightarrow e^{i\pi/2}(x_i^{0}-i\varepsilon_i)$, in which case the ordering $\varepsilon_i >\varepsilon_j > \cdots$ determines the ordering of the operator insertions in Lorentzian signature. This procedure defines Wightman correlation functions.

This implies that for any matrix elements of the Lorentzian time-ordered two-point function we can apply the following procedure. We go to a spacelike region away from singularities, drop the time-ordering prescription $i\varepsilon$, and then continue $x^0_{1}\rightarrow x^0_{1}\mp i\varepsilon$ while holding $x_2$ fixed, which yields the corresponding matrix elements of the Wightman functions with $1,2$-ordering for $-i\varepsilon$ and $2,1$-ordering for $+i\varepsilon$. 

It is useful to see an explicit example of how the Wightman and time-ordered two-point functions are related in position space. Working in $D=4$, using that $\Box (1/x_{12}^2) = 0$ away from coincident points, we can see that indeed
\begin{align}
\braket{0|T\{\phi(x_1) \phi(x_2)\}|0} &=  \theta(x^0_{12}) \braket{0|\phi(x_1) \phi(x_2)|0}+
\theta(x^0_{21})\braket{0|\phi(x_2) \phi(x_1)|0}
\\
&\propto ~
\theta(x^0_{12}) \frac{1}{(x^0_{12}-i\varepsilon)^2-\mathbf{x}_{12}^2} 
+
\theta(-x^0_{12}) \frac{1}{(x^0_{12}+i\varepsilon)^2-\mathbf{x}_{12}^2}
\\
&=
 \theta(x^0_{12}) \frac{1}{(x^0_{12})^2 -\mathbf{x}_{12}^2-i\varepsilon~ \text{sgn}(x^0_{12})} 
+
\theta(-x^0_{12}) \frac{1}{(x^0_{12})^2 -\mathbf{x}_{12}^2+i\varepsilon~ \text{sgn}(x^0_{12})} 
\\
&=
\frac{1}{(x^0_{12})^2 -\mathbf{x}_{12}^2-i\varepsilon},
\end{align}
as expected, where the overall numerical prefactor we have omitted is given in \Eq{app:int-a}.

In \Eq{eq:dressed GR 3pt position} we computed a tree-level matrix element of $T\{\Phi(x_1)\Phi(x_2)\}$, and obtained
\eq{
\big\langle 0 \big| T\{ \Phi(x_1) \Phi(x_2)\} \big| h(k) \big\rangle = \frac{\mathcal{N}(x_1,x_2,k,\epsilon)}{(-x_{12}^2+i\varepsilon)^{D/2}}\,,
}{}
where $\mathcal{N}(x_1,x_2,k,\epsilon)$ is nonsingular as a function of $x_1,x_2$. The corresponding matrix element of the commutator is then given by
\eq{
&\big\langle 0 \big| \left[\Phi(x_1), \Phi(x_2)\right] \big| h(k) \big\rangle \\
&= \mathcal{N}(x_1,x_2,k,\epsilon)
\left[\frac{1}{(-(x_1^0-i\varepsilon-x_2^0)^2+\mathbf{x}_{12}^2)^{D/2}}
-\frac{1}{(-(x_1^0+i\varepsilon-x_2^0)^2+\mathbf{x}_{12}^2)^{D/2}}\right] \\
&=2i\,\mathcal{N}(x_1,x_2,k,\epsilon)\,\textrm{Im}\frac{1}{(-x_{12}^2+i\varepsilon\,\textrm{sgn}(x^0_{12}))^{D/2}}\,.
}{}
In even dimensions this can be simplified using the following identity:
\eq{
\frac{1}{(-\sigma+i\varepsilon)^n}=\frac{1}{\Gamma(n)}\left(\frac{d}{d\sigma}\right)^{n-1}\frac{1}{-\sigma+i\varepsilon}\,.
}{}
The one-particle matrix element of the commutator is then given by
\eq{
&\big\langle 0 \big| \left[\Phi(x_1),  \Phi(x_2)\right] \big| h(k) \big\rangle =-i\frac{2\pi\,\textrm{sgn}(x^0_{12})}{\Gamma(D/2)}\mathcal{N}(x_1,x_2,k,\epsilon)\left(\frac{d}{dx_{12}^2}\right)^{D/2-1}\delta(x^2_{12})\,.
}{eq:matrixelementcommutator}
To compute the expectation value in the graviton coherent state we also need the matrix element $\langle h(k) | \left[\Phi(x_1), \Phi(x_2)\right] | 0 \rangle$. This can be calculated immediately from \Eq{eq:matrixelementcommutator} by crossing $\epsilon_{\mu\nu}(k)\rightarrow \epsilon_{\mu\nu}^{*}(k)$ and $k^{\mu}\rightarrow -k^{\mu}$. 

We now have all of the ingredients to compute the $\OO(\kappa)$ contribution to the commutator of dressed fields in a coherent graviton background. The leading order contribution to $\langle \gamma | \left[\Phi(x_1), \Phi(x_2)\right] | \gamma \rangle$ is the usual flat-space commutator,
\eq{
\big\langle 0 \big| \left[\Phi(x_1),  \Phi(x_2)\right] \big| 0\big\rangle \equiv i \, \textrm{sgn}(x^0_{12})\Delta(-x^2_{12}) = - i\frac{2\pi\,\textrm{sgn}(x^0_{12})}{4\pi^{D/2}}\left(\frac{d}{dx_{12}^2}\right)^{D/2-2}\delta(x^2_{12})\,,
}{}
and the first gravitational correction is
\eq{
\big\langle \gamma \big| \left[\Phi(x_1), \Phi(x_2)\right] \big| \gamma \big\rangle\bigg|_{\kappa} &= -i\frac{2\pi\,\textrm{sgn}(x^0_{12})}{\Gamma(D/2)} 
\left(\frac{d}{dx_{12}^2}\right)^{D/2-1}\delta(x^2_{12})\\
&\times\sum_\lambda\int\frac{d^d k}{(2\pi)^d}\frac{1}{2|k|}\bigg[\gamma^{\lambda}(k)\mathcal{N}(x_1,x_2,k,\epsilon^{\lambda})+\textrm{c.c.}\bigg]\,,
}{}
where we used the property (see \Eq{eq:dressed GR 3pt position}) that crossing the graviton in $\mathcal{N}$ just gives $\mathcal{N}^{*}$. 

To gain a physical understanding of this result, we will use the explicit expression for $\mathcal{N}(x_1,x_2,k,\epsilon)$ in \Eq{eq:dressed GR 3pt position} in the simplified case of a constant fluid velocity $u_1^{\mu}=u_2^{\mu}=u^{\mu}$,
\eq{
\big\langle \gamma \big| \left[\Phi(x_1), \Phi(x_2)\right] \big| \gamma \big\rangle\bigg|_{\kappa}=&-i\kappa\frac{2\pi\,\textrm{sgn}(x^0_{12})}{4\pi^{D/2}}
x_{12}^{\mu}x_{12}^{\nu}\left(\frac{d}{dx_{12}^2}\right)^{D/2-1}\delta(x^2_{12})\\
&\times\sum_\lambda \int\frac{d^d k}{(2\pi)^d}\frac{1}{2|k|}\,\bigg[P_{\mu a}P_{\nu b}\epsilon^{\lambda\,ab}(k)\gamma^{\lambda}(k)\frac{e^{-ik x_1}-e^{-ik  x_2}}{-ik\!\cdot\! x_{12}}
+\textrm{c.c.}\bigg]\,.
}{eq:coherentstatecommutator1}
Note that this difference in phases is just a simple parameter integral,
\eq{
\frac{e^{-ik x_1}-e^{-ik x_2}}{-ik\!\cdot\! x_{12}} = \int_{0}^{1}ds\,e^{-ik(x_2+s(x_1-x_2))}\,.
}{}
If we insert this in \Eq{eq:coherentstatecommutator1} above, we have
\eq{
\big\langle \gamma \big| \left[\Phi(x_1),  \Phi(x_2)\right] \big| \gamma \big\rangle\bigg|_{\kappa}=&-i\kappa\frac{2\pi\,\textrm{sgn}(x^0_{12})}{4\pi^{D/2}}
x_{12}^{\mu}x_{12}^{\nu}\left(\frac{d}{dx_{12}^2}\right)^{D/2-1}\delta(x^2_{12})\\
&\hspace{-20pt}\times\int_{0}^{1}ds\sum_\lambda \int\frac{d^d k}{(2\pi)^d}\frac{1}{2|k|}\,\bigg[P_{\mu a}P_{\nu b}\epsilon^{\lambda\,ab}(k)\gamma^{\lambda}(k)e^{-ik(x_2+s(x_1-x_2))}
+\textrm{c.c.}\bigg]\,.
}{eq:coherentstatecommutator2}
The projectors $P_{\mu a}$ in \Eq{eq:coherentstatecommutator2} serve only to map the graviton polarization into temporal gauge, for instance $u^{\mu}P_{\mu a}\epsilon^{\lambda\,ab}=0$. We can then identify the $k$ integral as the expectation value of the metric perturbation, $\gamma_{\mu\nu}(x)$ in synchronous gauge, evaluated along the path 
\eq{
\z^{\mu}(s)=x^\mu_2+s x_{12}^\mu\,.
}{}
The expression can then be organized in the suggestive form
\eq{
&\big\langle \gamma \big| \left[\Phi(x_1),  \Phi(x_2)\right] \big| \gamma \big\rangle\bigg|_{\kappa} \\
&= \textrm{sgn}(x^0_{12}) \left(\kappa x_{12}^{\mu}x_{12}^{\nu}\int_{0}^{1}ds\,\gamma_{\mu\nu}(z(s))\right) \,\frac{d}{d x_{12}^2}\bigg[-i\frac{2\pi\,}{4\pi^{D/2}}\left(\frac{d}{dx_{12}^2}\right)^{D/2-2}\delta(x^2_{12})\bigg]\,,
}{eq:coherentstatecommutatorfinal}
where the function in the square brackets, multiplied by $\textrm{sgn}(x^0_{12})$, is precisely the $\OO(\kappa^0)$ vacuum commutator of scalar fields.  As we will now demonstrate, this is exactly the same result one would obtain from the commutator of a scalar field propagating on a fixed, weakly curved, classical spacetime.

\subsubsection{Causality in Curved Backgrounds }

The causal Green's function for a scalar field in even-dimensional curved spacetime admits a Hadamard decomposition~\cite{PhysRevD.73.044027, Poisson:2011nh},
\eq{
G_{ret}(x_1,x_2) = \theta_{+}(x_1,x_2)\left[\sum_{n=0}^{D/2-2}U_{n}(x_1,x_2)\left(\frac{d}{d\sigma}\right)^{D/2-2-n}\delta(\sigma)+V(x_1,x_2)\theta(-\sigma)\right]\,,
}{}
where $\sigma$ is the squared geodesic distance between $x_1,x_2$ and the points are assumed to be in a normal convex neighborhood. The functions $U_{n},V$ have been computed, quite generally, in expansions about coincidence.\footnote{For example, see \cite{PhysRevD.73.044027} and references therein.} On a background which solves the vacuum Einstein equations, 
and is perturbatively close to Minkowski space, to linear order in the perturbation parameter the Hadamard decomposition simplifies dramatically: the tail terms $V$, the terms $U_{n}$ with $n>0$, and the inhomogeneous parts of $U_{0}$ all vanish. The only nontrivial contribution at $\OO(\kappa)$ arises in the modification to the geodesic distance relative to flat spacetime.

For a metric $g_{\mu\nu}=\eta_{\mu\nu}+\kappa \gamma_{\mu\nu}$, the geodesic distance between $x_1$ and $x_2$ at $\OO(\kappa)$ is
\eq{
\sigma^{1/2} = \int_{0}^{1}ds \bigg[-\left[\eta_{\mu\nu}+\kappa\gamma_{\mu\nu}(\z(s))\right]\frac{d\z^{\mu}(s)}{ds}\frac{d\z^{\nu}(s)}{ds}\bigg]^{1/2}\,.
}{}
The squared geodesic distance is then $\sigma = -x_{12}^{2}+\delta\sigma$, where
\eq{
\delta\sigma = -\kappa x_{12}^{\mu}x_{12}^{\nu}\int_{0}^{1}ds \gamma_{\mu\nu}(\z(s))+\OO(\kappa^2)\,.
}{} 
The scalar field commutator on a fixed classical background metric $g_{\mu\nu}=\eta_{\mu\nu}+\kappa \gamma_{\mu\nu}$, to $\OO(\kappa)$, is then given by
\eq{
i\,\textrm{sgn}(x^{0}_{12})\Delta(\sigma) &= i\,\textrm{sgn}(x^{0}_{12})\Delta(-x_{12}^{2})-i\,\textrm{sgn}(x^{0}_{12})\delta\sigma \frac{d}{dx^2_{12}}\Delta(-x_{12}^2) \\
&= \textrm{sgn}(x^0_{12})\left(1+\kappa x_{12}^{\mu}x_{12}^{\nu}\int_{0}^{1}ds\,\gamma_{\mu\nu}(z(s)) \,\frac{d}{d x_{12}^2}\right)\bigg[-i\frac{2\pi\,}{4\pi^{D/2}}\left(\frac{d}{dx_{12}^2}\right)^{D/2-2}\delta(x^2_{12})\bigg]\,,
}{}
which agrees exactly with the earlier coherent state calculation. Note that since we used geodesic dressing to invariantly define the locations of the operator insertions $\Phi(x_1)\Phi(x_2)$, the metric in \Eq{eq:coherentstatecommutatorfinal} comes out naturally in temporal gauge, $u^{\mu}\gamma_{\mu\nu}=0$.

To summarize, by using dressed operators we were able to ask questions about microcausality in quantum gravity in a diffeomorphism invariant approach. To leading order in perturbation theory we found that the commutator of scalar fields in a coherent background of gravitons is proportional to a light-cone singularity $\delta(\sigma)$, where the squared geodesic distance receives graviton corrections, $\sigma=-x_{12}^2 + \delta\sigma$. The commutator vanishes outside of the light cone as required by microcausality. However, the shape of the light cone itself is modified by the spacetime curvature, at least as parametrized by the invariant data that define the dressed operators.

The expression we obtain is precisely what is expected when performing a calculation on a fixed classical background, in temporal/synchronous coordinates. We are unable to make any claims about the sign of this correction, since the metric perturbation $\gamma_{\mu\nu}$ does not have definite sign. It would be very interesting, in future work, to repeat this analysis in the vacuum, but for the one-loop commutator.  In principle this requires evaluating a Fourier transform and taking an analytic continuation of the results computed in \Sec{sec:oneloopgravityresults}.

\subsection{Loop-Level Correlator}\label{sec:gravityloopcorrelator}

The tree-level matrix elements studied so far are nonzero at leading perturbative order, but they require turning on a graviton plane wave background. It is not easy to make universal claims about this contribution, such as positivity, because it is linear in the graviton polarization $\epsilon_{\mu\nu}$ and can thus be made positive, negative, or complex. 

In this section we will turn our attention to universal, unavoidable effects of geodesic dressing in quantum gravity. In particular, we will compute the vacuum two-point correlator $\langle \Omega| \Phi(x_1) \Phi(x_2)|\Omega\rangle$. This receives corrections at one loop from the correlated quantum fluctuations of the geodesics which, in turn, inherit their fluctuations from the underlying quantum fluctuations of the metric.

\subsubsection{Without Dressing}

Let us first focus on the calculation of $\langle \Omega| \phi(x_1) \phi(x_2)|\Omega\rangle$. As discussed extensively, without dressing this object is not gauge invariant. To set the stage for its dressed counterpart, we will investigate the undressed correlator's form in de Donder gauge. The correction to the correlator takes the form of a self-energy geometric series, as in \Eq{eq:QED_dressed_loop}. The diagrams which contribute have the same topology as in scalar QED, see \Eq{diag:QEDLoopUndressed}, and again diagram (b) is trivially vanishing in dimensional regularization. Thus, the leading self-energy correction is given by the one-loop Feynman integral,
\eq{
    \Sigma^{\textrm{de Donder}}(p)=-i \kappa^2 p_{a}p_{c}P_{4}^{\mu\nu a b}P_{4}^{\alpha\beta c d}\int_k\frac{P_{D\,\mu\nu\alpha\beta}}{k^{2}+i\varepsilon}\frac{(k+p)_b (k+p)_d}{(k+p)^{2}+i\varepsilon}\,.
}{grav_self-energy_undressed}
While not immediately obvious, this actually vanishes,
\begin{equation}\label{eq:dedonderselfenergy}
    \Sigma^{\textrm{de Donder}}(p)=0\,,
\end{equation}
as it also simplifies to a scaleless integral in dimensional regularization.

\subsubsection{With Dressing}

We will now compute the correlator of dressed scalars $\langle \Omega| \Phi(x_1)\Phi(x_2)|\Omega\rangle$ by summing dressing diagrams. The same diagram topologies as scalar QED, given in \Eq{diag:QEDLoopDressed}, contribute,

\eq{
\begin{array}{cccc}
\vcenter{\hbox{
\begin{tikzpicture}
\tikzfeynmanset{
every edge/.style={thick}
}
\begin{feynman}
  \vertex[dot, label=90:\(\phi(x_1)\)] (x1) {};
  \vertex[right=2 of x1, dot, label=90:\(\phi(x_2)\)] (x2) {};
  \vertex[below=2.4 of x1] (w1) {\(u_1\)};
  \vertex[below=2.4 of x2] (w2) {\(u_2\)};

  \vertex (m1) at ($(x1)!0.3!(x2)$);
  \vertex (m2) at ($(x1)!0.7!(x2)$);

  \diagram*{
    (x1) --[plain] (x2),
    (x1) --[scalar, blue] (w1),
    (x2) --[scalar, blue] (w2),
    (m1) --[graviton, out=270, in=270, looseness=2.0] (m2),
    (m1) --[draw=none, out=270, in=270, looseness=2.0, momentum'={[arrow shorten=0.35]\(k\)}] (m2),
  };
\end{feynman}
\end{tikzpicture}
}}
&
\vcenter{\hbox{
\begin{tikzpicture}
\tikzfeynmanset{
every edge/.style={thick}
}
\begin{feynman}
  \vertex[dot, label=90:\(\phi(x_1)\)] (x1) {};
  \vertex[right=2 of x1, dot, label=90:\(\phi(x_2)\)] (x2) {};
  \vertex[below=2.4 of x1] (w1) {\(u_1\)};
  \vertex[below=2.4 of x2] (w2) {\(u_2\)};

  \vertex (m) at ($(x1)!0.5!(x2)$);
  \vertex (g) at ($(x1)!0.5!(w1)$);

  \diagram*{
    (x1) --[plain] (x2),
    (x1) --[scalar, blue] (w1),
    (x2) --[scalar, blue] (w2),
    (g) --[draw=none, momentum'={[arrow shorten=0.25] \(k\)}] (m),
    (g) --[graviton] (m),
  };
\end{feynman}
\end{tikzpicture}
}}
&
\vcenter{\hbox{
\begin{tikzpicture}
\tikzfeynmanset{
every edge/.style={thick}
}
\begin{feynman}
  \vertex[dot, label=90:\(\phi(x_1)\)] (x1) {};
  \vertex[right=2 of x1, dot, label=90:\(\phi(x_2)\)] (x2) {};
  \vertex[below=2.4 of x1] (w1) {\(u_1\)};
  \vertex[below=2.4 of x2] (w2) {\(u_2\)};

  \vertex (m) at ($(x1)!0.5!(x2)$);
  \vertex (g) at ($(x2)!0.5!(w2)$);

  \diagram*{
    (x1) --[plain] (x2),
    (x1) --[scalar, blue] (w1),
    (x2) --[scalar, blue] (w2),
    (m) --[draw=none, momentum'={[arrow shorten=0.25] \(k\)}] (g),
    (m) --[graviton] (g),
  };
\end{feynman}
\end{tikzpicture}
}}
&
\vcenter{\hbox{
\begin{tikzpicture}
\tikzfeynmanset{
every edge/.style={thick}
}
\begin{feynman}
  \vertex[dot, label=90:\(\phi(x_1)\)] (x1) {};
  \vertex[right=2 of x1, dot, label=90:\(\phi(x_2)\)] (x2) {};
  \vertex[below=2.4 of x1] (w1) {\(u_1\)};
  \vertex[below=2.4 of x2] (w2) {\(u_2\)};
  \vertex (g1) at ($(x1)!0.5!(w1)$);
  \vertex (g2) at ($(x2)!0.5!(w2)$);

  \diagram*{
    (x1) --[plain] (x2),
    (x1) --[scalar, blue] (w1),
    (x2) --[scalar, blue] (w2),
    (g1) --[graviton] (g2),
    (g1) --[draw=none, momentum={[arrow shorten=0.3] \(k\)}] (g2),
  };
\end{feynman}
\end{tikzpicture}
}}
\\[2em]
\textbf{(I)} & \textbf{(IIa)} & \textbf{(IIb)} & \textbf{(III)}
\end{array}\,.
}{diag:GRLoopDressed}
Again, we have dropped additional diagrams with scaleless integrals.

We calculate these diagrams using the gravitational Feynman rules Eq.~\eqref{eq:grav-vertices}, Eq.~\eqref{eq:WL-graviton-feynman}, the scalar propagator $i/p^2$, and the de Donder-gauge graviton propagator $iP_{D}^{\alpha\beta\mu\nu}/k^2$. The four diagrams give
\eq{
\text{(I)} &= -\kappa^2\int_{p,k}e^{-ip(x_1-x_2)}\frac{1}{(p^2)^2} P_{4\,\mu\nu ab}P_{4\,\alpha\beta cd}P_{D}^{\alpha\beta\mu\nu}p^a p^c \frac{(p+k)^{b}(p+k)^{d}}{k^2(p+k)^2}\\
\text{(IIa)} &= -\kappa^2 \int_{p,k}e^{-ip(x_1-x_2)}\frac{1}{p^2}P_{4\,\alpha\beta c d}P_{D}^{\alpha\beta\mu\nu}p^{c}\frac{\OO_{b\mu\nu}(k,u_1)}{(-k\!\cdot\!u_1)^2}\frac{(p+k)^{b}(p+k)^{d}}{k^2(p+k)^2}\\
\text{(IIb)} &= -\kappa^2 \int_{p,k}e^{-ip(x_1-x_2)}\frac{1}{p^2}P_{4\,\alpha\beta c d}P_{D}^{\alpha\beta\mu\nu}p^{c}\frac{\OO_{b\mu\nu}(k,u_2)}{(k\!\cdot\!u_2)^2}\frac{(p+k)^{b}(p+k)^{d}}{k^2(p+k)^2}\\
\text{(III)} &= -\kappa^2 \int_{p,k}e^{-ip(x_1-x_2)}P_{D}^{\alpha\beta\mu\nu}\frac{\OO_{b\alpha\beta}(k,u_1)}{(-k\!\cdot\!u_1)^2}\frac{\OO_{d\mu\nu}(k,u_2)}{(k\!\cdot\!u_2)^2}\frac{(p+k)^{b}(p+k)^{d}}{k^2(p+k)^2}\,.
}{}
Summing these contributions, we can write the correction to the dressed correlator in terms of a self-energy,
\eq{
\Sigma^{\textrm{dressed}}(p) = -i\kappa^{2}\int_{k}P_{D}^{\alpha\beta\mu\nu}\frac{(p+k)^{b}(p+k)^{d}}{k^2(p+k)^2}\bigg[P_{4\,\mu\nu ab}\,p^{a}+p^2\frac{\OO_{b \mu\nu}(k,u_1)}{(-k\!\cdot\!u_1)^2}\bigg]
\bigg[P_{4\,\alpha\beta cd}\,p^{c}+p^2\frac{\OO_{d \alpha\beta}(k,u_2)}{(k\!\cdot\!u_2)^2}\bigg]\,.
}{eq:gravitydressedSE}

We can further massage this expression to see some familiar structures emerge. Performing a partial contraction, we find
\eq{
(p+k)^{b}\bigg[P_{4\,ab\mu\nu}p^{a}+p^2\frac{\OO_{b\mu\nu}(k,u_1)}{(-k\!\cdot\!u_1)^2}\bigg] &=(p+k)^{b}\,p^{a}P_{4\,\bar{\mu}\bar{\nu} ab} \overline{P^{\bar{\mu}(1)}_{\,\,\,\,\mu}}\,\overline{P^{\bar{\nu}(1)}_{\,\,\,\,\nu}} 
-(p+k)^2\frac{(p\!\cdot\!k)u_{1\,\mu}u_{1\,\nu}}{2(k\!\cdot\!u_1)^2} \\ 
&-(p+k)^2\frac{p_{\mu}u_{1\,\nu}+p_{\nu}u_{1\,\mu}}{2(-k\!\cdot\!u_1)}\,,
}{}
where the spatial projector is $P^{(i)}_{\mu \nu}=\eta_{\mu \nu}-\frac{k_{\mu}u_{i\,\nu}}{k\!\cdot\!u_{i}+i\varepsilon}$ as defined in Eq.~\eqref{eq:spatialprojector}, and the overline denotes complex conjugation. Contraction into the other square bracket yields the same structure, but in terms of $P^{(2)}$, $u_2$. The terms proportional to $(p+k)^2$ will pinch the scalar propagator in \Eq{eq:gravitydressedSE}, leading to a scaleless integral, and can thus be dropped. The dressed self-energy can then be written as
\eq{
    \Sigma^{\textrm{dressed}}(p)=-i\kappa^2 p_{a}p_{c}P_{4}^{\mu\nu a b}P_{4}^{\alpha\beta c d}\int_k \frac{N_{\mu\nu\alpha\beta}(k,u_1,u_2)}{k^{2}}\frac{(p+k)_b (p+k)_d}{(p+k)^{2}}\,,
}{eq:gravitydressingselfenergy}
where the effective graviton numerator is 
\eq{
    N_{\mu\nu\sigma\rho}(k,u_1,u_2)=\overline{P^{(1)}_{\mu a}}\,\overline{P^{(1)}_{\nu b}}P^{(2)}_{\sigma c}P^{(2)}_{\rho d}P^{abcd}_{D}\,.
}{eq:effectivegravitonpropagator}
In a homogeneous temporal gauge, corresponding to constant $u(x)=u$, the graviton propagator has exactly the form \Eq{eq:effectivegravitonpropagator} with $u_2=u_1=u.$ Here, in parallel with the QED case, the effective propagator in the self-energy has a sort of split temporal gauge form.

\subsubsection{Via Gauge Fixing}

As our last example showing the equivalence between calculations using geodesic dressing and calculations in $u(x)$ gauge, we will compute the correlator $\langle \Omega| \Phi(x_1) \Phi(x_2)|\Omega\rangle$ by evaluating only Diagram (I) of \Eq{diag:GRLoopDressed}, but with the internal graviton in $u(x)$ gauge.

The $\OO(\kappa^2)$, or one-loop contribution is
\eq{
\langle \Omega| \Phi(x_1) \Phi(x_2)|\Omega\rangle\bigg|_{\kappa^2} &= \kappa^2 \int_{y_1,y_2}\int_{p_1,p_2,l}e^{-ip_1\cdot(x_1-y_1)}e^{-ip_2\cdot(x_2-y_2)}e^{-il\cdot(y_1-y_2)}\frac{i}{p_1^2}\frac{i}{p_2^2}\frac{i}{l^2} \\
&\times\left(-ip_{1}^{a}l^{b}P_{4\,ab\mu_1\nu_1}\right)\,\left(ip_{2}^{c}l^{d}P_{4\,cd\mu_2\nu_2}\right) \,\,\langle 0|h^{\mu_2\nu_2}(y_2)h^{\mu_1\nu_1}(y_1)|0\rangle_{u(x)\text{-gauge}}\,,
}{}
and because the $u(x)$-gauge propagator is not translation invariant  we cannot immediately integrate over the internal vertices $y_1,y_2$. However, as was the case at tree level and in the QED calculations earlier, we will see that the $u(x)$ dependence gets localized to the operator insertions at $x_1,x_2$.

Inserting the $u(x)$-gauge propagator, \Eqs{eq:gravpropuxgauge}{eq:gravpropuxgauge2}, 
\eq{
&\langle \Omega| \Phi(x_1) \Phi(x_2)|\Omega\rangle\bigg|_{\kappa^2}\\
&= -i\kappa^2 \int_{y_1,y_2}\int_{p_1,p_2,l}\frac{e^{-ip_1(x_1-y_1)}e^{-ip_2(x_2-y_2)}e^{-il(y_1-y_2)}}{p_1^2 p_2^2 l^2} \left(p_{1}^{a}l^{b}P_{4\,ab\mu_1\nu_1}\right)\,
\left(p_{2}^{c}l^{d}P_{4\,cd\mu_2\nu_2}\right)  \\
&\times\int_{k}\frac{iP_{D\,\alpha_{2}\beta_{2}\alpha_{1}\beta_{1}}}{k^2}
\bigg[\eta^{\mu_2 \alpha_2}\eta^{\nu_2 \beta_2}e^{-ik y_2}-2i\left(\frac{\partial}{\partial y_2}\right)^{\mu_2}\left(\frac{\OO^{\nu_2 \alpha_2\beta_2}(k,u(y_2))e^{-ik y_2}}{[k\!\cdot\!u(y_2)]^2}\right)\bigg] \\
&\hspace{74pt}\times \bigg[\eta^{\mu_1 \alpha_1}\eta^{\nu_1 \beta_1}e^{ik y_1}+2i\left(\frac{\partial}{\partial y_1}\right)^{\mu_1}\left(\frac{\OO^{\nu_1 \alpha_1\beta_1}(k,u(y_1))e^{ik y_1}}{[-k\!\cdot\!u(y_1)]^2}\right)\bigg]\,.
}{}
As before, simplifications occur because the $u(y_{1}), u(y_{2})$-dependent terms are under a derivative, which we can readily integrate by parts to replace with momenta,
\eq{
&\langle \Omega| \Phi(x_1) \Phi(x_2)|\Omega\rangle\bigg|_{\kappa^2}\\
&= \kappa^2 \int_{y_1,y_2}\int_{p_1,p_2,l,k}e^{-ip_1 x_1}e^{-ip_2 x_2}\frac{e^{i(p_1-l+k) y_1}e^{i(p_2+l-k) y_2}}{p_1^2 p_2^2 l^2 k^2}\,P_{D\,\alpha_{2}\beta_{2}\alpha_{1}\beta_{1}}\left(p_{1}^{a}l^{b}P_{4\,ab\mu_1\nu_1}\right)\left(p_{2}^{c}l^{d}P_{4\,cd\mu_2\nu_2} \right) \\
&\times 
\bigg[\eta^{\mu_2 \alpha_2}\eta^{\nu_2 \beta_2}-2(p_2+l)^{\mu_2}\left(\frac{\OO^{\nu_2 \alpha_2\beta_2}(k,u(y_2))}{[k\!\cdot\!u(y_2)]^2}\right)\bigg]\bigg[\eta^{\mu_1 \alpha_1}\eta^{\nu_1 \beta_1}+2(p_1-l)^{\mu_1}\left(\frac{\OO^{\nu_1 \alpha_1\beta_1}(k,u(y_1))}{[-k\!\cdot\!u(y_1)]^2}\right)\bigg]\,.
}{eq:gravityoneloopfromgauge}
To see the simplification we contract the stress-tensor vertex into the total derivative, in other words into the coefficient of the $\OO$ tensor,
\eq{
\left(p_{1}^{a}l^{b}P_{4\,ab\mu_1\nu_1}\right)(p_1-l)^{\mu_1} = \frac{p_1^2}{2}l_{\nu_1}-\frac{l^2}{2}p_{1\,\nu_1}\,.
}{}
The term proportional to $l^2$ will pinch the internal scalar propagator in \Eq{eq:gravityoneloopfromgauge}, leaving a scaleless integral which we can then drop.  The term proportional to $p_1^2$ will instead pinch an external scalar propagator, allowing us to integrate $p_1$ and obtain $\delta(x_1-y_1)$ which localizes the dust field to the matter insertion, $u(y_1)\rightarrow u(x_1)=u_1$. The same mechanism occurs when contracting the other stress-tensor vertex into the other set of square brackets, and the upshot is again that we can simply replace $u(y_2)\rightarrow u(x_2)=u_2$ in the integrand. 

The $u(y_{1}), u(y_{2})$-dependent terms enter the graviton propagator with a total derivative in the form of a gauge transformation, and stress tensors are conserved except at the locations of matter insertions; these facts in concert imply the $u(y_{1,2})$ dependence necessarily gets localized to matter insertions. Altogether, we can then write the correlator as
\eq{
&\langle \Omega| \Phi(x_1) \Phi(x_2)|\Omega\rangle\bigg|_{\kappa^2}\\
&= \kappa^2 \int_{y_1,y_2}\int_{p_1,p_2,l,k}e^{-ip_1 x_1}e^{-ip_2 x_2}\frac{e^{i(p_1-l+k) y_1}e^{i(p_2+l-k) y_2}}{p_1^2 p_2^2 l^2 k^2}\,P_{D\,\alpha_{2}\beta_{2}\alpha_{1}\beta_{1}} l_{b}l_{d} \\
&\times 
\bigg[P_{4}^{cd\alpha_2 \beta_2}p_{2\,c}-p_2^2\left(\frac{\OO^{d \alpha_2\beta_2}(k,u_2)}{[k\!\cdot\!u_2]^2}\right)\bigg]
\bigg[P_{4}^{ab\alpha_1\beta_1}p_{1\,a}+p_1^{2}\left(\frac{\OO^{b \alpha_1\beta_1}(k,u_1)}{[-k\!\cdot\!u_1]^2}\right)\bigg]\,.
}{}
Now we can finally perform the $y_{1}, y_{2}$ integrals to obtain the momentum conserving delta functions.  Doing so fixes $p_{2}=-p_{1}$ and $l=p_1+k$, and we see that
\eq{
&\langle \Omega| \Phi(x_1) \Phi(x_2)|\Omega\rangle\bigg|_{\kappa^2} = \int_{p}e^{-ip\cdot(x_1-x_2)}\left(\frac{i}{p^2}\right)^2 i\Sigma^{\textrm{dressed}}(p)\,,
}{}
where $\Sigma^{\textrm{dressed}}(p)$ is indeed the result of the dressing calculation, \Eq{eq:gravitydressedSE}.

\subsubsection{Loop Integration}\label{sec:oneloopgravityresults}

We will now explicitly evaluate the dressed correlation function $\langle \Omega|\Phi(x_1)\Phi(x_2)|\Omega \rangle$ at one loop. As we have demonstrated above, the calculation is most efficiently done by computing a single self-energy diagram, but with the graviton in the gauge $u^{\mu}(x)h_{\mu\nu}=0$. Despite the inhomogeneous gauge condition, the correction still respects translation invariance, and is given by a single self-energy integral 
\eq{
    \Sigma^{\textrm{dressed}}(p)=-i\kappa^2 p_{a}p_{c}P_{4}^{\mu\nu a b}P_{4}^{\sigma\rho c d}\mu^{4-D}\int_k \frac{N_{\mu\nu\sigma\rho}(k,u_1,u_2)}{k^{2}+i\varepsilon}\frac{(p-k)_b (p-k)_d}{(p-k)^{2}+i\varepsilon}\,,
}{grav_self-energy}
where the effective graviton numerator in $u(x)$ gauge is
\begin{equation}
    N_{\mu\nu\sigma\rho}(k,u_1,u_2)=P^{(1)}_{\mu a}P^{(1)}_{\nu b}\overline{P^{(2)}_{\sigma c}}\,\overline{P^{(2)}_{\rho d}}P^{abcd}_{D}\,.
\end{equation}
$P^{(i)}_{\mu\nu}$ is the projector in \Eq{eq:spatialprojector} and we have rerouted the direction of the loop momentum $k$ to conform with more conventional Feynman integral tables~\cite{Smirnov2012AnalyticTools}.

Just as in QED we will not attempt to evaluate this integral for general $u_1, u_2$. There is a pinch singularity at $u_1=u_2$ which is more severe in gravity than QED due to double poles. As in the QED case, we retreat to the configuration $u_2=-u_1$, corresponding to one operator being connected by a geodesic to past infinity and the other to future infinity. Admittedly, this configuration seems less physical than an experiment conducted using probes in simultaneously free-falling laboratories, all initialized in the past. However, we defer exploring this important case, where all $u_j$ are future directed, to future work.

The evaluation of the integral in $\Sigma^{\textrm{dressed}}(p)$ proceeds just like its QED counterpart, with details given in \App{app:one_loop}. In four dimensions, the dressed gravitational self-energy is

\eq{
    &\Sigma^{\textrm{dressed}}(p)\\
    &= \frac{\kappa^2}{96\pi^2}\bigg[
    \frac{17p^4+10p^2\omega^2}{4-D} 
    - p^4\left(8\log\left(\frac{-p^2}{\tilde{\mu}^2}\right)+\log\left(\frac{-\omega}{\tilde{\mu}}\right)\right)
    -p^2 \omega^2\left(\log\left(\frac{-p^2}{\tilde{\mu}^2}\right)+8\log\left(\frac{-\omega}{\tilde{\mu}}\right)\right) \\
    &\hspace{45pt}-5p^4+\frac{19}{4} p^2 \omega^2-\frac{3p^2\omega^2 \sin^{-1}(\sqrt{y})}{\sqrt{y(1-y)}}\left(2+\log\left(\frac{-p^2}{\omega^2}\right)\right)+\frac{3\pi^2(-p^2)^{3/2}\omega}{\sqrt{1-y}} \\
    &\hspace{45pt}+3p^2 \omega^2 {}_2F_{1}^{(0,1,0,0)}(1,1,\tfrac{3}{2},y)+3p^2 \omega^2 {}_2F_{1}^{(0,0,1,0)}(1,1,\tfrac{3}{2},y)+\OO(D-4)
    \bigg]\, ,
}{eq:gravity_dressed_selfenergy}
where, again, the branch cuts are determined by the $i\varepsilon$ prescriptions $-p^2\rightarrow-(p^2+i\varepsilon)$,  $\omega=2p\cdot u_1+i\varepsilon$, and $y=(p\cdot u_1+i\varepsilon)^2/(p^2+i\varepsilon)$. The dressed self-energy is considerably more interesting than the vacuous de Donder gauge result, \Eq{eq:dedonderselfenergy}. It has novel UV divergences, with concomitant log-running, as well as novel analytic structure. The implications of this for the position space expression will be left for future work. 

\subsection{Dressing with Quantum Fields}\label{sec:gravity_physical_interpetation}

The one-loop gravitational self-energy is superficially similar to its scalar QED analog in \Eq{eq:QED1loopselfenergy}.  However, there is a critical technical distinction that actually has physical import, which relates to counterterms.  As we will see, this reveals an important caveat about dynamical dressing fields.

\subsubsection{Counterterms}

While our gauge and gravity calculations involve the same types of functions, the latter is distinguished by a Lorentz-violating, UV divergent term proportional to $p^2\omega^2$.  The UV divergences in \Eq{eq:gravity_dressed_selfenergy} are absorbed by the counterterm Lagrangian,
\eq{
    \delta S=\int d^4x\,\sqrt{-g} \bigg[\frac{1}{2}c_1(\nabla_{\mu} \nabla^{\mu}\phi)^2 +\frac{1}{2}c_2 (\nabla_\mu \nabla^\mu \phi)(u^\lambda u^\nu \nabla_\lambda\nabla_\nu \phi) \bigg]\,,
}{}
where $c_1$ and $c_2$ are running Wilson coefficients.   Here the fluid four-velocity $u^{\mu}(x)$ necessarily appears in the counterterm Lagrangian. This lies in contrast with the case of scalar QED, where according to \Eq{eq:QED1loopselfenergy}, no such Lorentz-violating divergence appears. 

On general grounds, one expects Lorentz-violating counterterms to appear in Lorentz-violating gauges.  In this situation, however, the fluid velocity $u^{\mu}$ is a dynamical spurion for Lorentz violation.   Considering this as a \textit{physical} dynamical field rather than a fixed structure changes the interpretation of this counterterm in a crucial way. Said another way, Lorentz invariance is no longer broken explicitly, but spontaneously. 

\subsubsection{Uncertain Clocks and Rulers}

The requirement of a counterterm reveals an important physical point: to render local observables sensible, we must locate them relationally to dynamical clocks and rulers---but in a quantum world these objects exhibit uncertainty.

In the present context, this subtlety emerges because we entirely neglected the action for the fluid itself. Moreover, by imposing the equation of motion $u^{\nu}\nabla_{\nu}u^{\mu}=0$, we assumed that the fluid can be treated semiclassically. Both of these simplifications, which are almost always made in the literature on geodesic dressing, require scrutiny.
In QED this is justified by taking the mass of the dust particles to infinity so that they simply follow classical straight-line paths through Minkowski spacetime. In quantum gravity the story is different, and there is a challenge which warrants careful thought. In particular, there are opposing limitations coming from quantum mechanics on one side and from general relativity on the other.

In particular, increasing the mass of the fluid particles incurs gravitational backreaction. Taking this to the extreme, the dust particles will eventually collapse into black holes and destroy the very system we are interested in studying. To avoid backreaction we must then assume that the test particles are sufficiently light, and in the ideal case they should be arbitrarily light. In classical general relativity this poses no issue and we understand geodesics as describing the motion of test particles in spacetime. 

In quantum gravity, however, we are not permitted to take the masses of particles to be arbitrarily small. The fluid is a dynamical system coupled to a quantum mechanical metric, and so it is also quantum mechanical.  For a pressureless dust, the fluid consists of a collection of free-falling point particles. Quantum mechanically, this is then a collection of freely dispersing wavepackets. A particle with mass $m$ which has been freely dispersing as a Gaussian wavepacket for time $t$ has, at minimum, a position spread $\delta x \sim \sqrt{\hbar t/m}$. Given a sufficiently long time, any free wavepacket describing a finite-mass particle will diffuse throughout space. For an arbitrarily light particle, this diffusion time is arbitrarily short. 

In practice this does not impose a significant restriction on the mass of the dust particles. Suppose we want our dressing to insert the operator within a resolution $\Delta x$. The mass of each dust particle must not be too big, $m < \Delta x/G$, otherwise the particle will form a black hole larger than $\Delta x$. The mass also cannot be too small, $m>t/(\Delta x)^2$, since otherwise wavepacket spread will dominate. This implies a weak but nonzero bound on the resolution. Restoring $c$ and $\hbar$, we obtain
\eq{
\Delta x > (ct L^{2}_{\rm Pl})^{1/3} \approx 10^{-18}\,\textrm{meters}\times\left(\frac{t}{\textrm{year}}\right)^{1/3}\,.
}{}
While this is not a practical limitation, it is an interesting limitation in principle.  For any choice of mass the semiclassical geodesic description breaks down after sufficiently long times. In this work---as in essentially all treatments in the literature---we made the assumption that fields can be reliably dressed ``to infinity'' via geodesics. If the geodesics are truly describing the semiclassical motion of a quantum mechanical system, then after an infinite time they will no longer be localized at all in the bulk. 

In summary, gravity necessarily reacts to the finite masses of the free-falling clocks, while quantum mechanics mandates the spread of their corresponding wavefunctions.  Together, these fundamental principles impose a universal IR cutoff on the propagation time for any given geodesic.  The initial data surface, from which the geodesics emanate, is then required to lie at a finite time interval in the past.
This effect implies that even the most idealized reference system is inevitably bounded by a finite lifetime before it hopelessly delocalizes, putting local measurements out of reach even in principle. One can view this phenomenon as the IR partner of the UV obstruction in our opening parable, and it appears just as inescapable: gravity forbids not only local observations but also eternally local observers. 

Having the initial data surface at a finite time, rather than asymptotic infinity, implies an important change in our physical interpretation. When we dress local operators ``to infinity'', they become gauge invariant because diffeomorphisms are inactive at infinity. When we dress to a surface finitely in the past, the resulting dressed operators are only invariant under diffeomorphisms that act trivially on the initial data surface. One then needs a justification for {\it i}) why diffeomorphisms do not act on the initial time surface, or equivalently {\it ii}) why the coordinates on the initial time surface have physical meaning.

A somewhat formal justification for point {\it i}) is given in \cite{Goeller:2022rsx}, where the importance of splitting gauge transformations into ``small'' and ``large'' components is emphasized, alongside a discussion of how the space of initial data and their symmetries under ``large'' gauge transformations must be understood as a part of the specification of the physical system.

Here we will instead give a more physical intuition for point {\it ii}). The initial data surface can be understood as a model for a period of time during which an experimenter performs a series of measurements and operations that ascertain how the geodesics were initialized. Our dressed observables are then defined relationally with respect to the laboratory of this experimentalist. 

It is tempting to argue that the dressing geodesics are fictitious and not to be taken seriously as physical systems. By this logic, neither backreaction nor quantum fluctuations need be considered. However, this perspective seems inherently flawed. Firstly, if we are dressing to a system which can never be realized physically, then these observables have no connection to real world measurements. Secondly, the requirement of $u^{\mu}$-dependent counterterms suggests that we should treat the fluid as dynamical degrees of freedom. Thirdly, the counterterm structure shows that interaction terms between the fluid and other dynamical fields are generated in the gravitational path integral by renormalization group flow. Taken together, these points imply that we have to place the fluid on the same physical footing as these fields, specifically by including the action for the dust field in the path integral and treating the dust as a proper dynamical quantum field.

While one can work semiclassically with an IR cutoff, the tension between the semiclassical geodesic dressing and the infinite time limit appears inevitable. To confront such quantum effects head-on, one can instead dress to fully \textit{quantum} systems. In this case one does not make a gauge choice based on a classical equation of motion, for instance harmonic gauge $\Box x^{\mu}=0$ or synchronous gauge $u^{\alpha}\nabla_{\alpha}u^{\beta}=0$, but instead anchors the coordinates to a physical quantum mechanical subsystem that is actually present in the theory. Such quantum reference frames are well studied for quantum systems \cite{Bartlett:2006tzx} and have accumulated much recent attention for their role in describing observers and observables in quantum gravity, for example see \cite{Chandrasekaran:2022cip, Susskind:2023rxm, Maldacena:2024spf, Ivo:2025yek, DeVuyst:2024khu, DeVuyst:2024fxc, Harlow:2025pvj, Abdalla:2025gzn, Chen:2024rpx, Kudler-Flam:2024psh, Kolchmeyer:2024fly}. A particularly exciting line of inquiry is the potential importance of quantum reference frames and dressed observables to the black hole information paradox~\cite{Raju:2020smc, Geng:2021hlu, Geng:2024dbl, Antonini:2025sur, Geng:2026asi}.

\section{Discussion}

Dressed observables are more than just formal constructions---they are eminently calculable in perturbation theory.  On the one hand, we have shown how to introduce dressing via a dynamical fluid that evolves and reacts with the gauge and gravitational fields.  Here the effects of dressing are computed using simple Feynman rules. On the other hand, we have shown how the very same physics can be entirely reshuffled into a peculiar choice of gauge fixing using a dynamical reference vector. These dual approaches are a fairly general consequence of gauge invariance that we expect applies to other forms of dressing \cite{Giddings:2025xym}.

The surprising simplicity of these dressed observables augurs well for studying other low-energy quantum gravity phenomena in perturbation theory.  This might include a variety of bulk physics that probes causality or black holes.  It is then natural to comment on some avenues for future inquiry.

The present work is just the tip of the iceberg for what is perturbatively calculable in the world of dressed bulk observables.  
For this reason, it would first and foremost  be interesting to calculate other dressed observables, either at higher points or at higher loops.  Here we have centered on two-point correlators up to one loop and three-point correlators at tree level, so there is potential to learn much more from explicit calculation.  
The impetus for this effort is the modern amplitudes program, where the existence of concrete mathematical expressions has been an engine for new insights.  Dressed observables are strongly reminiscent of on-shell scattering amplitudes, so it stands to reason that they might similarly exhibit interesting hidden structures. Furthermore, the computational ease afforded by flat space momenta may also extend to dressed observables in curved spacetime using differential operator kinematics \cite{Cheung:2022pdk, Herderschee:2022ntr}.

Second of all, it would be interesting to construct dressed observables from the perspective of the bootstrap, rather than direct Feynman diagrammatic expansion.     
At present, essentially all efforts in the bootstrap have been directed at on-shell scattering amplitudes or correlators in conformal field theories.  In contrast with these asymptotic observables, dressed observables perturbatively approximate local operators, making them a novel class of bootstrap candidates.

Thirdly, there is the intriguing question of how to connect formal dressed observables to bona fide quantum gravitational measurement. Even without gravity, this relationship is highly nontrivial \cite{Moult:2025nhu}. While dressed scalar correlators are not themselves yet experimental observables, they are a dry run for the more realistic quantities. For instance, dressed stress tensor operators could be a finite-distance generalization of the asymptotic detector operators in \cite{Herrmann:2024yai}. Experimental probes of the interface between quantum mechanics and gravity can require a more detailed model of measurement than detectors following worldlines, but we expect to ultimately measure dressed observables. For recent examples, see \cite{Aziz2025, Vermeulen:2024vgl, Bengyat:2023hxs, Overstreet:2021hea, Carney:2024wnp,Carney:2022dku, Carney:2021yfw, Tobar:2023ksi}.

Last but not least is the question of whether geometry survives the presence of dynamical gravitons.  Many proposals for quantum notions of focusing, area, and time delay either suppress graviton fluctuations or treat them only indirectly. This leaves a basic gap in our understanding, since gravitons are the degrees of freedom that most directly probe the low-energy tension between diffeomorphism invariance and locality.  The framework developed here offers a direct perturbative handle on this problem. We already saw this in the dressed commutator, which encodes the gravitational deformation of the light cone. More broadly, the same logic has been used to construct bulk observables that directly diagnose light-cone bending in quantum gravity \cite{Carney:2024wnp, Sivaramakrishnan:2025srr}. Other geometric quantities have also been lifted into the quantum regime: area \cite{Engelhardt:2014gca} and focusing \cite{Bousso:2015mna} can be formulated using generalized entropy, and graviton corrections to area have recently been computed in AdS/CFT \cite{Colin-Ellerin:2025dgq}. Related definitions of subregions, causal diamonds, and their observable algebras have also been constructed by dressing to an observer \cite{Witten:2023qsv, Jensen:2023yxy}.  We leave to future work whether perturbative dressing can offer any insight into these bulk gravitational phenomena.

\begin{center} 
{\bf Acknowledgments}
\end{center}

We thank participants of the “Observables in Quantum Gravity:
from Theory to Experiment” conference, held in January
2025 at the Aspen Center for Physics, for illuminating discussions. J.W.-G. is supported by a President's Postdoctoral Fellowship and by the Department of Energy (Grant No. DE-FG02-04ER41338 and FG02-06ER41449).
C.C., L.Z., and A.S. are supported by the D.O.E., Office of High Energy Physics, under Award No. DE-SC0011632, by
the Walter Burke Institute for Theoretical Physics, and by the Leinweber Forum for Theoretical Physics. A.S. is also supported by the Heising-Simons Foundation “Observational Signatures of Quantum Gravity” collaboration grant 2021-2817. 

\noindent

\pagebreak 

\appendix

\section{Conventions}
\label{app:conventions}

The symbolic conventions of this paper are as follows:
\begin{itemize}
\item Metric signature: $(+, -, -, -)$.
\item Momentum conserving delta functions: $\delta(\sum_j p_j) = (2\pi)^D \delta^{(D)}(\sum_j p^{\mu}_j)$
\item Fourier transform and inverse transform:
    \eq{
    f(x)&=\int_{p} e^{-ipx}\tilde{f}(p)= \int\frac{d^Dp}{(2\pi)^D}\, e^{-ipx}\tilde{f}(p),\\
    \tilde{f}(p)&=\int_x e^{ipx}f(x)=\int d^Dx\, e^{ipx}f(x).
    }{}
In Feynman diagrams, incoming momentum arrows represent
    \eq{
        \begin{tikzpicture}[baseline=-\the\dimexpr\fontdimen22\textfont2\relax] 
             \tikzfeynmanset{
             every edge/.style={black, thick}
             }
            \begin{feynman}[inline=(x)]
              \vertex[dot, label=180:\(f(x)\)] (x) {}; 
              \vertex[right=1 of x] (e);
              \diagram*{
                (x) --[rmomentum=\(p\)] (e),
              };
              \draw[fill=black] (x) circle(0.075);
            \end{feynman}
        \end{tikzpicture} &= \int_p e^{-ipx} \tilde{f}(p),\\ 
        \begin{tikzpicture}[baseline=-\the\dimexpr\fontdimen22\textfont2\relax] 
             \tikzfeynmanset{
             every edge/.style={black, thick}
             }
            \begin{feynman}[inline=(x)]
              \vertex[dot, label=180:\(f^\dagger(x)\)] (x) {}; 
              \vertex[right=1 of x] (e);
              \diagram*{
                (x) --[rmomentum=\(p\)] (e),
              };
              \draw[fill=black] (x) circle(0.075);
            \end{feynman}
        \end{tikzpicture} &= \int_p e^{-ipx} \tilde{f}^\dagger(-p),
     }{eq:momentum-arrow}
while outgoing arrows correspond to reversing the sign of the momentum, $p\to -p$.

\item  Wick contraction rules:
    \eq{
    \langle \tilde{\phi}(p)\tilde{\phi}^\dagger(-q)\rangle &= \frac{i}{p^2 + i\varepsilon} \times \delta(p+q),\qquad\text{(complex scalar)}
    \\
    \langle \tilde{\phi}(p)\tilde{\phi}(q)\rangle &= \frac{i}{p^2 + i\varepsilon} \times \delta(p+q),\qquad\text{(real scalar)}
    \\
    \langle \tilde{A_\mu}(p)\tilde{A}_\nu(q)\rangle &= \frac{-i\eta_{\mu\nu}}{p^2 + i\varepsilon} \times \delta(p+q),\qquad \text{(Feynman gauge)}
    \\
    \langle \tilde{h}_{\mu\nu}(p) \tilde{h}_{\rho\sigma}(q)\rangle & = \frac{i P_{D\,\mu\nu\rho\sigma}}{p^2 + i\varepsilon} \times \delta(p+q), \qquad\!\!\!\text{(de Donder gauge)}
    }{eq:X-X-contraction}
    where $P_{D\,\mu\nu\rho\sigma} = \frac{1}{2} (\eta_{\mu\rho}\eta_{\nu\sigma}+\eta_{\mu\sigma}\eta_{\nu\rho})-\frac{1}{D-2}\eta_{\mu\nu}\eta_{\rho\sigma}$. 
\item Operator convention: 
\eq{
A_\mu(x)=\sum_{\lambda}\int\frac{d^3\bm{p}}{(2\pi)^3 (2|\bm{p}|)^{1/2}}\left[\epsilon^{\lambda}_\mu(p)\hat{a}_{\lambda,p} e^{-ipx}+\epsilon_\mu^{\lambda\,\ast}(p)\hat{a}^\dagger_{\lambda,p} e^{ipx}\right],
}{}
for the photon field operator, where $p\equiv(|\bm{p}|,\bm{p})$ is on shell and $[a_{\lambda,p},a^{\dagger}_{\lambda^{\prime}\,p^\prime}]=\delta_{\lambda,\lambda^{\prime}}(2\pi)^3\delta^{(3)}(\bm{p}-\bm{p^\prime})$.
For on-shell external photons, $|A_{\lambda}(k)\rangle\equiv (2|\bm{k}|)^{1/2}\,a_{\lambda,k}^\dagger|0\rangle$ with $k\equiv(|\bm{k}|,\bm{k})$, this implies
\eq{
\langle 0 | A_\mu(x)| A_{\lambda}(k)\rangle &= \epsilon^{\lambda}_\mu(k)e^{-ikx},\\
\langle A_{\lambda}(k) | A_\mu(x)| 0\rangle &= \epsilon_\mu^{\lambda\,\ast}(k) e^{ikx},\\
\langle 0 | \tilde{A}_\mu(p)| A_{\lambda}(k)\rangle &= \epsilon^{\lambda}_\mu(k)\delta(p-k),\\
\langle A_{\lambda}(k) | \tilde{A}_\mu(p)| 0\rangle &= \epsilon_\mu^{\lambda\,\ast}(k)\delta(p+k).
}{eq:photon-plr}
Expressions for the graviton field operator  are identical except with each vector polarization $\epsilon^\lambda_\mu(k)$ replaced by the tensor polarization $\epsilon^\lambda_{\mu\nu}(k)$. For notational simplicity, we often suppress the helicity label $\lambda$ for both external states and polarization tensors, writing $|A(k)\rangle$, $\epsilon_{\mu}(k)$, and so on.

\end{itemize}

\section{Tree-Level Fourier Integrals}
\label{app:tree_fourier}

In order to calculate the tree-level dressed correlators $\langle \Omega | \Phi^\dagger(x_1)\Phi(x_2) | A(k)\rangle$ in scalar QED, and $\langle \Omega |\Phi(x_1) \Phi(x_2) | h(k)\rangle$ in gravity, all we need are the following Fourier integrals:
\eq{
\int_{p} e^{-ip x} &\times \left\{ \frac{1}{p^2},\  \frac{p^\mu}{p^2(k-p)^2} \right\} \quad \text{(QED)},\\
\int_{p} e^{-ip x} &\times \left\{ \frac{p^\mu}{p^2},\  \frac{p^\mu p^\nu}{p^2(k-p)^2} \right\} \quad \text{(GR)},
}{}
where the denominators are implicitly defined with the prescription $p^2\to p^2+i\varepsilon$. The relevant integrals can be obtained from the following family of integrals,
\begin{equation}
G(\lambda_1,\lambda_2; x) = \int_{p}
\frac{e^{-i p  x}}{(p^2 + i\varepsilon)^{\lambda_1} ((k-p)^2+i\varepsilon)^{\lambda_2}}\,.
\end{equation}
Specifically, we will need $G(1,0;x)$, $G(2,0;x)$, and $G(1,1;x)$. The momenta in the numerators can be easily obtained by acting with $i \partial_\mu$. 

The first basic integral $G(\lambda,0;x)$ is
\eq{
G(\lambda,0;x) =  \int_{p}
\frac{e^{-i p  x}}{(p^2 + i\varepsilon)^\lambda}
=
 \frac{i\, (-1)^\lambda}{4^\lambda \pi^{D/2}}\frac{\Gamma\!\left(D/2 - \lambda\right)}
{\Gamma(\lambda)}
\frac{1}{(-x^2 + i\varepsilon)^{D/2 - \lambda}}\,,
}{}
given for example in Smirnov~\cite{Smirnov2012AnalyticTools} in mostly-minus signature. The familiar case $\braket{0|T(\phi(x)\phi(0))|0} $ $=$ $ i G(1,0;x)$ is 
\eq{
G(1,0,x) = \int_{p}
\frac{e^{-i p  x}}{p^2 + i\varepsilon}
=
 \frac{-i\,\Gamma\!\left(D/2 - 1\right)}
{4\pi^{D/2}}
\frac{1}{(-x^2 + i\varepsilon)^{D/2 - 1}}\,,
}{app:int-a}
\noindent
which in $D=4$ gives the standard result 
$\frac{-i}{4\pi^{2}}
\frac{1}{-x^2 + i\varepsilon}.
$
We then have
\eq{
i\partial^\mu_x ~G(1,0;x) = \int_{p}
\frac{p^\mu \, e^{-i p  x}}{p^2 + i\varepsilon}
=\frac{\Gamma\!\left(D/2\right)}
{2\pi^{D/2}}
\frac{x^\mu}{(-x^2 + i\varepsilon)^{D/2}}\,.
}{app:int-b}
We will also need 
\eq{
G(2,0;x) = 
\int_{p}
\frac{e^{-i p  x}}{(p^2 + i\varepsilon)^2}
=
\frac{i\,\Gamma\!\left(D/2 - 2\right)}
{16\pi^{D/2}}
\frac{1}{(-x^2 + i\varepsilon)^{D/2 - 2}}\,.
}{app:int-c0}
As an aside, note that we can obtain the $D=4$ result as the $\delta\to 0$ limit of $D=4-2\delta$. With UV cutoff $a$,
\eq{
\int \frac{d^4 p}{(2\pi)^4}
\frac{e^{-i p  x}}{(p^2 + i\varepsilon)^2}
=\lim_{\delta\to 0}\frac{-i}{16\pi^2}\frac{1+\delta\ln(-x^2+i\varepsilon)}{\delta}=\frac{-i}{16\pi^2}\ln\frac{-x^2+i\varepsilon}{a^2}\,.
}{}
Continuing on, we can give the integrand of $G(\lambda_1,\lambda_2;x)$ nontrivial numerators by taking derivatives,
\eq{
i \partial^\mu_x ~ G(2,0;x) = \int_{p}
\frac{p^\mu\, e^{-i p  x}}{(p^2 + i\varepsilon)^2}
=\frac{-\,\Gamma\!\left(D/2 - 1\right)}{8\pi^{D/2}}\frac{x^\mu}{(-x^2 + i\varepsilon)^{D/2 - 1}}\,,
}{app:int-c}
\noindent
and
\eq{
(i \partial^\mu_x)(i\partial^\nu_x)~ G(2,0;x) =
\int_{p}
\frac{p^\mu p^\nu\, e^{-i p  x}}{(p^2 + i\varepsilon)^2}
= \frac{i\,\Gamma\!\left(D/2 - 1\right)}
{8\pi^{D/2}}
\frac{x^2 \eta^{\mu\nu} - (D-2) x^\mu x^\nu}
{(-x^2 + i\varepsilon)^{D/2}}\,.
}{app:int-d}

Next, we compute $G(1,1;x)$,
\begin{equation}
G(1,1;x)=\int_{p} \frac{e^{-i p x}}{(p^2+i\varepsilon)((k-p)^2+i\varepsilon)}\,,
\end{equation}
with $k^2=0$. We compute this by first using Feynman parameters,
\begin{equation}
G(1,1;x)=\int_{p} \frac{e^{-i p x}}{(p^2+i\varepsilon)((k-p)^2+i\varepsilon)} = \int_{p} \int_0^1 \frac{du ~ e^{-i p x}}{(u p^2+(1-u)(p-k)^2 + i\varepsilon)^2 }\,.
\end{equation}
Simplifying and shifting $p \rightarrow p+(1-u)k$, we have
\begin{equation}
G(1,1;x)=\int_{p}\int_0^1 \frac{du ~ e^{-i (p+(1-u)k)x}}{(p^2+i\varepsilon)^2} = \int_0^1 du e^{-i(1-u)k x} ~G(2,0;x) \,,
\end{equation}
which gives
\begin{equation}
G(1,1;x)=\frac{1-e^{-i k x}}{i kx} G(2,0;x) =  \frac{\Gamma(D/2-2) }{16 \pi^{D/2}} \frac{1}{(-x^2+i\varepsilon)^{D/2-2}}\frac{1-e^{-i k x}}{k\!\cdot\! x} \, .
\end{equation}

\paragraph{Scalar QED} We now have the ingredients to calculate the Fourier transforms in Sec.~\ref{subsec:QED_dressed_tree}:
\eq{
\big \langle \Omega \big| \Phi^\dagger(x_1) \Phi(x_2) \big| A(k) \big\rangle
&= \begin{aligned}[t]
  & i g \int_{p_1\,p_2}\, e^{-i p_1 x_1 - i p_2 x_2}\,
\frac{\delta(p_1 + p_2 - k)}{p_1^2\, p_2^2}\\
&\times \Bigg[
(p_2- p_1)\!\cdot\!\epsilon
+ p_1^2\,\frac{u_1\!\cdot\!\epsilon}{u_1\!\cdot\!k}
- p_2^2\,\frac{u_2\!\cdot\!\epsilon}{u_2\!\cdot\!k}
\Bigg] \, .
\end{aligned}
}{}
Using $k\cdot\epsilon=0$, the first term in the square bracket is
\eq{
&ig\,e^{-ik x_2}
\int_{p_1} \,
e^{-ip_1 x_{12}}\,
\frac{-2p_1\!\cdot\! \epsilon}{(p_1^2+i\varepsilon)\,((k-p_1)^2+i\varepsilon)}
=
-2ig\,e^{-ik x_2}
(i \epsilon \!\cdot\! \partial_{x_{12}})
G(1,1;x_{12})
\\
&= \frac{-g\Gamma(D/2-1)}{4\pi^{D/2}}\,
\frac{\epsilon\!\cdot\!x_{12}}{(-x_{12}^2+i\varepsilon)^{D/2-1}}\,
\frac{e^{-ik x_1}-e^{-ik x_2}}{k\!\cdot\!x_{12}} \, .
}{}
The integrals in the other two terms simplify to $G(1,0;x)$, and so altogether we have
\eq{
  & \big\langle \Omega\big| \Phi^\dagger(x_1) \Phi(x_2)\big| A(k)\big\rangle  
  = \\
  & \frac{-g \Gamma(D/2-1)}{4\pi^{D/2} (-x_{12}^{2}+i\varepsilon)^{D/2-1}}
  \left[
  \left(\frac{\epsilon \!\cdot\! x_{12}}{k \!\cdot\!x_{12}} - \frac{\epsilon \!\cdot\!u_{1}}{k \!\cdot\!u_{1}}\right) e^{-i k x_{1}}
  -
  \left(\frac{\epsilon \!\cdot\!x_{12}}{k \!\cdot\!x_{12}} - \frac{\epsilon \!\cdot\!u_{2}}{k \!\cdot\!u_{2}}\right) e^{-i k  x_{2}}
  \right]\,.
}{}

\paragraph{Gravity} Now we calculate the Fourier integral in Sec.~\ref{subsec:GR_dressed_tree}:
\eq{
  &\big\langle 0 \big|  \Phi(x_1) \Phi(x_2) \big|  h(k) \big\rangle 
  = -i\kappa \int_{p_1\,p_2}\, e^{-ip_1x_1-ip_2x_2}\times {\delta(p_1+p_2-k)}\\
  &\times \bigg\{
  \frac{(p_1\!\cdot\!\epsilon\!\cdot\! p_2)}{p_1^2p_2^2}
  +
  \frac{1}{p_2^2}
  \frac{-2(p_2\!\cdot\!\epsilon \!\cdot\!u_1)(k\!\cdot\! u_1)+(p_2\!\cdot\! k)(u_1\!\cdot\!\epsilon \!\cdot\!u_1)}{2(k\!\cdot\! u_1)^2}
  +
  \frac{1}{p_1^2}
  \frac{-2(p_1\!\cdot\!\epsilon \!\cdot\!u_2)(k\!\cdot\! u_2)+(p_1\!\cdot\! k)(u_2\!\cdot\!\epsilon\!\cdot\! u_2)}{2(k\!\cdot\!u_2)^2}
  \bigg\}.
}{}
The first term in the brace is, using $k \cdot \epsilon=0$ and $\epsilon^\mu_{\ \mu}=0$,
\eq{
& {-i\kappa} \, e^{-ikx_2}\int_{p_1} 
e^{-i p_1  x_{12}}
\frac{-p_1 \!\cdot\! \epsilon \!\cdot\! p_1}
{(p_1^2 + i\epsilon)((k-p_1)^2 + i\epsilon)}
=
{i\kappa} \, e^{-ikx_2}
(i\partial_{x_{12}} )\!\cdot\!\epsilon \!\cdot\!(i \partial_{x_{12}}) G(1,1;x_{12})
\\
&= -i\kappa \frac{\Gamma\!\left(D/2\right)}
{4\pi^{D/2}} \,
\frac{e^{-ik x_2} -e^{-i k x_1}}{k \!\cdot\!x_{12}}
\frac{(x_{12}\!\cdot\!\epsilon \!\cdot\!x_{12})}
{(-x_{12}^2 + i\epsilon)^{D/2}}\,.
}{}
The integrals in the other two terms reduce to $i \partial_{x} G(1,0;x)$. The full result is
\eq{
&\big\langle 0 \big|  \Phi(x_1) \Phi(x_2) \big| h(k) \big\rangle \\
&=
\begin{aligned}[t]
\frac{i\kappa \Gamma(D/2)}{4\pi^{D/2}\,(-x_{12}^{2}+i\varepsilon)^{D/2}}
\Bigg[\,
&e^{-i k x_{1}}
\left(
 \frac{(x_{12}\!\cdot\!\epsilon\!\cdot\! x_{12})}{k\!\cdot\! x_{12}}
-\frac{2(u_{1}\!\cdot\!\epsilon\!\cdot\! x_{12})}{u_{1}\!\cdot\! k}
+\frac{(k\!\cdot\! x_{12})\,(u_{1}\!\cdot\!\epsilon\!\cdot\!u_{1})}{(u_{1}\!\cdot\! k)^{2}}
\right)\\
-\,
&e^{-i k x_{2}}
\left(
 \frac{(x_{12}\!\cdot\!\epsilon \!\cdot\!x_{12})}{k\!\cdot\! x_{12}}
-\frac{2\,(u_{2}\!\cdot\!\epsilon \!\cdot\!x_{12})}{u_{2}\!\cdot\! k}
+\frac{(k\!\cdot\! x_{12})\,(u_{2}\!\cdot\!\epsilon \!\cdot\!u_{2})}{(u_{2}\!\cdot\! k)^{2}}
\right)\,
\Bigg]\,.
\end{aligned}
}{}

\section{One-Loop Feynman Integrals}
\label{app:one_loop}

The self-energy result depends on three master integrals: two of these are textbook single-scalar integrals which can be found in the reference by Smirnov~\cite{Smirnov2012AnalyticTools}, and the third integral requires a little more work, as it depends on both $p^2$ and $(p\cdot u)^2$. Such scalar integrals have been encountered before in temporal or axial gauge Yang-Mills calculations, and it is known that they can be evaluated via recursion identities \cite{Alekseev1991}.

Contracting the indices in the dressed quantum gravity self-energy in \Eq{grav_self-energy}, we obtain a sum of 23 scalar integrals in the family
\begin{equation}
G(\lambda_{1},\lambda_{2},\lambda_{3})=\int \frac{d^{D}k}{i\pi^{D/2}}\frac{1}{(k^{2}+i\varepsilon)^{\lambda_{1}}[(k-p)^{2}+i\varepsilon]^{\lambda_{2}}(2k\!\cdot\!u+i\varepsilon)^{\lambda_{3}}}\,,
\end{equation}
where we have already discarded a number of integrals which are scaleless and thus vanishing in dimensional regularization.  

In general the integral is a function of both $p^2$ and $p\cdot u$, and since our momentum is off shell, we require the general case of this integral where both of these invariants are nonzero.\footnote{To keep expressions tidy, in what follows we omit the $i\varepsilon$'s  and restore them in the final result.} This general kinematic dependence makes the integral more involved than its specific cases, e.g. when $p\cdot u=0$. There are limiting cases which are accessible via standard Feynman parameter integration, and we'll make use of these later:
\begin{itemize}
    \item  For $\lambda_{1}=0$, we can shift $k\rightarrow k+p$ to obtain an integral in the form of (10.25) in \cite{Smirnov2012AnalyticTools}.
    \item For $\lambda_{2}=0$ the integrals are scaleless.
    \item For $\lambda_{3}=0$ this is a standard bubble integral, and is given in (10.5) of \cite{Smirnov2012AnalyticTools}.
    \item For general values of $\lambda_{j}$, but with $p\cdot u=0$, the integral is given by (10.27) in \cite{Smirnov2012AnalyticTools}.
\end{itemize}

Despite the slight complications, this integral family can be reduced via IBP identities to a simple set of masters. Using the LiteRed program~\cite{Lee:2013mka}  we obtain the masters
\begin{equation}
    \{G(1,1,0), G(0,1,1), G(1,1,1)\}\,.
\end{equation}
The first two integrals are known, as just mentioned above, and so the only remaining unknown scalar integral is $G(1,1,1)$.  In terms of these masters the dressed gravitational self-energy is written
\eq{
\Sigma^{\textrm{dressed}}(p) &= \frac{\kappa^2 }{(4\pi)^2}
(\tilde{\mu}^2e^{\gamma_{E}})^{\epsilon}\bigg[
\frac{p^2\left[p^2(1-2\epsilon)^2+(pu)^2(32-6\epsilon-24\epsilon^2+8\epsilon^3)\right]}{12p\!\cdot \!u}G(0,1,1) \\
&+\frac{p^2\,(1-2\epsilon)(2-\epsilon)\left[2p^2+(1-2\epsilon)(pu)^2\right]}{3}G(1,1,0) \\
&-\frac{p^2(pu)(1-2\epsilon)(2-\epsilon)\left[3p^2-2(pu)^2\epsilon\right]}{3}G(1,1,1)
\bigg]\,.
}{}
with the modified renormalization scale related to the original as usual, $\tilde{\mu}^{2}e^{\gamma_{E}} = 4\pi \mu^{2}$. Since there is no risk of confusion with polarization vectors, here we use $D=4-2\epsilon$.

The known masters are given by
\begin{equation}
    G(1,1,0)=c_{1}(\epsilon)\,(-p^2-i\varepsilon)^{-\epsilon}\,,
\end{equation}
\begin{equation}
    G(0,1,1)=c_{2}(\epsilon)\,(-2p\cdot u-i\varepsilon)^{1-2\epsilon}\,,
\end{equation}
where 
\begin{equation}
    c_1(\epsilon)=\frac{\Gamma(1-\epsilon)^2\Gamma(\epsilon)}{\Gamma(2-2\epsilon)}=\frac{1}{\epsilon} +2-\gamma_{E}+\mathcal{O}(\epsilon^{1})\,,
\end{equation}
\begin{equation}
    c_2(\epsilon)=\Gamma(1-\epsilon)\Gamma(-1+2\epsilon)=-\frac{1}{2\epsilon} -1+\frac{\gamma_{E}}{2}+\mathcal{O}(\epsilon^{1})\,,
\end{equation}
and it remains to compute $G(1,1,1)$ for ourselves. 

This integral was presented in~\cite{Alekseev1991}, and we confirmed their expression with a different method. We evaluate $G(1,1,1)$ by differentiating with respect to the challenging kinematic parameter and using IBP relations to simplify the result,
\begin{equation}
    \frac{d}{dx}G(1,1,1)=\frac{2\epsilon-1}{-p^2+x^2}\left(-G(1,1,0)+\frac{G(0,1,1)}{x}+x G(1,1,1)\right)\,.
\end{equation}
where $x=p\cdot u$. Since the first two masters are known, this is a straightforward first-order linear ODE and can readily be solved. One must be careful though, because $G(0,1,1)$ also depends on $p\cdot u$, so one needs to insert the solution for these known integrals before integrating with respect to $x$.

We obtain a solution to the differential equation,
\eq{
    G(1,1,1)&=C_{0}(p^2,\epsilon)(-p^2)^{-\tfrac{1}{2}+\epsilon}(1-y)^{-\tfrac{1}{2}+\epsilon}+x(-p^2)^{-1-\epsilon}c_1(\epsilon)(1-2\epsilon)\,{}_2F_{1}(1,1-\epsilon,\tfrac{3}{2},y)\\
    &+x(-p^2)^{-1-\epsilon}(-y)^{-\epsilon}c_2(\epsilon){}_2F_{1}(1,1-2\epsilon,\tfrac{3}{2}-\epsilon,y)\,,
}{}
where we have defined $y=x^2/p^2$. The coefficient of integration $C_0(p^2,\epsilon)$ is fixed by taking the $x\rightarrow0$ limit, and matching to the known expression in this limit, (10.27) of \cite{Smirnov2012AnalyticTools}. All together then, our result for the unknown master integral is
\eq{
    G(1,1,1)&=c_{3}(\epsilon)(-p^2)^{-\tfrac{1}{2}-\epsilon}(1-y)^{-\tfrac{1}{2}+\epsilon}+x(-p^2)^{-1-\epsilon}c_1(\epsilon)(1-2\epsilon)\,{}_2F_{1}(1,1-\epsilon,\tfrac{3}{2},y)\\
    &+x(-p^2)^{-1-\epsilon}(-y)^{-\epsilon}c_2(\epsilon){}_2F_{1}(1,1-2\epsilon,\tfrac{3}{2}-\epsilon,y)\,,
}{}
with 
\begin{equation}
    c_3(\epsilon)=-\frac{\Gamma(\frac{1}{2}-\epsilon)^2\Gamma(\frac{1}{2}+\epsilon)\Gamma(\tfrac{1}{2})}{2\Gamma(1-2\epsilon)}=-\frac{\pi^2}{2}+\OO(\epsilon^{1})\,.
\end{equation}
The integral $G(1,1,1)$ is finite in the limit to four dimensions, $\epsilon\rightarrow 0$. The $i\varepsilon$'s can be restored in the above by the replacement rule,
\eq{
-p^2 &\rightarrow -(p^2 +i\varepsilon)\,,\\
x &\rightarrow p\cdot u + i\varepsilon\,,\\
y&\rightarrow \frac{(p\cdot u+i\varepsilon)^2}{p^2+i\varepsilon}\,.
}{}

\section{Relationship between Gauge Conditions}
\label{app:Gammauu_vs_uh}

\noindent
Below we show the equivalence between the gauge conditions
\begin{equation}
\Gamma^\mu{}_{\alpha\beta}(x)\,u^\alpha(x)u^\beta(x)=0
\label{eq:Gammauu_zero_app}
\end{equation}
and
\begin{equation}
u^\mu(x) h_{\mu\nu}(x)=0
\label{eq:uh_zero_app}
\end{equation}
under vanishing boundary conditions for $h_{\mu\nu}$ and the zeroth order geodesic condition $u^\alpha\partial_\alpha u^\mu=0$. For convenience, we start with a few definitions. Let
\begin{equation}
q_\mu(x)=u^\alpha(x) h_{\alpha\mu}(x),
\qquad
s(x)=u^\mu(x)q_\mu(x)=u^\mu(x)u^\nu(x)h_{\mu\nu}(x),
\label{eq:q_s_def_app}
\end{equation}
and define the derivative along the flow of $u^\mu$ by $D=u^\alpha\partial_\alpha$, so that the geodesic condition reads
\begin{equation}
D u^\mu=0.
\label{eq:u_geodesic_app}
\end{equation}

\paragraph{$\bm{u h=0 \ \Rightarrow \ \Gamma u  u=0.}$} 
At linear order, the Levi-Civita connection is
\begin{equation}
\Gamma^\mu{}_{\alpha\beta}
=
\frac{\kappa}{2}
\Big(
\partial_\alpha h^\mu{}_\beta
+
\partial_\beta h^\mu{}_\alpha
-
\partial^\mu h_{\alpha\beta}
\Big).
\label{eq:linear_connection_app}
\end{equation}
Contracting with $u^\alpha u^\beta$ gives
\begin{equation}
\begin{aligned}
\Gamma^\mu{}_{\alpha\beta}u^\alpha u^\beta
& =
\kappa\,u^\alpha u^\beta\partial_\alpha h^\mu{}_\beta
-\frac{\kappa}{2}\,u^\alpha u^\beta\partial^\mu h_{\alpha\beta}\\
& =
\kappa \left[D q^\mu
+ (\partial^\mu u^\alpha)q_\alpha
-\frac{1}{2}\partial^\mu s
\right]
-\kappa(Du^\beta)h^\mu_\beta
.
\end{aligned}
\label{eq:Gammauu_q_s_app}
\end{equation}
The gauge condition $u^\mu h_{\mu\nu}=0$ gives $q^\mu=0$, $s=0$, and together with the geodesic condition $Du^\beta=0$, we have $\Gamma^\mu{}_{\alpha\beta}u^\alpha u^\beta = 0$. 

\paragraph{$\bm{ \Gamma u  u=0 \ \Rightarrow\ u h=0.}$} With $Du^\beta=0$ imposed, contracting \eqref{eq:Gammauu_q_s_app} with $u_\mu$ yields
\begin{equation}
u_\mu \Gamma^\mu{}_{\alpha\beta}u^\alpha u^\beta = \frac{\kappa}{2}Ds.
\label{eq:contract_u_app}
\end{equation}
Now imposing $\Gamma^\mu_{\ \alpha\beta} u^\alpha u^\beta = 0$, we find
\begin{equation}
Ds=0,
\label{eq:Ds_zero_app}
\end{equation}
which means $s$ is a constant along each integral curve of $u^\mu$. With the boundary condition that $h_{\mu\nu}$ vanishes at infinity, it follows that
\begin{equation}
s(x)=0.
\label{eq:s_zero_app}
\end{equation}
Plugging this back into Eq.~\eqref{eq:Gammauu_q_s_app}, the condition $\Gamma^\mu_{\ \alpha\beta} u^\alpha u^\beta = 0$ then reduces to
\begin{equation}
D q^\mu + (\partial^\mu u^\alpha)q_\alpha =0.
\label{eq:q_transport_app}
\end{equation}
This is a homogeneous first-order transport equation along the flow of $u^\mu$. Since $q^\mu$ also vanishes at infinity by virtue of the flat boundary condition, the unique solution is
\begin{equation}
q^\mu(x)=0,
\label{eq:q_zero_app}
\end{equation}
in other words, $u^\mu(x)h_{\mu\nu}(x)=0$.

Therefore, under the geodesic condition $Du^\mu=0$ and the boundary condition that $h_{\mu\nu}$ vanishes at infinity, the gauge conditions $\Gamma^\mu{}_{\alpha\beta}u^\alpha u^\beta=0$ and $u^\mu h_{\mu\nu}=0$ are equivalent. Note that the geodesic condition plays a crucial role in this equivalence: if $D u^\mu \neq 0$, an extra term proportional to $(D u^\beta) h^\mu{}_\beta$ appears in Eq.~\eqref{eq:Gammauu_q_s_app}, and the argument above no longer applies.

\vspace{\baselineskip}

\bibliographystyle{utphys}

\bibliography{references}

\end{document}